\documentclass[aps,prb,twocolumn,amsmath,amssymb,superscriptaddress,letterpaper,longbibliography]{revtex4-1}
\usepackage{times}
\usepackage{amsfonts}
\usepackage{mathrsfs}
\usepackage{graphicx}
\usepackage{dcolumn}
\usepackage{bm}
\usepackage{color}

\usepackage[colorlinks,bookmarks=false,citecolor=blue,linkcolor=red,urlcolor=blue]{hyperref}
\usepackage{ulem} 

\def\be{\begin{equation}}       \def\ee{\end{equation}}
\def\bea{\begin{eqnarray}}      \def\eea{\end{eqnarray}}

\newcommand{\kk}{{\bf k}}
\newcommand{\ket}[1]{\left| #1 \right>} 

\begin{document}

\title{Boundary-obstructed topological high-T$_c$ superconductivity in iron pnictides}


\author{Xianxin Wu}
\affiliation{Department of Physics, the Pennsylvania State University, University Park, PA, 16802, USA}
\affiliation{Beijing National Laboratory for Condensed Matter Physics,
and Institute of Physics, Chinese Academy of Sciences, Beijing 100190, China}
\email{xianxinwu@gmail.com}

\author{Wladimir A. Benalcazar}
\affiliation{Department of Physics, the Pennsylvania State University, University Park, PA, 16802, USA}

\author{Yinxiang Li}
\affiliation {Tin Ka-Ping College of Science, University of Shanghai for Science and Technology, Shanghai, 200093, China}

\author{Ronny Thomale}
\affiliation{Institut f\"ur Theoretische Physik und Astrophysik,
  Julius-Maximilians-Universit\"at W\"urzburg, 97074 W\"urzburg,
  Germany}

\author{Chao-Xing Liu}
\email{cxl56@psu.edu}
\affiliation{Department of Physics, the Pennsylvania State University, University Park, PA, 16802, USA}

\author{Jiangping Hu}
\email{jphu@iphy.ac.cn}
\affiliation{Beijing National Laboratory for Condensed Matter Physics,
and Institute of Physics, Chinese Academy of Sciences, Beijing 100190, China}
\affiliation{CAS Center of Excellence in Topological Quantum Computation and Kavli Institute of Theoretical Sciences,
	University of Chinese Academy of Sciences, Beijing 100190, China}

\date{\today}

\begin{abstract}

Non-trivial topology and unconventional pairing are two central guiding principles in the contemporary search for and analysis of superconducting materials and heterostructure compounds. Previously, a topological superconductor has been predominantly conceived to result from a topologically non-trivial band subject to intrinsic or external superconducting proximity effect. Here, we propose a new class of topological superconductors which are uniquely induced by unconventional pairing. They exhibit a boundary-obstructed higher-order topological character and, depending on their dimensionality,  feature unprecedently robust Majorana bound states or hinge modes protected by chiral symmetry. We predict the 112-family of iron pnictides, such as Ca$_{1-x}$La$_x$FeAs$_2$, to be a highly suited material candidate for our proposal, which can be tested by edge spectroscopy. Because of the boundary-obstruction, the topologically non-trivial feature of the 112 pnictides does not reveal itself for a bulk-only torus band analysis without boundaries, and as such had evaded previous investigations. Our proposal not only opens a new arena for highly stable Majorana modes in high-temperature superconductors, but also provides the smoking gun evidence for extended s-wave order in the iron pnictides.

\end{abstract}

\maketitle

Iron-based high temperature (high T$_c$) superconductors have recently appeared as an exciting platform to realize topological superconductivity at high temperatures\cite{Hao2014,Wu2016,Wang2015,XuPRL2016,Zhang2018,Zhang2019NP,Shi2017,Peng2019,HaoNSR2019}. Due to an intrinsic superconducting proximity effect, the surfaces in these materials can host Majorana zero modes (MZMs), evidence of which has been observed in the vortex core of Fe(Te,Se) and (Li$_{1-x}$Fe$_x$)OHFeSe crystals\cite{YinJX2015,WangDF2018,LiuQ2018,KongLY2019,Machida2019}. That is, a band inversion between the Fe $d$ and the ligand $p$ orbitals is found, which then culminates with the superconducting proximity imposed by the particle-particle instability at the Fermi level. While this band inversion appears to be rather generic for the pnictides, current experimental evidence suggests that the topological superconducting phase necessitates significant tuning of Fermi level and chemical composition.

In unconventional high T$_c$ superconductors, the pairing symmetry is as essential as it can be difficult to directly identify it. In cuprates, only several years after their discovery, a $d$-wave pairing symmetry has been unambiguously proved by detecting the $\pi$ phase shift  in corner junction interferometer experiments\cite{HarlingenRMP1995}. In iron-based superconductors, the pairing symmetry has been subject of a long-lasting debate ever since their discovery a decade ago. An $s_{\pm}$-wave pairing, possessing a sign-reversed gap on hole and electron pockets in momentums space,  has been  proposed for iron-based superconductors with some indirect evidences in neutron scattering and scanning tunneling microscopy \cite{Hirschfeld2011,Christianson2008,Lumsden2009,Christianson2009,QiuYM2009,Hanaguri2010,GrotheS2012}. So far, however, no decisive experiment has been proposed to distinguish $s_{\pm}$-wave from a sign-unchanged $s$-wave pairing because both states share the same A$_{1g}$ symmetry character. The $s_{\pm}$-wave pairing was suggested to realize DIII topological superconductors\cite{ZhangF2013}. As we will show in this work, $s_{\pm}$-wave and $s$-wave, despite their identical symmetry character, can give rise to different topological phases, and thus, we claim that the topological aspects in iron-based high T$_c$ superconductors can shine a light on this outstanding problem\cite{ZhangF2013}.

Higher-order topological phases\cite{ZhangF2013-1,benalcazar2014,benalcazar2017quad,benalcazar2017quadPRB,song2017,langbehn2017} are a new family of phases of matter with the defining property of hosting fractional charges or topological states at corners or hinges of the material. In 2D insulators, these phases manifest fractional corner charges protected by crystalline symmetries and are directly related to the positions of the Wannier centers of the occupied bands\cite{song2017} or to the topology of its Wannier bands~\cite{benalcazar2014,benalcazar2017quad}. In superconductors, despite the absence of a Wannier description, particle-hole and/or chiral symmetries can also protect the existence of corner-localized MZMs. Recently, several proposals have been put forward for the realization of higher-order topological superconductors (HOTSC) in 2D or 3D~\cite{benalcazar2014,YanZB2018,WangQY2018,chen2018,WangYX2018,zhu2018tunable,geier2018second,khalaf2018higher,pan2018lattice,wu2019higher,ZhangRX2019PRL,Volpez2019,Ghorashi2019,wu2019inplane,ZhangRX-PRL20192,Wufetese2019},
some of which include certain iron-based superconductor compounds\cite{WangQY2018,YanZB2018,ZhangRX2019PRL,Wufetese2019,ZhangRX-PRL20192}.
Setting aside that the complete characetrization of HOTSC by topological invariants is still missing, there is, in addition, a significant disconnect between the toy models studied in those works and actual material candidates.

In order to understand the new class of superconductors we propose in this work, we develop a novel fused perspective on the current fields of unconventional superconductivity and higher-order topological states of matter. Specifically, we show that $s_{\pm}$-wave symmetry pairing, together with the topological properties in the 112-family of iron pnictides, drives the material into a HOTSC that hosts a Kramers pair of MZMs at each corner of each unit-layer. Different from previous proposals, the MZMs are particularly robust, as they do not depend on crystal symmetries, but are protected by chiral symmetry.  The demonstration of high-order topology in this family of compounds will provide a "smoking-gun" evidence for $s_{\pm}$-wave pairing. Remarkably, the topological phase we find for this material has the property of being adiabatically connected to the trivial phase in the absence of boundaries, but, once boundaries are introduced, an edge-localized obstruction topologically separates it from the trivial phase. This property, referred to as boundary topological obstruction, was originally identified in the minimal model for a higher-order topological insulator (HOTI) hosting a quantized quadrupole moment protected by reflection symmetries~\cite{benalcazar2017quad,benalcazar2017quadPRB}. In insulators, boundary topological obstructions have recently been explained in terms of the Wannier centers of the occupied bands~\cite{khalaf2019boundary}. Here, we generalize the concept of boundary topological obstructions to superconductors -- for which a Wannier description is absent -- and identify the 112-family of iron pnictides as the first intrinsic material realization of boundary-obstructed HOTSC. The existence of MZMs in boundary-obstructed HOTSCs is a clear signature of its nontrivial topology. Upon a phase transition into a trivial phase, the bulk remains gapped, only the edges become gapless, providing one-dimensional channels for the MZMs to hybridize as they disappear into the trivial phase.

\begin{figure}[tb]
\centerline{\includegraphics[width=1.0\columnwidth]{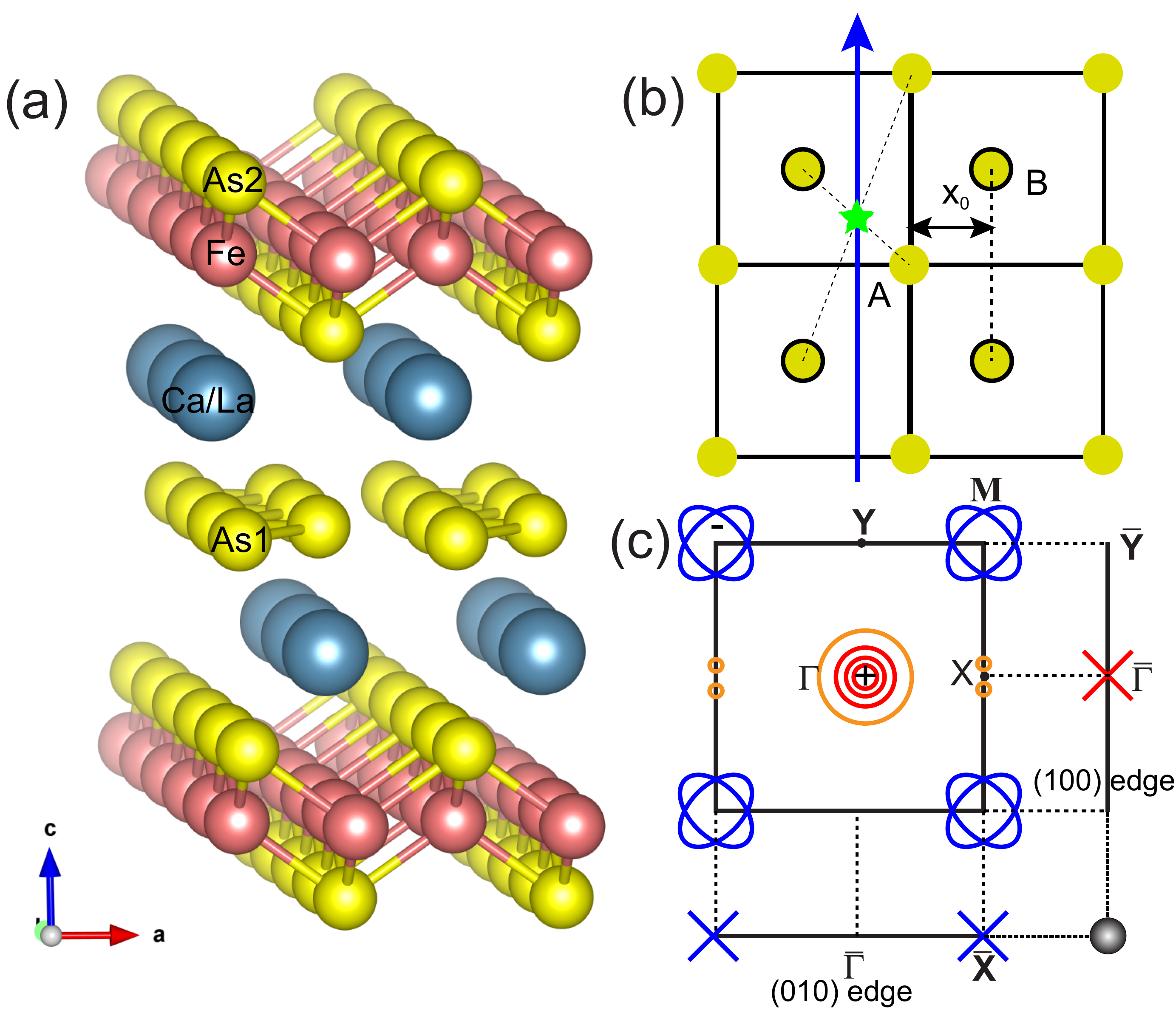}}
\caption{ Crystal structure and Fermi surfaces for (Ca,La)FeAs$_2$. (a) Crystal structure for (Ca,La)FeAs$_2$. (b) Lattice model for the As1 layers, where As1 atoms form a zigzag chain along $y$. (c) Fermi surfaces and pairing gap functions for (Ca,La)FeAs$_2$. The red and blue curves represent the Fermi surfaces from FeAs layers in the presence of $s_{\pm}$ pairing. The superconducting gaps possess a sign change between hole pockets around ${\bf \Gamma}$ and electron pockets around ${\bf M}$. The orange curves are Fermi surfaces from As1 layers. The edge Dirac cone from As1 layers acquire a positive (negative) superconducting gap for the (100) edge ( (010) edge) in proximity to the bulk hole (electron) pockets around ${\bf \Gamma}$ (${\bf M}$). There is a Majorana Kramers pair (gray circles) at each corner where two edges meet.
 \label{crystal} }
\end{figure}

The detection of MZMs in this material would be a decisive evidence for $s_{\pm}$-wave paring in iron-based superconductors for the following reason: The 112-family of iron pnictides, including Ca$_{1-x}$La$_x$FeAs$_2$\cite{Katayama2013} and (Ca,Pr)FeAs$_2$\cite{Yakita2013}, with T$_c$ up to 47 K\cite{Kudo2014}, are intrinsic topological insulator/high T$_c$ superconductor heterostructures\cite{WuXX2014,WuXX2015-PRB} with a staggered intercalation between zigzag chainlike As1 layers with the quantum spin Hall state and the superconducting Fe-As layers along the $c$ axis, as shown in Fig.\ref{crystal}(a). The edge Dirac cones from the As1 layers at two orthogonal (100) and (010) edges are in proximity to projections of bulk pockets around ${\bf \Gamma}$ and ${\bf M}$ from adjacent FeAs layers, respectively. The $s_{\pm}$-wave pairing with opposite gap functions on pockets around ${\bf \Gamma}$ and ${\bf M}$ points will create the Majorana Kramers pairs at corners, as demonstrated in Fig.\ref{crystal}(c).

In what follows, we first investigate $s_{\pm}$ pairing in Ca$_{1-x}$La$_x$FeAs$_2$ and relate the bulk spectrum to that of its edges, which gives rise to the mechanism that realizes the corner MZMs. We then demonstrate that under $s_{\pm}$ pairing, the topological phase in this material is boundary-obstructed by showing that across the topological phase transition (TPT) between the topological and trivial phases, the energy gap only closes at the boundary of a slab configuration, while it remains open in the bulk. Finally, since the two phases are protected by chiral symmetry, we propose a new quantity, the edge winding number, as the invariant that captures the corresponding topological obstruction, and show that this invariant jumps by an integer across a TPT.

\textit{s$_{\pm}$ pairing in Ca$_{1-x}$La$_x$FeAs$_2$ --} We take CaFeAs$_2$ as a typical example in 112-type. Besides the hole pockets around ${\bf \Gamma}$ and electron pockets around ${\bf M}$ from FeAs layers, there are additional Fermi surfaces in CaFeAs$_2$ attributed to the zig-zag As1 layers (Fig.\ref{crystal}(c)). As the correlation effect is relatively weak in As atoms, the pairing state of CaFeAs$_2$ is expected to be dominantly determined by the Fermi surfaces from FeAs layers. Here we adopt a five-band model whose band structure fits well with those in DFT calculations [see Sec. I in the supplementary material (SM)]. Fig.\ref{edgebands}(a) displays the spin susceptibility. The peak around $(\pi,0)$ is attributed to the Fermi surface nesting between the hole and electron pockets. Consequently, the dominant pairing is $s_{\pm}$-wave from the effective repulsive electron-electron interactions mediated by spin fluctuations\cite{Hirschfeld2011,Thomale2011}. Fig.\ref{edgebands}(b) shows the typical gap function of $s_{\pm}$ pairing from random phase approximation (RPA) calculations for CaFeAs$_2$, revealing a sign change in superconducting gaps between hole and electron pockets (see Sec. II in SM). The sign change and gap size can also be well described by a simple form factor $\cos k_x\times \cos k_y$ in one-Fe unit cell originating from the next-nearest neighbor antiferromagnetic exchange coupling\cite{Seo2008}. Unlike the $d$-wave state in cuprates, in which a sign change occurs in orthogonal directions, $s_{\pm}$ pairing with a sign change in momentum space is extremely difficult to be detected via Josephson interferometry. Although some evidence for the existence of sign change in pairing states have been provided in  inelastic neutron scattering and scanning tunneling microscopy measurements\cite{Hirschfeld2011,Christianson2008,Lumsden2009,Christianson2009,QiuYM2009,Hanaguri2010,GrotheS2012}, a decisive signature for $s_{\pm}$-pairing is still missing in iron pnictides. In the following, we show that $s_\pm$ pairing gives rise to a HOTSC phase in CaFeAs$_2$ with Majorana corner states, which can be regarded as the direct evidence for $s_{\pm}$-pairing.

 \begin{figure}[tb]
\centerline{\includegraphics[width=1.0\columnwidth]{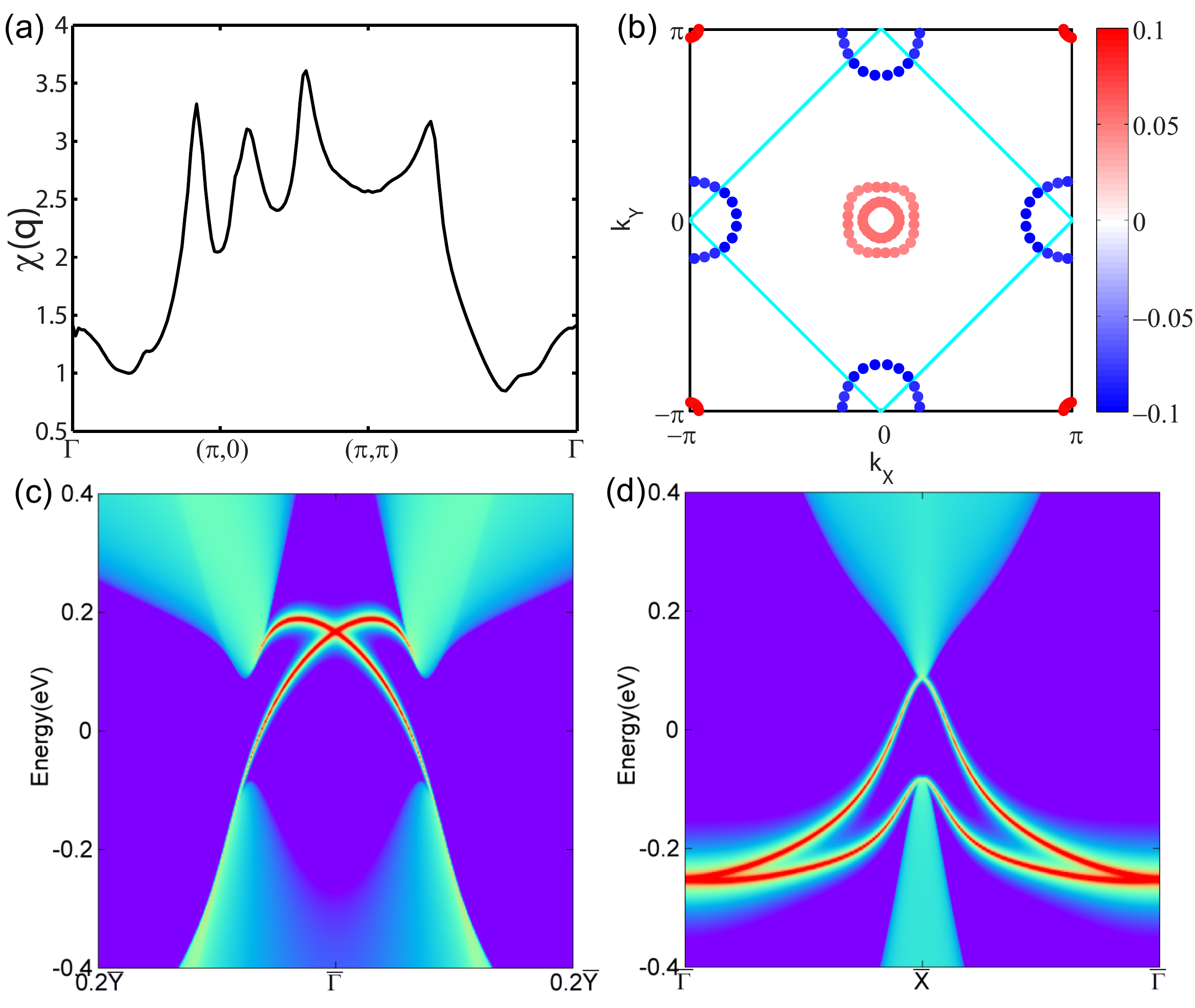}}
\caption{ Spin susceptibility and $s_{\pm}$-wave gap function for CaFeAs$_2$ and edge states for As1 layers. (a) Distribution of the largest eigenvalues for RPA spin susceptibility matrices. (b) Dominant $s_{\pm}$-wave sate from RPA calculations in one-Fe Brillouin zone. The interaction parameters are $U=1.4$ eV and $J/U=0.1$ with Kanamori relations $U=U'+2J$ and $J=J'$. The blue lines represent the Brillouin zone for two-Fe unit cell. Edge states for As1 layers in CaFeAs$_2$: (c) (100) edge, (d) (010) edge.
 \label{edgebands} }
\end{figure}

\textit{Effective Model for As1 layers --} We start with the tight-binding model for the As1 layers. A two-dimensional four-band model on a square lattice (Fig.\ref{crystal}(b)) has been derived to capture the band structure attributed to $p_x$ and $p_y$ orbitals of two As1 atoms in one unit cell.  We introduce the operator $\psi^\dag_{\textbf{k}\sigma}=(c^\dag_{Ax\sigma}(\textbf{k}),c^\dag_{Ay\sigma}(\textbf{k}),c^\dag_{Bx\sigma}(\textbf{k}),c^\dag_{By\sigma}(\textbf{k}))$, where $c^\dag_{\alpha\eta\sigma}(\textbf{k})$ is a Fermionic creation operator with  $\sigma$,  $\eta$ and $\alpha$ being spin, orbital and sublattice indices respectively. The tight-binding Hamiltonian reads
\begin{eqnarray}
\mathcal{H}_0=\sum_{\textbf{k}\sigma}\psi^\dag_{\textbf{k}\sigma}h(\textbf{k})\psi_{\textbf{k}\sigma}.
\end{eqnarray}
The matrix elements in the Hamiltonian $h(\textbf{k})$ are provided in Sec. III in the SM. A band inversion occurs at the ${\bf X}$ point and it generates two gapless Dirac cones in the bulk dispersion along ${\bf X}$-${\bf M}$ path without spin-orbit coupling (SOC), protected by the screw axis along $y$ \cite{WuXX2014,WuXX2015-PRB}. Two small pockets appear around the ${\bf X}$ point deriving from these cones, demonstrated by orange circles in Fig.\ref{crystal}(c) and supported by ARPES experiments\cite{LiMY2015,JiangS2016,LiuZT2016}. Further including SOC opens a gap in the Dirac cones and the As1 layers becomes $\mathbb{Z}_2$ topologically nontrivial, leading to an intrinsic topological insulator/high T$_c$ superconductor heterostructure in CaFeAs$_2$. Around the ${\bf X}$ point, the bulk dispersion of As1 layers can be described by an effective Hamiltonian $\mathcal{H}^{\bf X}_\text{eff}=\sum_{\bm{k}}\tilde{\psi}^\dag_{\bm{k}}h_\text{eff}(\mathbf{k})\tilde{\psi}_{\bm{k}} $ with basis $\tilde{\psi}^\dag_{\bm{k}}=(c^\dag_{X\bm{k},-\uparrow},c^\dag_{X\bm{k},+\uparrow},c^\dag_{X\bm{k},-\downarrow},c^\dag_{X\bm{k},+\downarrow})$, where "+/-" denotes the eigenvalue of $C_{2z}$ for eigenstates at the ${\bf X}$ point and
\begin{eqnarray}
 h_\text{eff}(\bm{k})=\epsilon_0(\mathbf{k}) + M(\mathbf{k})\sigma_z-A_1k_xs_0\sigma_2 + A_2k_ys_3\sigma_1.
\end{eqnarray}
Here $\epsilon_0(\mathbf{k})=C_0+C_1k^2_x+C_2k^2_y$ and $M(k)=M_0-B_1k^2_x-B_2k^2_y$, and $\bm{\sigma}$ and $\bm{s}$ are Pauli matrices in orbital and spin degrees of freedom. The parameters in the model are given in Sec. IV of SM. The edge states can be obtained by solving a semi-infinite system along $x$ or $y$ direction within tight-binding model. For the (100) edge, a distorted Dirac cone appears around the $\bar{\Gamma}$ point located above the Fermi level, as shown in Fig.\ref{edgebands}(c), while, for the (010) edge, a Dirac cone occurs around the $\bar{X}$ point and is embedded in the bulk conduction bands, as shown in Fig.\ref{edgebands}(d). We emphasize that the Fermi level for both edges only crosses the lower part of the edge Dirac cone within a relatively large electron doping region from the substitution of La/Pr for Ca in experiments.

 Below T$_c$, the As1 layers become superconducting through the proximity effect to the adjacent FeAs layers. We model the superconducting pairing on As1 layers the same way as in FeAs layers and consider a spin-singlet intra-orbital pairing within the same sublattice. The corresponding pairing Hamiltonian reads
 \begin{eqnarray}
\mathcal{H}_\text{SC}&\!\!=&\!\!\!\sum_{\alpha \nu\sigma\bm{k}}\!\!\! \sigma[\Delta_0+2\Delta_1(cosk_x+cosk_y)]\nonumber\\
&&c^\dag_{\alpha\nu\sigma}(\bm{k})c^\dag_{\alpha\nu\bar{\sigma}}(-\bm{k})
+h.c.,
\end{eqnarray}
where $\Delta_0$ and $\Delta_1$ are the on-site pairing and pairing between the next nearest neighbor sites, respectively. The $\Delta_1$ term gives rise to the well-known $s_{\pm}$-wave pairing in iron based superconductors. Owing to the absence of spin-flip SOC terms, the Bogoliubov-de Gennes (BdG) tight-binding Hamiltonian $\mathcal{H}_\text{BdG}=\mathcal{H}_0+\mathcal{H}_\text{SC}$ can be written as two block diagonal parts $\mathcal{H}_\text{BdG}^{\uparrow\downarrow}$ and $\mathcal{H}_\text{BdG}^{\downarrow\uparrow}$ (see Sec. V in SM). In each block, time reversal and particle-hole symmetries are absent but chiral symmetry is preserved, thus, each block belongs to class AIII in the ten fold classification.

 \begin{figure*}[t]
\centerline{\includegraphics[width=1.7\columnwidth]{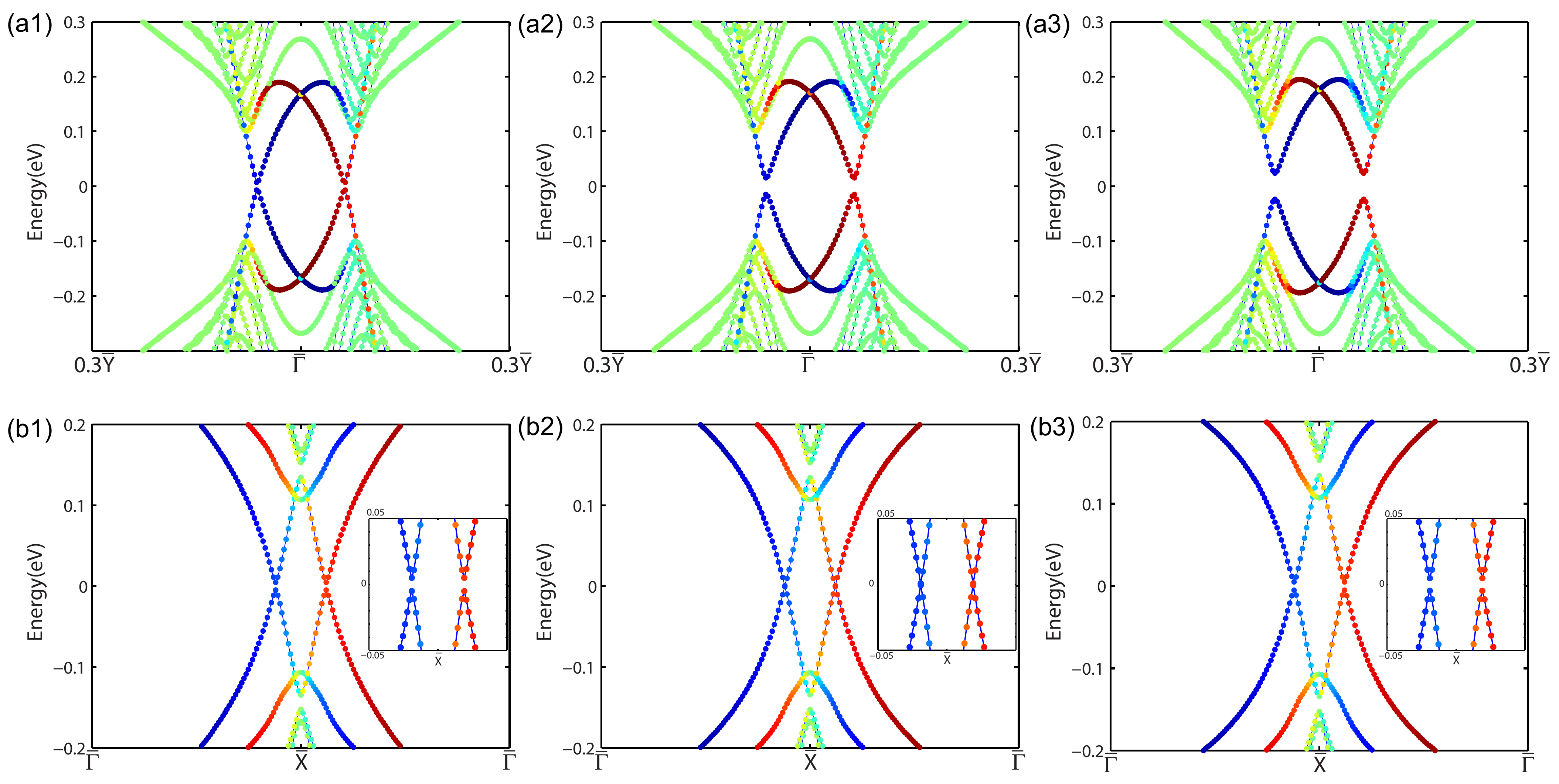}}
\caption{ Evolution of band structures for (100) edge (a) and (010) edge (b) as a function of the $s_{\pm}$ pairing $\Delta_1$. The plots in each column are generated with the same parameters and the adopted parameters in three columns (from left to right) are: $\Delta_1= 0$, $\Delta_1= 51$ meV and $\Delta_1=100$ meV with a fixed $\Delta_0=5$ meV. The color code shows the average position of each state. The blue and red circles denote the left (top) and right (bottom) localized edge states at the (100) ((010)) edges, respectively. The green circles represent extended bulk states.
 \label{edgetransition} }
\end{figure*}

\textit{Boundary-obstruction and phase transitions --} To study the edge properties of the As1 layers, we consider a slab configuration for the above BdG Hamiltonian $\mathcal{H}^{\uparrow\downarrow}_\text{BdG}$ with an open boundary along either $x$ or $y$ directions. With the on-site pairing term $\Delta_0$, the edge states from As1 layers open a gap, as shown in Fig.\ref{edgetransition} (a1) and (b1). Now we investigate the effect of $s_{\pm}$-wave pairing on these edge states. At the (100) edge, the gap of edge states around $\bar{\Gamma}$ monotonously increases with increasing $\Delta_1$, as shown in Fig.\ref{edgetransition} (a1) to (a3). For the (010) edge, however, the gap around $\bar{X}$ exhibits a rather different behavior. Upon $\Delta_1$ increasing to 0.051 eV, the gap closes and a pair of gapless modes with linear dispersion appears [Fig. \ref{edgetransition} (b2)]. Further increasing $\Delta_1$ reopens the gap, suggesting a TPT that separates a $\Delta_0$-dominated phase from a $\Delta_1$-dominated phase. Note that throughout this transition there is no gap closing in the bulk states (see Sec. VIII in SM). In fact, we show in Sec. VII of the SM that all the symmetry indicator invariants due to the $C_{2z}$ and reflection symmetries in the lattice structure of $\mathcal{H}_\text{BdG}$ identically vanish due to time-reversal symmetry, which is a necessary condition for the existence of boundary-obstructed phases.

To calculate the effective pairing at the edges, we first analytically obtain the wavefunctions of the edge states by solving $\mathcal{H}_\text{eff}$ with open boundary conditions and then projecting the bulk pairing on the edge states\cite{YanZB2018,ZhangSB2019}. The obtained pairings at the Dirac points on the (100) and (010) edges are
\begin{eqnarray}
\Delta^{(100)}_\text{eff}&=&\Delta_0+2\Delta_1\frac{M_0}{B_1},\\
\Delta^{(010)}_\text{eff}&=&\Delta_0-2\Delta_1\frac{M_0}{B_2}.
\end{eqnarray}
 We find that $\Delta_0$ provides the same pairing at the two edges while the $s_{\pm}$-wave pairing $\Delta_1$ provides an opposite pairing sign. This can be heuristically understood from Fig.\ref{crystal}(c). The edge Dirac states appear around $\bar{\Gamma}$ and $\bar{X}$ for (100) and (010) edges; the momentum independent term proportional to $\Delta_0$ induces the same positive superconducting gap at both edges; the $s_{\pm}$-wave pairing, on the other hand, induces superconducting gaps with opposite signs as the corresponding Dirac cones are in proximity to the bulk superconducting gap around ${\bf \Gamma}$ and ${\bf M}$ at the (100) and (010) edges, respectively (see Fig.\ref{crystal}(c)). At a fixed non-zero value of $\Delta_0$, with increasing $\Delta_1$, $\Delta^{(100)}_\text{eff}$ monotonously increases while $\Delta^{(010)}_\text{eff}$ first decreases to zero, followed by increasing in amplitude albeit with opposite sign, which is consistent with the aforementioned numerical calculations. The finite chemical potential at the edges with respect to Dirac points should be taken into consideration but their relative small values have a negligible effect on the effective pairing (see Sec. VI in SM).

\begin{figure}[tb]
\centerline{\includegraphics[width=1.0\columnwidth]{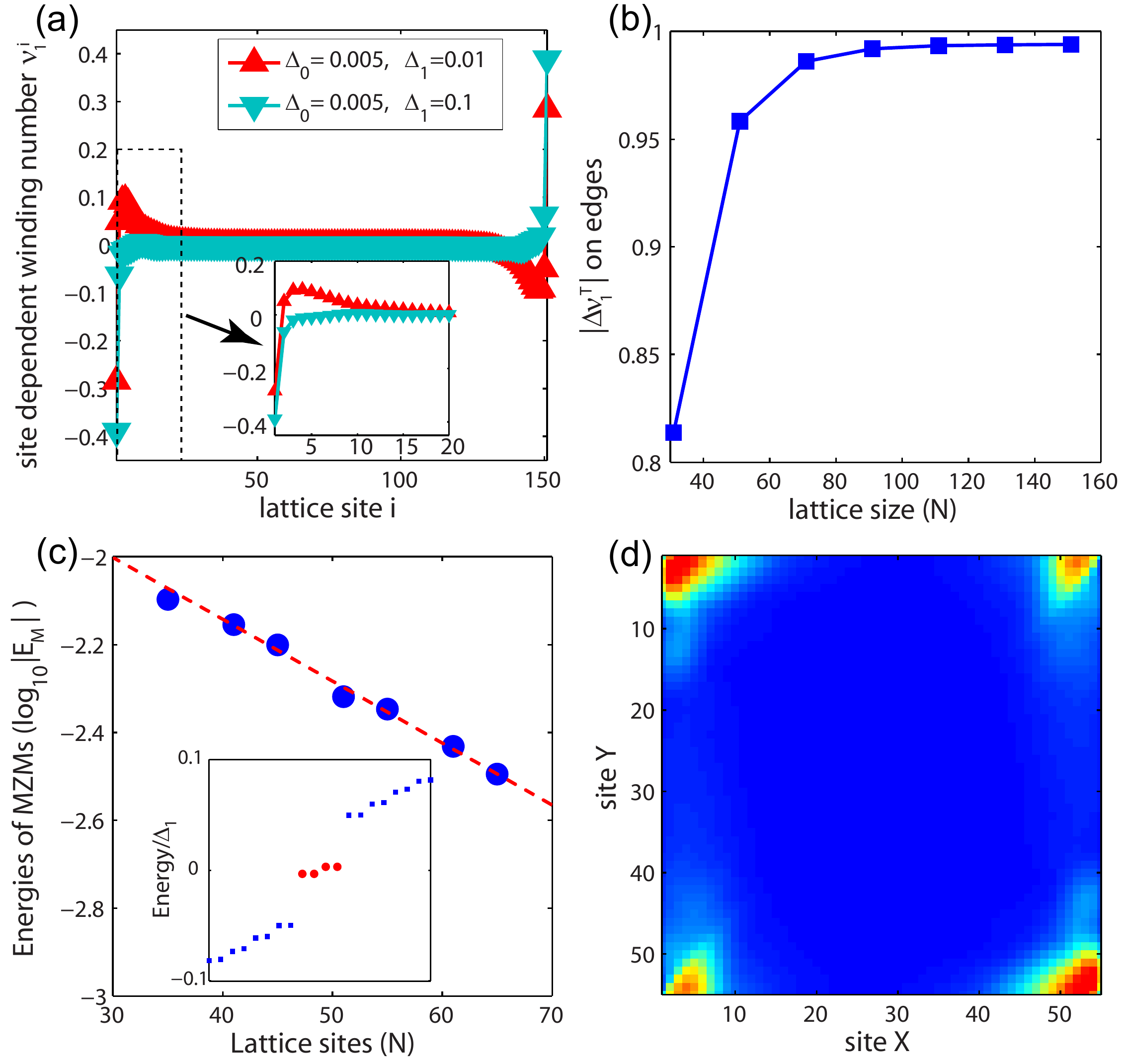}}
\caption{ Edge winding number and energy and spatial pattern for MZMs. (a) Site-resolved winding numbers for the (010) edge with $N= 150$ unit cells along $y$ in the two topologically distinct phases. (b) Winding number difference $\Delta \nu^T_1$ as a function of the lattice size $N$. (c) Energy of the MZMs as a function of the lattice size for a square geometry. The inset shows a typical energy spectra for the full lattice. (d) The probability density functions of the MZMs for a square geometry with eight MZMs (each corner hosts a Kramers pair of MZM) with $65\times65$ lattice sites.
 \label{MZMwav} }
\end{figure}

To characterize the topological nature of the TPT, let us focus on the blocks $\mathcal{H}_\text{BdG}^{\uparrow\downarrow}$, and $\mathcal{H}_\text{BdG}^{\downarrow\uparrow}$, each of which belongs to class AIII. Hamiltonians in this class are topologically characterized by the 1D winding number\cite{Chiu2016}, defined by $\nu_1=\frac{i}{2\pi}\int_\text{BZ}dk Tr[q^{-1}_{\bm{k}}\partial_kq_{\bm{k}}]$. Here the unitary $q_k$ matrix is the off-diagonal part of the so-called $Q$ matrix, given by
$Q_{\bm{k}}=\bm{1}-2P_{\bm{k}}=\left(
\begin{array}{cc}
0 & q_{\bm{k}}  \\
q^\dag_{\bm{k}}  & 0
\end{array}
\right)$ within the eigenbasis of chiral symmetry, where $P_{\bm{k}}$ is the projection operator of the Hamiltonian for a slab model $\mathcal{H}^{\uparrow\downarrow}_{BdG,slab}$ with $N$ lattice sites. We first consider the winding number on a slab configuration with open boundary along the (010) direction. The total winding number $\nu_1$ is zero across the TPT (see Sec. VIII in SM).  Motivated by the fact that the bands close at the (010) edges of the slab during the TPT, we define a site-resolved winding number by projecting the total winding number $\nu_1$ into the lattice site basis as
\begin{eqnarray}
\nu^i_1=\frac{i}{2\pi}\sum_{\gamma}\int_\text{BZ}dk[q^{-1}_{\bm{k}}\partial_kq_{\bm{k}}]_{i\gamma,i\gamma},
\end{eqnarray}
such that $\nu_1=\sum_{i=1}^N\nu^i_1$. Here, $i$ is the index for lattice site and $\gamma$ denotes the sublattice or orbital index (details are given Sec. VIII in SM). This site-resolved winding number resembles the site-resolved polarization defined in Ref.~\onlinecite{benalcazar2017quadPRB} to calculate the edge-localized dipole moments in insulators, including the boundary-obstructed quadrupole topological insulator~\cite{benalcazar2017quad}. The dependence of $\nu^i_1$ on the site $i$ for the HOTSC and trivial phases is displayed by the blue and red curves in Fig.\ref{MZMwav}(a). For both curves, one can see that the contribution to the winding number mainly comes from edges, with the opposite edges having opposite contributions. We also notice that the profile of $\nu^i_1$ near one edge (see the zoom-in Fig.\ref{MZMwav}(a)) has a substantial difference between the HOTSC and trivial phases. We define the edge winding number $\nu^T_1$ and $\nu^B_1$ for the top and bottom edges, respectively, by summing $\nu^i_1$ for the upper half part ($i=1,...,N/2$) and lower half part ($i=N/2+1,...,N$) of the slab and examine the winding number change $|\Delta\nu^{T/B}_{1}|$ between two phases for the upper and lower half parts as a function of the lattice size $N$. As shown in Fig.\ref{MZMwav}(b), we find that the winding number change $|\Delta\nu^{T}_{1}|$ approaches 1 in the thermodynamic limit ($N\rightarrow \infty$). Thus, the edge winding number $\nu^{T/B}_1$ characterizes the different topologies across the TPT in our new class of superconductor.

\textit{Majorana corner states --} In the $\Delta_1$ dominated phase, the (100) and (010) edges have the opposite superconducting gaps, belonging to topologically distinct phases. As a consequence, MZMs are expected to occur at the corners where they meet. We adopted the Hamiltonian $\mathcal{H}_\text{BdG}$ with both open boundary conditions along $x$ and $y$ directions, and performed calculations with several lattice sizes. There are eight mid-gap states and their energies are zero up to finite-size effects. We examine the energy splitting due to the hybridization of the MZMs at different corners by plotting the energies $E_\text{M}$ of mid-gaps states as a function of lattice size in Fig.\ref{MZMwav}(c). The linear relationship between $log_{10}(E_m)$ and lattice size suggests that $E_\text{M}$ goes to zero in the thermodynamic limit, compatible with the existence of zero-energy states exponentially localized at the corners of the lattice, as shown in Fig.\ref{MZMwav}(d). At each corner, there are two zero-energy states forming a Majorana Kramers pair. The appearance of MZMs at the corners supports our prediction of a HOTSC phase with boundary-obstruction in CaFeAs$_2$.

\textit{Discussion --}
As the 112-family of iron pnictides is 3D and the interlayer coupling is relatively weak, the helical Majorana states, localized at the hinges between (100) and (010) surfaces, have a weak dispersion along $k_z$. MZMs can also appear at corners/hinges between non-orthogonal surfaces, as long as the corresponding surface states have the opposite effective superconducting gap in proximity to the bulk $s_{\pm}$ pairing. Therefore, the $s_{\pm}$ pairing is directly manifested by the appearance of MZMs at hinges, protected by chiral symmetry and irrespective of crystalline symmetries. We contrast our proposal with that of Ref.~\onlinecite{ZhangF2013}. In it, a heteroustructure couples Rashba semiconductors with nodeless iron-based superconductors. Unlike our proposal, the realization of such a structure would require a large Rashba SOC. It is also worth noting that the toy model in Ref. \onlinecite{WangQY2018}, realizes a bulk-and boundary-obstructed HOTSC with and without the fourfold rotational symmetry, respectively. The 112-family of iron pnictides to some extent can be considered as a realization of the toy model in Ref. \onlinecite{WangQY2018} with boundary-obstructed HOTSC. The MZMs we find are robust against impurities and disorders in real material scenarios that always break crystalline symmetries. In realistic materials, the symmetries are reduced compared with the adopted model. Under this scenario, SOC terms could in principle appear that flip the spin sectors, however, the Majorana Kramers pairs will survive as they are protected by time reversal symmetry. The detection of hinge MZMs in CaFeAs$_2$ hence provides a ``smoking-gun" evidence for the $s_{\pm}$ pairing in iron pnictides.  As the Fermi level only crosses the lower part of edge Dirac cones for edges in As1 layers within a wide electron-doping range, the MZMs can survive in electron-doped compounds Ca$_{1-x}$La$_x$FeAs$_2$, providing a new high-temperature platform for MZMs without fine-tuning.

In particular, due to the weak interlayer coupling in 112-family of iron pnictides, the cleavage occurs between Ca/La layers and As1 layers or FeAs layers, generating step edges on (001) surfaces. At the ends of step edges, which can be viewed as corners of As1 layers, MZMs can give rise to a zero-bias peak in the  scanning tunneling microscopy measurements (see Sec. VIII in SM) and their localized nature can be manifested in spatial zero-energy mapping (see Fig.\ref{MZMwav}(d)). Hinge MZMs are expected to induce a zero-bias anomaly in transport measurements with junctions in contact with the corresponding hinge\cite{Gray2019}. The Majorana Kramers pair can split under an in-plane magnetic field. As the Sb doping can enhance both T$_c$ and the quantum spin Hall gap of the As1 layer\cite{Kudo2014}, Ca$_{1-x}$La$_x$Fe(As$_{1-y}$Sb$_y$)$_2$ can be a good choice for samples in experimental measurements.

\textit{Conclusion --} We propose the 112 family of iron pnicitides as the first material realization of boundary-obstructed topological superconductivity, owing to their intrinsic $s_{\pm}$ pairing and effective topological insulator/high-T$_c$ superconductor heterostructure profile. The edge topological obstruction, indepedent of crystalline symmetries and uniquely characterized by the edge winding number, provides a robust platform for the realization of MZMs, which en passant also constitutes decisive evidence for  $s_{\pm}$-wave pairing in the 112 pnictides.

{\it Acknowledgements:}
We thank  J. Yu and N. Hao for helpful discussions. W.A.B. thanks the support of the Eberly Postdoctoral Fellowship at the Pennsylvania State University. C.X.Liu acknowledges the support of the Office of Naval Research (Grant No. N00014-18-1-2793), DOE grant (DE-SC0019064) and Kaufman New Initiative research grant of the Pittsburgh Foundation.  J. Hu is supported by the Ministry of Science and Technology of China 973 program (Grant No.~2017YFA0303100), National Science Foundation of China (Grant No. NSFC-11888101), and the Strategic Priority Research Program of CAS (Grant No. XDB28000000). The work in W\"urzburg is funded by the Deutsche
Forschungsgemeinschaft (DFG, German Research Foundation) through
Project-ID 258499086 - SFB 1170 and through the W\"urzburg-Dresden Cluster of Excellence on Complexity and Topology in Quantum Matter --\textit{ct.qmat} Project-ID 39085490 - EXC 2147.

\bibliography{HOSCspm}

\begin{thebibliography}{67}%
\makeatletter
\providecommand \@ifxundefined [1]{%
 \@ifx{#1\undefined}
}%
\providecommand \@ifnum [1]{%
 \ifnum #1\expandafter \@firstoftwo
 \else \expandafter \@secondoftwo
 \fi
}%
\providecommand \@ifx [1]{%
 \ifx #1\expandafter \@firstoftwo
 \else \expandafter \@secondoftwo
 \fi
}%
\providecommand \natexlab [1]{#1}%
\providecommand \enquote  [1]{``#1''}%
\providecommand \bibnamefont  [1]{#1}%
\providecommand \bibfnamefont [1]{#1}%
\providecommand \citenamefont [1]{#1}%
\providecommand \href@noop [0]{\@secondoftwo}%
\providecommand \href [0]{\begingroup \@sanitize@url \@href}%
\providecommand \@href[1]{\@@startlink{#1}\@@href}%
\providecommand \@@href[1]{\endgroup#1\@@endlink}%
\providecommand \@sanitize@url [0]{\catcode `\\12\catcode `\$12\catcode
  `\&12\catcode `\#12\catcode `\^12\catcode `\_12\catcode `\%12\relax}%
\providecommand \@@startlink[1]{}%
\providecommand \@@endlink[0]{}%
\providecommand \url  [0]{\begingroup\@sanitize@url \@url }%
\providecommand \@url [1]{\endgroup\@href {#1}{\urlprefix }}%
\providecommand \urlprefix  [0]{URL }%
\providecommand \Eprint [0]{\href }%
\providecommand \doibase [0]{http://dx.doi.org/}%
\providecommand \selectlanguage [0]{\@gobble}%
\providecommand \bibinfo  [0]{\@secondoftwo}%
\providecommand \bibfield  [0]{\@secondoftwo}%
\providecommand \translation [1]{[#1]}%
\providecommand \BibitemOpen [0]{}%
\providecommand \bibitemStop [0]{}%
\providecommand \bibitemNoStop [0]{.\EOS\space}%
\providecommand \EOS [0]{\spacefactor3000\relax}%
\providecommand \BibitemShut  [1]{\csname bibitem#1\endcsname}%
\let\auto@bib@innerbib\@empty
\bibitem [{\citenamefont {Hao}\ and\ \citenamefont {Hu}(2014)}]{Hao2014}%
  \BibitemOpen
  \bibfield  {author} {\bibinfo {author} {\bibfnamefont {Ningning}\
  \bibnamefont {Hao}}\ and\ \bibinfo {author} {\bibfnamefont {Jiangping}\
  \bibnamefont {Hu}},\ }\bibfield  {title} {\enquote {\bibinfo {title}
  {{Topological Phases in the Single-Layer FeSe}},}\ }\href {\doibase
  10.1103/PhysRevX.4.031053} {\bibfield  {journal} {\bibinfo  {journal} {Phys.
  Rev. X}\ }\textbf {\bibinfo {volume} {4}},\ \bibinfo {pages} {031053}
  (\bibinfo {year} {2014})}\BibitemShut {NoStop}%
\bibitem [{\citenamefont {Wu}\ \emph {et~al.}(2016)\citenamefont {Wu},
  \citenamefont {Qin}, \citenamefont {Liang}, \citenamefont {Fan},\ and\
  \citenamefont {Hu}}]{Wu2016}%
  \BibitemOpen
  \bibfield  {author} {\bibinfo {author} {\bibfnamefont {Xianxin}\ \bibnamefont
  {Wu}}, \bibinfo {author} {\bibfnamefont {Shengshan}\ \bibnamefont {Qin}},
  \bibinfo {author} {\bibfnamefont {Yi}~\bibnamefont {Liang}}, \bibinfo
  {author} {\bibfnamefont {Heng}\ \bibnamefont {Fan}}, \ and\ \bibinfo {author}
  {\bibfnamefont {Jiangping}\ \bibnamefont {Hu}},\ }\bibfield  {title}
  {\enquote {\bibinfo {title} {{Topological characters in
  Fe(Te$_{1-x}$Se$_{x}$) thin films}},}\ }\href {\doibase
  10.1103/PhysRevB.93.115129} {\bibfield  {journal} {\bibinfo  {journal} {Phys.
  Rev. B}\ }\textbf {\bibinfo {volume} {93}},\ \bibinfo {pages} {115129}
  (\bibinfo {year} {2016})}\BibitemShut {NoStop}%
\bibitem [{\citenamefont {Wang}\ \emph {et~al.}(2015)\citenamefont {Wang},
  \citenamefont {Zhang}, \citenamefont {Xu}, \citenamefont {Zeng},
  \citenamefont {Miao}, \citenamefont {Xu}, \citenamefont {Qian}, \citenamefont
  {Weng}, \citenamefont {Richard}, \citenamefont {Fedorov}, \citenamefont
  {Ding}, \citenamefont {Dai},\ and\ \citenamefont {Fang}}]{Wang2015}%
  \BibitemOpen
  \bibfield  {author} {\bibinfo {author} {\bibfnamefont {Zhijun}\ \bibnamefont
  {Wang}}, \bibinfo {author} {\bibfnamefont {P.}~\bibnamefont {Zhang}},
  \bibinfo {author} {\bibfnamefont {Gang}\ \bibnamefont {Xu}}, \bibinfo
  {author} {\bibfnamefont {L.~K.}\ \bibnamefont {Zeng}}, \bibinfo {author}
  {\bibfnamefont {H.}~\bibnamefont {Miao}}, \bibinfo {author} {\bibfnamefont
  {Xiaoyan}\ \bibnamefont {Xu}}, \bibinfo {author} {\bibfnamefont
  {T.}~\bibnamefont {Qian}}, \bibinfo {author} {\bibfnamefont {Hongming}\
  \bibnamefont {Weng}}, \bibinfo {author} {\bibfnamefont {P.}~\bibnamefont
  {Richard}}, \bibinfo {author} {\bibfnamefont {A.~V.}\ \bibnamefont
  {Fedorov}}, \bibinfo {author} {\bibfnamefont {H.}~\bibnamefont {Ding}},
  \bibinfo {author} {\bibfnamefont {Xi}~\bibnamefont {Dai}}, \ and\ \bibinfo
  {author} {\bibfnamefont {Zhong}\ \bibnamefont {Fang}},\ }\bibfield  {title}
  {\enquote {\bibinfo {title} {{Topological nature of the
  FeSe$_{0.5}$Te$_{0.5}$ superconductor}},}\ }\href {\doibase
  10.1103/PhysRevB.92.115119} {\bibfield  {journal} {\bibinfo  {journal} {Phys.
  Rev. B}\ }\textbf {\bibinfo {volume} {92}},\ \bibinfo {pages} {115119}
  (\bibinfo {year} {2015})}\BibitemShut {NoStop}%
\bibitem [{\citenamefont {Xu}\ \emph {et~al.}(2016)\citenamefont {Xu},
  \citenamefont {Lian}, \citenamefont {Tang}, \citenamefont {Qi},\ and\
  \citenamefont {Zhang}}]{XuPRL2016}%
  \BibitemOpen
  \bibfield  {author} {\bibinfo {author} {\bibfnamefont {Gang}\ \bibnamefont
  {Xu}}, \bibinfo {author} {\bibfnamefont {Biao}\ \bibnamefont {Lian}},
  \bibinfo {author} {\bibfnamefont {Peizhe}\ \bibnamefont {Tang}}, \bibinfo
  {author} {\bibfnamefont {Xiao-Liang}\ \bibnamefont {Qi}}, \ and\ \bibinfo
  {author} {\bibfnamefont {Shou-Cheng}\ \bibnamefont {Zhang}},\ }\bibfield
  {title} {\enquote {\bibinfo {title} {{Topological Superconductivity on the
  Surface of Fe-Based Superconductors}},}\ }\href {\doibase
  10.1103/PhysRevLett.117.047001} {\bibfield  {journal} {\bibinfo  {journal}
  {Phys. Rev. Lett.}\ }\textbf {\bibinfo {volume} {117}},\ \bibinfo {pages}
  {047001} (\bibinfo {year} {2016})}\BibitemShut {NoStop}%
\bibitem [{\citenamefont {Zhang}\ \emph {et~al.}(2018)\citenamefont {Zhang},
  \citenamefont {Yaji}, \citenamefont {Hashimoto}, \citenamefont {Ota},
  \citenamefont {Kondo}, \citenamefont {Okazaki}, \citenamefont {Wang},
  \citenamefont {Wen}, \citenamefont {Gu}, \citenamefont {Ding},\ and\
  \citenamefont {Shin}}]{Zhang2018}%
  \BibitemOpen
  \bibfield  {author} {\bibinfo {author} {\bibfnamefont {Peng}\ \bibnamefont
  {Zhang}}, \bibinfo {author} {\bibfnamefont {Koichiro}\ \bibnamefont {Yaji}},
  \bibinfo {author} {\bibfnamefont {Takahiro}\ \bibnamefont {Hashimoto}},
  \bibinfo {author} {\bibfnamefont {Yuichi}\ \bibnamefont {Ota}}, \bibinfo
  {author} {\bibfnamefont {Takeshi}\ \bibnamefont {Kondo}}, \bibinfo {author}
  {\bibfnamefont {Kozo}\ \bibnamefont {Okazaki}}, \bibinfo {author}
  {\bibfnamefont {Zhijun}\ \bibnamefont {Wang}}, \bibinfo {author}
  {\bibfnamefont {Jinsheng}\ \bibnamefont {Wen}}, \bibinfo {author}
  {\bibfnamefont {G.~D.}\ \bibnamefont {Gu}}, \bibinfo {author} {\bibfnamefont
  {Hong}\ \bibnamefont {Ding}}, \ and\ \bibinfo {author} {\bibfnamefont {Shik}\
  \bibnamefont {Shin}},\ }\bibfield  {title} {\enquote {\bibinfo {title}
  {{Observation of topological superconductivity on the surface of an
  iron-based superconductor}},}\ }\href {\doibase 10.1126/science.aan4596}
  {\bibfield  {journal} {\bibinfo  {journal} {Science}\ }\textbf {\bibinfo
  {volume} {360}},\ \bibinfo {pages} {182--186} (\bibinfo {year}
  {2018})}\BibitemShut {NoStop}%
\bibitem [{\citenamefont {Zhang}\ \emph
  {et~al.}(2019{\natexlab{a}})\citenamefont {Zhang}, \citenamefont {Wang},
  \citenamefont {Wu}, \citenamefont {Yaji}, \citenamefont {Ishida},
  \citenamefont {Kohama}, \citenamefont {Dai}, \citenamefont {Sun},
  \citenamefont {Bareille}, \citenamefont {Kuroda}, \citenamefont {Kondo},
  \citenamefont {Okazaki}, \citenamefont {Kindo}, \citenamefont {Wang},
  \citenamefont {Jin}, \citenamefont {Hu}, \citenamefont {Thomale},
  \citenamefont {Sumida}, \citenamefont {Wu}, \citenamefont {Miyamoto},
  \citenamefont {Okuda}, \citenamefont {Ding}, \citenamefont {Gu},
  \citenamefont {Tamegai}, \citenamefont {Kawakami}, \citenamefont {Sato},\
  and\ \citenamefont {Shin}}]{Zhang2019NP}%
  \BibitemOpen
  \bibfield  {author} {\bibinfo {author} {\bibfnamefont {Peng}\ \bibnamefont
  {Zhang}}, \bibinfo {author} {\bibfnamefont {Zhijun}\ \bibnamefont {Wang}},
  \bibinfo {author} {\bibfnamefont {Xianxin}\ \bibnamefont {Wu}}, \bibinfo
  {author} {\bibfnamefont {Koichiro}\ \bibnamefont {Yaji}}, \bibinfo {author}
  {\bibfnamefont {Yukiaki}\ \bibnamefont {Ishida}}, \bibinfo {author}
  {\bibfnamefont {Yoshimitsu}\ \bibnamefont {Kohama}}, \bibinfo {author}
  {\bibfnamefont {Guangyang}\ \bibnamefont {Dai}}, \bibinfo {author}
  {\bibfnamefont {Yue}\ \bibnamefont {Sun}}, \bibinfo {author} {\bibfnamefont
  {Cedric}\ \bibnamefont {Bareille}}, \bibinfo {author} {\bibfnamefont {Kenta}\
  \bibnamefont {Kuroda}}, \bibinfo {author} {\bibfnamefont {Takeshi}\
  \bibnamefont {Kondo}}, \bibinfo {author} {\bibfnamefont {Kozo}\ \bibnamefont
  {Okazaki}}, \bibinfo {author} {\bibfnamefont {Koichi}\ \bibnamefont {Kindo}},
  \bibinfo {author} {\bibfnamefont {Xiancheng}\ \bibnamefont {Wang}}, \bibinfo
  {author} {\bibfnamefont {Changqing}\ \bibnamefont {Jin}}, \bibinfo {author}
  {\bibfnamefont {Jiangping}\ \bibnamefont {Hu}}, \bibinfo {author}
  {\bibfnamefont {Ronny}\ \bibnamefont {Thomale}}, \bibinfo {author}
  {\bibfnamefont {Kazuki}\ \bibnamefont {Sumida}}, \bibinfo {author}
  {\bibfnamefont {Shilong}\ \bibnamefont {Wu}}, \bibinfo {author}
  {\bibfnamefont {Koji}\ \bibnamefont {Miyamoto}}, \bibinfo {author}
  {\bibfnamefont {Taichi}\ \bibnamefont {Okuda}}, \bibinfo {author}
  {\bibfnamefont {Hong}\ \bibnamefont {Ding}}, \bibinfo {author} {\bibfnamefont
  {G.~D.}\ \bibnamefont {Gu}}, \bibinfo {author} {\bibfnamefont {Tsuyoshi}\
  \bibnamefont {Tamegai}}, \bibinfo {author} {\bibfnamefont {Takuto}\
  \bibnamefont {Kawakami}}, \bibinfo {author} {\bibfnamefont {Masatoshi}\
  \bibnamefont {Sato}}, \ and\ \bibinfo {author} {\bibfnamefont {Shik}\
  \bibnamefont {Shin}},\ }\bibfield  {title} {\enquote {\bibinfo {title}
  {{Multiple topological states in iron-based superconductors}},}\ }\href
  {\doibase 10.1038/s41567-018-0280-z} {\bibfield  {journal} {\bibinfo
  {journal} {Nat. Phys.}\ }\textbf {\bibinfo {volume} {15}},\ \bibinfo {pages}
  {41--47} (\bibinfo {year} {2019}{\natexlab{a}})}\BibitemShut {NoStop}%
\bibitem [{\citenamefont {Shi}\ \emph {et~al.}(2017)\citenamefont {Shi},
  \citenamefont {Han}, \citenamefont {Richard}, \citenamefont {Wu},
  \citenamefont {Peng}, \citenamefont {Qian}, \citenamefont {Wang},
  \citenamefont {Hu}, \citenamefont {Sun},\ and\ \citenamefont
  {Ding}}]{Shi2017}%
  \BibitemOpen
  \bibfield  {author} {\bibinfo {author} {\bibfnamefont {Xun}\ \bibnamefont
  {Shi}}, \bibinfo {author} {\bibfnamefont {Zhi-Qing}\ \bibnamefont {Han}},
  \bibinfo {author} {\bibfnamefont {Pierre}\ \bibnamefont {Richard}}, \bibinfo
  {author} {\bibfnamefont {Xian-Xin}\ \bibnamefont {Wu}}, \bibinfo {author}
  {\bibfnamefont {Xi-Liang}\ \bibnamefont {Peng}}, \bibinfo {author}
  {\bibfnamefont {Tian}\ \bibnamefont {Qian}}, \bibinfo {author} {\bibfnamefont
  {Shan-Cai}\ \bibnamefont {Wang}}, \bibinfo {author} {\bibfnamefont
  {Jiang-Ping}\ \bibnamefont {Hu}}, \bibinfo {author} {\bibfnamefont {Yu-Jie}\
  \bibnamefont {Sun}}, \ and\ \bibinfo {author} {\bibfnamefont {Hong}\
  \bibnamefont {Ding}},\ }\bibfield  {title} {\enquote {\bibinfo {title}
  {{FeTe$_{1-x}$Se$_x$ monolayer films: towards the realization of
  high-temperature connate topological superconductivity}},}\ }\href {\doibase
  https://doi.org/10.1016/j.scib.2017.03.010} {\bibfield  {journal} {\bibinfo
  {journal} {Sci. Bull.}\ }\textbf {\bibinfo {volume} {62}},\ \bibinfo {pages}
  {503--507} (\bibinfo {year} {2017})}\BibitemShut {NoStop}%
\bibitem [{\citenamefont {Peng}\ \emph {et~al.}(2019)\citenamefont {Peng},
  \citenamefont {Li}, \citenamefont {Wu}, \citenamefont {Deng}, \citenamefont
  {Shi}, \citenamefont {Fan}, \citenamefont {Li}, \citenamefont {Huang},
  \citenamefont {Qian}, \citenamefont {Richard}, \citenamefont {Hu},
  \citenamefont {Pan}, \citenamefont {Mao}, \citenamefont {Sun},\ and\
  \citenamefont {Ding}}]{Peng2019}%
  \BibitemOpen
  \bibfield  {author} {\bibinfo {author} {\bibfnamefont {X.~L.}\ \bibnamefont
  {Peng}}, \bibinfo {author} {\bibfnamefont {Y.}~\bibnamefont {Li}}, \bibinfo
  {author} {\bibfnamefont {X.~X.}\ \bibnamefont {Wu}}, \bibinfo {author}
  {\bibfnamefont {H.~B.}\ \bibnamefont {Deng}}, \bibinfo {author}
  {\bibfnamefont {X.}~\bibnamefont {Shi}}, \bibinfo {author} {\bibfnamefont
  {W.~H.}\ \bibnamefont {Fan}}, \bibinfo {author} {\bibfnamefont
  {M.}~\bibnamefont {Li}}, \bibinfo {author} {\bibfnamefont {Y.~B.}\
  \bibnamefont {Huang}}, \bibinfo {author} {\bibfnamefont {T.}~\bibnamefont
  {Qian}}, \bibinfo {author} {\bibfnamefont {P.}~\bibnamefont {Richard}},
  \bibinfo {author} {\bibfnamefont {J.~P.}\ \bibnamefont {Hu}}, \bibinfo
  {author} {\bibfnamefont {S.~H.}\ \bibnamefont {Pan}}, \bibinfo {author}
  {\bibfnamefont {H.~Q.}\ \bibnamefont {Mao}}, \bibinfo {author} {\bibfnamefont
  {Y.~J.}\ \bibnamefont {Sun}}, \ and\ \bibinfo {author} {\bibfnamefont
  {H.}~\bibnamefont {Ding}},\ }\bibfield  {title} {\enquote {\bibinfo {title}
  {{Observation of topological transition in high-${T}_{c}$ superconducting
  monolayer ${\mathrm{FeTe}}_{1\ensuremath{-}x}{\mathrm{Se}}_{x}$ films on
  ${\mathrm{SrTiO}}_{3}(001)$}},}\ }\href {\doibase
  10.1103/PhysRevB.100.155134} {\bibfield  {journal} {\bibinfo  {journal}
  {Phys. Rev. B}\ }\textbf {\bibinfo {volume} {100}},\ \bibinfo {pages}
  {155134} (\bibinfo {year} {2019})}\BibitemShut {NoStop}%
\bibitem [{\citenamefont {Hao}\ and\ \citenamefont {Hu}(2018)}]{HaoNSR2019}%
  \BibitemOpen
  \bibfield  {author} {\bibinfo {author} {\bibfnamefont {Ning}\ \bibnamefont
  {Hao}}\ and\ \bibinfo {author} {\bibfnamefont {Jiangping}\ \bibnamefont
  {Hu}},\ }\bibfield  {title} {\enquote {\bibinfo {title} {{Topological quantum
  states of matter in iron-based superconductors: from concept to material
  realization}},}\ }\href@noop {} {\bibfield  {journal} {\bibinfo  {journal}
  {Natl. Sci. Rev.}\ }\textbf {\bibinfo {volume} {6}},\ \bibinfo {pages}
  {213--226} (\bibinfo {year} {2018})}\BibitemShut {NoStop}%
\bibitem [{\citenamefont {Yin}\ \emph {et~al.}(2015)\citenamefont {Yin},
  \citenamefont {Wu}, \citenamefont {Wang}, \citenamefont {Ye}, \citenamefont
  {Gong}, \citenamefont {Hou}, \citenamefont {Shan}, \citenamefont {Li},
  \citenamefont {Liang}, \citenamefont {Wu}, \citenamefont {Li}, \citenamefont
  {Ting}, \citenamefont {Wang}, \citenamefont {Hu}, \citenamefont {Hor},
  \citenamefont {Ding},\ and\ \citenamefont {Pan}}]{YinJX2015}%
  \BibitemOpen
  \bibfield  {author} {\bibinfo {author} {\bibfnamefont {J.~X.}\ \bibnamefont
  {Yin}}, \bibinfo {author} {\bibfnamefont {Zheng}\ \bibnamefont {Wu}},
  \bibinfo {author} {\bibfnamefont {J.~H.}\ \bibnamefont {Wang}}, \bibinfo
  {author} {\bibfnamefont {Z.~Y.}\ \bibnamefont {Ye}}, \bibinfo {author}
  {\bibfnamefont {Jing}\ \bibnamefont {Gong}}, \bibinfo {author} {\bibfnamefont
  {X.~Y.}\ \bibnamefont {Hou}}, \bibinfo {author} {\bibfnamefont {Lei}\
  \bibnamefont {Shan}}, \bibinfo {author} {\bibfnamefont {Ang}\ \bibnamefont
  {Li}}, \bibinfo {author} {\bibfnamefont {X.~J.}\ \bibnamefont {Liang}},
  \bibinfo {author} {\bibfnamefont {X.~X.}\ \bibnamefont {Wu}}, \bibinfo
  {author} {\bibfnamefont {Jian}\ \bibnamefont {Li}}, \bibinfo {author}
  {\bibfnamefont {C.~S.}\ \bibnamefont {Ting}}, \bibinfo {author}
  {\bibfnamefont {Z.~Q.}\ \bibnamefont {Wang}}, \bibinfo {author}
  {\bibfnamefont {J.~P.}\ \bibnamefont {Hu}}, \bibinfo {author} {\bibfnamefont
  {P.~H.}\ \bibnamefont {Hor}}, \bibinfo {author} {\bibfnamefont
  {H.}~\bibnamefont {Ding}}, \ and\ \bibinfo {author} {\bibfnamefont {S.~H.}\
  \bibnamefont {Pan}},\ }\bibfield  {title} {\enquote {\bibinfo {title}
  {{Observation of a robust zero-energy bound state in iron-based
  superconductor Fe(Te,Se)}},}\ }\href {\doibase 10.1038/nphys3371} {\bibfield
  {journal} {\bibinfo  {journal} {Nat. Phys.}\ }\textbf {\bibinfo {volume}
  {11}},\ \bibinfo {pages} {543} (\bibinfo {year} {2015})}\BibitemShut
  {NoStop}%
\bibitem [{\citenamefont {Wang}\ \emph
  {et~al.}(2018{\natexlab{a}})\citenamefont {Wang}, \citenamefont {Kong},
  \citenamefont {Fan}, \citenamefont {Chen}, \citenamefont {Zhu}, \citenamefont
  {Liu}, \citenamefont {Cao}, \citenamefont {Sun}, \citenamefont {Du},
  \citenamefont {Schneeloch}, \citenamefont {Zhong}, \citenamefont {Gu},
  \citenamefont {Fu}, \citenamefont {Ding},\ and\ \citenamefont
  {Gao}}]{WangDF2018}%
  \BibitemOpen
  \bibfield  {author} {\bibinfo {author} {\bibfnamefont {Dongfei}\ \bibnamefont
  {Wang}}, \bibinfo {author} {\bibfnamefont {Lingyuan}\ \bibnamefont {Kong}},
  \bibinfo {author} {\bibfnamefont {Peng}\ \bibnamefont {Fan}}, \bibinfo
  {author} {\bibfnamefont {Hui}\ \bibnamefont {Chen}}, \bibinfo {author}
  {\bibfnamefont {Shiyu}\ \bibnamefont {Zhu}}, \bibinfo {author} {\bibfnamefont
  {Wenyao}\ \bibnamefont {Liu}}, \bibinfo {author} {\bibfnamefont
  {Lu}~\bibnamefont {Cao}}, \bibinfo {author} {\bibfnamefont {Yujie}\
  \bibnamefont {Sun}}, \bibinfo {author} {\bibfnamefont {Shixuan}\ \bibnamefont
  {Du}}, \bibinfo {author} {\bibfnamefont {John}\ \bibnamefont {Schneeloch}},
  \bibinfo {author} {\bibfnamefont {Ruidan}\ \bibnamefont {Zhong}}, \bibinfo
  {author} {\bibfnamefont {Genda}\ \bibnamefont {Gu}}, \bibinfo {author}
  {\bibfnamefont {Liang}\ \bibnamefont {Fu}}, \bibinfo {author} {\bibfnamefont
  {Hong}\ \bibnamefont {Ding}}, \ and\ \bibinfo {author} {\bibfnamefont
  {Hong-Jun}\ \bibnamefont {Gao}},\ }\bibfield  {title} {\enquote {\bibinfo
  {title} {{Evidence for Majorana bound states in an iron-based
  superconductor}},}\ }\href {\doibase 10.1126/science.aao1797} {\bibfield
  {journal} {\bibinfo  {journal} {Science}\ }\textbf {\bibinfo {volume}
  {362}},\ \bibinfo {pages} {333} (\bibinfo {year}
  {2018}{\natexlab{a}})}\BibitemShut {NoStop}%
\bibitem [{\citenamefont {Liu}\ \emph {et~al.}(2018)\citenamefont {Liu},
  \citenamefont {Chen}, \citenamefont {Zhang}, \citenamefont {Peng},
  \citenamefont {Yan}, \citenamefont {Wen}, \citenamefont {Lou}, \citenamefont
  {Huang}, \citenamefont {Tian}, \citenamefont {Dong}, \citenamefont {Wang},
  \citenamefont {Bao}, \citenamefont {Wang}, \citenamefont {Yin}, \citenamefont
  {Zhao},\ and\ \citenamefont {Feng}}]{LiuQ2018}%
  \BibitemOpen
  \bibfield  {author} {\bibinfo {author} {\bibfnamefont {Qin}\ \bibnamefont
  {Liu}}, \bibinfo {author} {\bibfnamefont {Chen}\ \bibnamefont {Chen}},
  \bibinfo {author} {\bibfnamefont {Tong}\ \bibnamefont {Zhang}}, \bibinfo
  {author} {\bibfnamefont {Rui}\ \bibnamefont {Peng}}, \bibinfo {author}
  {\bibfnamefont {Ya-Jun}\ \bibnamefont {Yan}}, \bibinfo {author}
  {\bibfnamefont {Chen-Hao-Ping}\ \bibnamefont {Wen}}, \bibinfo {author}
  {\bibfnamefont {Xia}\ \bibnamefont {Lou}}, \bibinfo {author} {\bibfnamefont
  {Yu-Long}\ \bibnamefont {Huang}}, \bibinfo {author} {\bibfnamefont
  {Jin-Peng}\ \bibnamefont {Tian}}, \bibinfo {author} {\bibfnamefont {Xiao-Li}\
  \bibnamefont {Dong}}, \bibinfo {author} {\bibfnamefont {Guang-Wei}\
  \bibnamefont {Wang}}, \bibinfo {author} {\bibfnamefont {Wei-Cheng}\
  \bibnamefont {Bao}}, \bibinfo {author} {\bibfnamefont {Qiang-Hua}\
  \bibnamefont {Wang}}, \bibinfo {author} {\bibfnamefont {Zhi-Ping}\
  \bibnamefont {Yin}}, \bibinfo {author} {\bibfnamefont {Zhong-Xian}\
  \bibnamefont {Zhao}}, \ and\ \bibinfo {author} {\bibfnamefont {Dong-Lai}\
  \bibnamefont {Feng}},\ }\bibfield  {title} {\enquote {\bibinfo {title}
  {{Robust and Clean Majorana Zero Mode in the Vortex Core of High-Temperature
  Superconductor
  $\mathbf{(}{\mathrm{Li}}_{0.84}{\mathrm{Fe}}_{0.16}\mathbf{)}\mathrm{OHFeSe}$}},}\
  }\href {\doibase 10.1103/PhysRevX.8.041056} {\bibfield  {journal} {\bibinfo
  {journal} {Phys. Rev. X}\ }\textbf {\bibinfo {volume} {8}},\ \bibinfo {pages}
  {041056} (\bibinfo {year} {2018})}\BibitemShut {NoStop}%
\bibitem [{\citenamefont {Kong}\ \emph {et~al.}(2019)\citenamefont {Kong},
  \citenamefont {Zhu}, \citenamefont {Papaj}, \citenamefont {Chen},
  \citenamefont {Cao}, \citenamefont {Isobe}, \citenamefont {Xing},
  \citenamefont {Liu}, \citenamefont {Wang}, \citenamefont {Fan}, \citenamefont
  {Sun}, \citenamefont {Du}, \citenamefont {Schneeloch}, \citenamefont {Zhong},
  \citenamefont {Gu}, \citenamefont {Fu}, \citenamefont {Gao},\ and\
  \citenamefont {Ding}}]{KongLY2019}%
  \BibitemOpen
  \bibfield  {author} {\bibinfo {author} {\bibfnamefont {Lingyuan}\
  \bibnamefont {Kong}}, \bibinfo {author} {\bibfnamefont {Shiyu}\ \bibnamefont
  {Zhu}}, \bibinfo {author} {\bibfnamefont {Micha{\l}}\ \bibnamefont {Papaj}},
  \bibinfo {author} {\bibfnamefont {Hui}\ \bibnamefont {Chen}}, \bibinfo
  {author} {\bibfnamefont {Lu}~\bibnamefont {Cao}}, \bibinfo {author}
  {\bibfnamefont {Hiroki}\ \bibnamefont {Isobe}}, \bibinfo {author}
  {\bibfnamefont {Yuqing}\ \bibnamefont {Xing}}, \bibinfo {author}
  {\bibfnamefont {Wenyao}\ \bibnamefont {Liu}}, \bibinfo {author}
  {\bibfnamefont {Dongfei}\ \bibnamefont {Wang}}, \bibinfo {author}
  {\bibfnamefont {Peng}\ \bibnamefont {Fan}}, \bibinfo {author} {\bibfnamefont
  {Yujie}\ \bibnamefont {Sun}}, \bibinfo {author} {\bibfnamefont {Shixuan}\
  \bibnamefont {Du}}, \bibinfo {author} {\bibfnamefont {John}\ \bibnamefont
  {Schneeloch}}, \bibinfo {author} {\bibfnamefont {Ruidan}\ \bibnamefont
  {Zhong}}, \bibinfo {author} {\bibfnamefont {Genda}\ \bibnamefont {Gu}},
  \bibinfo {author} {\bibfnamefont {Liang}\ \bibnamefont {Fu}}, \bibinfo
  {author} {\bibfnamefont {Hong-Jun}\ \bibnamefont {Gao}}, \ and\ \bibinfo
  {author} {\bibfnamefont {Hong}\ \bibnamefont {Ding}},\ }\bibfield  {title}
  {\enquote {\bibinfo {title} {{Half-integer level shift of vortex bound states
  in an iron-based superconductor}},}\ }\href {\doibase
  10.1038/s41567-019-0630-5} {\bibfield  {journal} {\bibinfo  {journal} {Nat.
  Phys.}\ }\textbf {\bibinfo {volume} {15}},\ \bibinfo {pages} {1181} (\bibinfo
  {year} {2019})}\BibitemShut {NoStop}%
\bibitem [{\citenamefont {Machida}\ \emph {et~al.}(2019)\citenamefont
  {Machida}, \citenamefont {Sun}, \citenamefont {Pyon}, \citenamefont {Takeda},
  \citenamefont {Kohsaka}, \citenamefont {Hanaguri}, \citenamefont {Sasagawa},\
  and\ \citenamefont {Tamegai}}]{Machida2019}%
  \BibitemOpen
  \bibfield  {author} {\bibinfo {author} {\bibfnamefont {T.}~\bibnamefont
  {Machida}}, \bibinfo {author} {\bibfnamefont {Y.}~\bibnamefont {Sun}},
  \bibinfo {author} {\bibfnamefont {S.}~\bibnamefont {Pyon}}, \bibinfo {author}
  {\bibfnamefont {S.}~\bibnamefont {Takeda}}, \bibinfo {author} {\bibfnamefont
  {Y.}~\bibnamefont {Kohsaka}}, \bibinfo {author} {\bibfnamefont
  {T.}~\bibnamefont {Hanaguri}}, \bibinfo {author} {\bibfnamefont
  {T.}~\bibnamefont {Sasagawa}}, \ and\ \bibinfo {author} {\bibfnamefont
  {T.}~\bibnamefont {Tamegai}},\ }\bibfield  {title} {\enquote {\bibinfo
  {title} {Zero-energy vortex bound state in the superconducting topological
  surface state of fe(se,te)},}\ }\href {\doibase 10.1038/s41563-019-0397-1}
  {\bibfield  {journal} {\bibinfo  {journal} {Nat. Mater.}\ }\textbf {\bibinfo
  {volume} {18}},\ \bibinfo {pages} {811--815} (\bibinfo {year}
  {2019})}\BibitemShut {NoStop}%
\bibitem [{\citenamefont {Van~Harlingen}(1995)}]{HarlingenRMP1995}%
  \BibitemOpen
  \bibfield  {author} {\bibinfo {author} {\bibfnamefont {D.~J.}\ \bibnamefont
  {Van~Harlingen}},\ }\bibfield  {title} {\enquote {\bibinfo {title}
  {{Phase-sensitive tests of the symmetry of the pairing state in the
  high-temperature superconductors---Evidence for
  ${d}_{{x}^{2}\ensuremath{-}{y}^{2}}$ symmetry}},}\ }\href {\doibase
  10.1103/RevModPhys.67.515} {\bibfield  {journal} {\bibinfo  {journal} {Rev.
  Mod. Phys.}\ }\textbf {\bibinfo {volume} {67}},\ \bibinfo {pages} {515}
  (\bibinfo {year} {1995})}\BibitemShut {NoStop}%
\bibitem [{\citenamefont {Hirschfeld}\ \emph {et~al.}(2011)\citenamefont
  {Hirschfeld}, \citenamefont {Korshunov},\ and\ \citenamefont
  {Mazin}}]{Hirschfeld2011}%
  \BibitemOpen
  \bibfield  {author} {\bibinfo {author} {\bibfnamefont {P.~J.}\ \bibnamefont
  {Hirschfeld}}, \bibinfo {author} {\bibfnamefont {M.~M.}\ \bibnamefont
  {Korshunov}}, \ and\ \bibinfo {author} {\bibfnamefont {I.~I.}\ \bibnamefont
  {Mazin}},\ }\bibfield  {title} {\enquote {\bibinfo {title} {{Gap symmetry and
  structure of Fe-based superconductors}},}\ }\href@noop {} {\bibfield
  {journal} {\bibinfo  {journal} {Rep. Prog. Phys.}\ }\textbf {\bibinfo
  {volume} {74}},\ \bibinfo {pages} {124508} (\bibinfo {year}
  {2011})}\BibitemShut {NoStop}%
\bibitem [{\citenamefont {Christianson}\ \emph {et~al.}(2008)\citenamefont
  {Christianson}, \citenamefont {Goremychkin}, \citenamefont {Osborn},
  \citenamefont {Rosenkranz}, \citenamefont {Lumsden}, \citenamefont
  {Malliakas}, \citenamefont {Todorov}, \citenamefont {Claus}, \citenamefont
  {Chung}, \citenamefont {Kanatzidis}, \citenamefont {Bewley},\ and\
  \citenamefont {Guidi}}]{Christianson2008}%
  \BibitemOpen
  \bibfield  {author} {\bibinfo {author} {\bibfnamefont {A.~D.}\ \bibnamefont
  {Christianson}}, \bibinfo {author} {\bibfnamefont {E.~A.}\ \bibnamefont
  {Goremychkin}}, \bibinfo {author} {\bibfnamefont {R.}~\bibnamefont {Osborn}},
  \bibinfo {author} {\bibfnamefont {S.}~\bibnamefont {Rosenkranz}}, \bibinfo
  {author} {\bibfnamefont {M.~D.}\ \bibnamefont {Lumsden}}, \bibinfo {author}
  {\bibfnamefont {C.~D.}\ \bibnamefont {Malliakas}}, \bibinfo {author}
  {\bibfnamefont {I.~S.}\ \bibnamefont {Todorov}}, \bibinfo {author}
  {\bibfnamefont {H.}~\bibnamefont {Claus}}, \bibinfo {author} {\bibfnamefont
  {D.~Y.}\ \bibnamefont {Chung}}, \bibinfo {author} {\bibfnamefont {M.~G.}\
  \bibnamefont {Kanatzidis}}, \bibinfo {author} {\bibfnamefont {R.~I.}\
  \bibnamefont {Bewley}}, \ and\ \bibinfo {author} {\bibfnamefont
  {T.}~\bibnamefont {Guidi}},\ }\bibfield  {title} {\enquote {\bibinfo {title}
  {{Unconventional superconductivity in Ba$_{0.6}$K$_{0.4}$Fe$_2$As$_2$ from
  inelastic neutron scattering}},}\ }\href {\doibase 10.1038/nature07625}
  {\bibfield  {journal} {\bibinfo  {journal} {Nature}\ }\textbf {\bibinfo
  {volume} {456}},\ \bibinfo {pages} {930} (\bibinfo {year}
  {2008})}\BibitemShut {NoStop}%
\bibitem [{\citenamefont {Lumsden}\ \emph {et~al.}(2009)\citenamefont
  {Lumsden}, \citenamefont {Christianson}, \citenamefont {Parshall},
  \citenamefont {Stone}, \citenamefont {Nagler}, \citenamefont {MacDougall},
  \citenamefont {Mook}, \citenamefont {Lokshin}, \citenamefont {Egami},
  \citenamefont {Abernathy}, \citenamefont {Goremychkin}, \citenamefont
  {Osborn}, \citenamefont {McGuire}, \citenamefont {Sefat}, \citenamefont
  {Jin}, \citenamefont {Sales},\ and\ \citenamefont {Mandrus}}]{Lumsden2009}%
  \BibitemOpen
  \bibfield  {author} {\bibinfo {author} {\bibfnamefont {M.~D.}\ \bibnamefont
  {Lumsden}}, \bibinfo {author} {\bibfnamefont {A.~D.}\ \bibnamefont
  {Christianson}}, \bibinfo {author} {\bibfnamefont {D.}~\bibnamefont
  {Parshall}}, \bibinfo {author} {\bibfnamefont {M.~B.}\ \bibnamefont {Stone}},
  \bibinfo {author} {\bibfnamefont {S.~E.}\ \bibnamefont {Nagler}}, \bibinfo
  {author} {\bibfnamefont {G.~J.}\ \bibnamefont {MacDougall}}, \bibinfo
  {author} {\bibfnamefont {H.~A.}\ \bibnamefont {Mook}}, \bibinfo {author}
  {\bibfnamefont {K.}~\bibnamefont {Lokshin}}, \bibinfo {author} {\bibfnamefont
  {T.}~\bibnamefont {Egami}}, \bibinfo {author} {\bibfnamefont {D.~L.}\
  \bibnamefont {Abernathy}}, \bibinfo {author} {\bibfnamefont {E.~A.}\
  \bibnamefont {Goremychkin}}, \bibinfo {author} {\bibfnamefont
  {R.}~\bibnamefont {Osborn}}, \bibinfo {author} {\bibfnamefont {M.~A.}\
  \bibnamefont {McGuire}}, \bibinfo {author} {\bibfnamefont {A.~S.}\
  \bibnamefont {Sefat}}, \bibinfo {author} {\bibfnamefont {R.}~\bibnamefont
  {Jin}}, \bibinfo {author} {\bibfnamefont {B.~C.}\ \bibnamefont {Sales}}, \
  and\ \bibinfo {author} {\bibfnamefont {D.}~\bibnamefont {Mandrus}},\
  }\bibfield  {title} {\enquote {\bibinfo {title} {{Two-dimensional resonant
  magnetic excitation in
  ${\mathrm{BaFe}}_{1.84}{\mathrm{Co}}_{0.16}{\mathrm{As}}_{2}$}},}\ }\href
  {\doibase 10.1103/PhysRevLett.102.107005} {\bibfield  {journal} {\bibinfo
  {journal} {Phys. Rev. Lett.}\ }\textbf {\bibinfo {volume} {102}},\ \bibinfo
  {pages} {107005} (\bibinfo {year} {2009})}\BibitemShut {NoStop}%
\bibitem [{\citenamefont {Christianson}\ \emph {et~al.}(2009)\citenamefont
  {Christianson}, \citenamefont {Lumsden}, \citenamefont {Nagler},
  \citenamefont {MacDougall}, \citenamefont {McGuire}, \citenamefont {Sefat},
  \citenamefont {Jin}, \citenamefont {Sales},\ and\ \citenamefont
  {Mandrus}}]{Christianson2009}%
  \BibitemOpen
  \bibfield  {author} {\bibinfo {author} {\bibfnamefont {A.~D.}\ \bibnamefont
  {Christianson}}, \bibinfo {author} {\bibfnamefont {M.~D.}\ \bibnamefont
  {Lumsden}}, \bibinfo {author} {\bibfnamefont {S.~E.}\ \bibnamefont {Nagler}},
  \bibinfo {author} {\bibfnamefont {G.~J.}\ \bibnamefont {MacDougall}},
  \bibinfo {author} {\bibfnamefont {M.~A.}\ \bibnamefont {McGuire}}, \bibinfo
  {author} {\bibfnamefont {A.~S.}\ \bibnamefont {Sefat}}, \bibinfo {author}
  {\bibfnamefont {R.}~\bibnamefont {Jin}}, \bibinfo {author} {\bibfnamefont
  {B.~C.}\ \bibnamefont {Sales}}, \ and\ \bibinfo {author} {\bibfnamefont
  {D.}~\bibnamefont {Mandrus}},\ }\bibfield  {title} {\enquote {\bibinfo
  {title} {{Static and Dynamic Magnetism in Underdoped Superconductor
  ${\mathrm{BaFe}}_{1.92}{\mathrm{Co}}_{0.08}{\mathrm{As}}_{2}$}},}\ }\href
  {\doibase 10.1103/PhysRevLett.103.087002} {\bibfield  {journal} {\bibinfo
  {journal} {Phys. Rev. Lett.}\ }\textbf {\bibinfo {volume} {103}},\ \bibinfo
  {pages} {087002} (\bibinfo {year} {2009})}\BibitemShut {NoStop}%
\bibitem [{\citenamefont {Qiu}\ \emph {et~al.}(2009)\citenamefont {Qiu},
  \citenamefont {Bao}, \citenamefont {Zhao}, \citenamefont {Broholm},
  \citenamefont {Stanev}, \citenamefont {Tesanovic}, \citenamefont
  {Gasparovic}, \citenamefont {Chang}, \citenamefont {Hu}, \citenamefont
  {Qian}, \citenamefont {Fang},\ and\ \citenamefont {Mao}}]{QiuYM2009}%
  \BibitemOpen
  \bibfield  {author} {\bibinfo {author} {\bibfnamefont {Yiming}\ \bibnamefont
  {Qiu}}, \bibinfo {author} {\bibfnamefont {Wei}\ \bibnamefont {Bao}}, \bibinfo
  {author} {\bibfnamefont {Y.}~\bibnamefont {Zhao}}, \bibinfo {author}
  {\bibfnamefont {Collin}\ \bibnamefont {Broholm}}, \bibinfo {author}
  {\bibfnamefont {V.}~\bibnamefont {Stanev}}, \bibinfo {author} {\bibfnamefont
  {Z.}~\bibnamefont {Tesanovic}}, \bibinfo {author} {\bibfnamefont {Y.~C.}\
  \bibnamefont {Gasparovic}}, \bibinfo {author} {\bibfnamefont
  {S.}~\bibnamefont {Chang}}, \bibinfo {author} {\bibfnamefont {Jin}\
  \bibnamefont {Hu}}, \bibinfo {author} {\bibfnamefont {Bin}\ \bibnamefont
  {Qian}}, \bibinfo {author} {\bibfnamefont {Minghu}\ \bibnamefont {Fang}}, \
  and\ \bibinfo {author} {\bibfnamefont {Zhiqiang}\ \bibnamefont {Mao}},\
  }\bibfield  {title} {\enquote {\bibinfo {title} {{Spin Gap and Resonance at
  the Nesting Wave Vector in Superconducting
  ${\mathrm{FeSe}}_{0.4}{\mathrm{Te}}_{0.6}$}},}\ }\href {\doibase
  10.1103/PhysRevLett.103.067008} {\bibfield  {journal} {\bibinfo  {journal}
  {Phys. Rev. Lett.}\ }\textbf {\bibinfo {volume} {103}},\ \bibinfo {pages}
  {067008} (\bibinfo {year} {2009})}\BibitemShut {NoStop}%
\bibitem [{\citenamefont {Hanaguri}\ \emph {et~al.}(2010)\citenamefont
  {Hanaguri}, \citenamefont {Niitaka}, \citenamefont {Kuroki},\ and\
  \citenamefont {Takagi}}]{Hanaguri2010}%
  \BibitemOpen
  \bibfield  {author} {\bibinfo {author} {\bibfnamefont {T.}~\bibnamefont
  {Hanaguri}}, \bibinfo {author} {\bibfnamefont {S.}~\bibnamefont {Niitaka}},
  \bibinfo {author} {\bibfnamefont {K.}~\bibnamefont {Kuroki}}, \ and\ \bibinfo
  {author} {\bibfnamefont {H.}~\bibnamefont {Takagi}},\ }\bibfield  {title}
  {\enquote {\bibinfo {title} {{Unconventional $s$-Wave Superconductivity in
  Fe(Se,Te)}},}\ }\href {\doibase 10.1126/science.1187399} {\bibfield
  {journal} {\bibinfo  {journal} {Science}\ }\textbf {\bibinfo {volume}
  {328}},\ \bibinfo {pages} {474} (\bibinfo {year} {2010})}\BibitemShut
  {NoStop}%
\bibitem [{\citenamefont {Grothe}\ \emph {et~al.}(2012)\citenamefont {Grothe},
  \citenamefont {Chi}, \citenamefont {Dosanjh}, \citenamefont {Liang},
  \citenamefont {Hardy}, \citenamefont {Burke}, \citenamefont {Bonn},\ and\
  \citenamefont {Pennec}}]{GrotheS2012}%
  \BibitemOpen
  \bibfield  {author} {\bibinfo {author} {\bibfnamefont {S.}~\bibnamefont
  {Grothe}}, \bibinfo {author} {\bibfnamefont {Shun}\ \bibnamefont {Chi}},
  \bibinfo {author} {\bibfnamefont {P.}~\bibnamefont {Dosanjh}}, \bibinfo
  {author} {\bibfnamefont {Ruixing}\ \bibnamefont {Liang}}, \bibinfo {author}
  {\bibfnamefont {W.~N.}\ \bibnamefont {Hardy}}, \bibinfo {author}
  {\bibfnamefont {S.~A.}\ \bibnamefont {Burke}}, \bibinfo {author}
  {\bibfnamefont {D.~A.}\ \bibnamefont {Bonn}}, \ and\ \bibinfo {author}
  {\bibfnamefont {Y.}~\bibnamefont {Pennec}},\ }\bibfield  {title} {\enquote
  {\bibinfo {title} {{Bound states of defects in superconducting LiFeAs studied
  by scanning tunneling spectroscopy}},}\ }\href {\doibase
  10.1103/PhysRevB.86.174503} {\bibfield  {journal} {\bibinfo  {journal} {Phys.
  Rev. B}\ }\textbf {\bibinfo {volume} {86}},\ \bibinfo {pages} {174503}
  (\bibinfo {year} {2012})}\BibitemShut {NoStop}%
\bibitem [{\citenamefont {Zhang}\ \emph
  {et~al.}(2013{\natexlab{a}})\citenamefont {Zhang}, \citenamefont {Kane},\
  and\ \citenamefont {Mele}}]{ZhangF2013}%
  \BibitemOpen
  \bibfield  {author} {\bibinfo {author} {\bibfnamefont {Fan}\ \bibnamefont
  {Zhang}}, \bibinfo {author} {\bibfnamefont {C.~L.}\ \bibnamefont {Kane}}, \
  and\ \bibinfo {author} {\bibfnamefont {E.~J.}\ \bibnamefont {Mele}},\
  }\bibfield  {title} {\enquote {\bibinfo {title} {Time-reversal-invariant
  topological superconductivity and majorana kramers pairs},}\ }\href {\doibase
  10.1103/PhysRevLett.111.056402} {\bibfield  {journal} {\bibinfo  {journal}
  {Phys. Rev. Lett.}\ }\textbf {\bibinfo {volume} {111}},\ \bibinfo {pages}
  {056402} (\bibinfo {year} {2013}{\natexlab{a}})}\BibitemShut {NoStop}%
\bibitem [{\citenamefont {Zhang}\ \emph
  {et~al.}(2013{\natexlab{b}})\citenamefont {Zhang}, \citenamefont {Kane},\
  and\ \citenamefont {Mele}}]{ZhangF2013-1}%
  \BibitemOpen
  \bibfield  {author} {\bibinfo {author} {\bibfnamefont {Fan}\ \bibnamefont
  {Zhang}}, \bibinfo {author} {\bibfnamefont {C.~L.}\ \bibnamefont {Kane}}, \
  and\ \bibinfo {author} {\bibfnamefont {E.~J.}\ \bibnamefont {Mele}},\
  }\bibfield  {title} {\enquote {\bibinfo {title} {Surface state magnetization
  and chiral edge states on topological insulators},}\ }\href {\doibase
  10.1103/PhysRevLett.110.046404} {\bibfield  {journal} {\bibinfo  {journal}
  {Phys. Rev. Lett.}\ }\textbf {\bibinfo {volume} {110}},\ \bibinfo {pages}
  {046404} (\bibinfo {year} {2013}{\natexlab{b}})}\BibitemShut {NoStop}%
\bibitem [{\citenamefont {Benalcazar}\ \emph {et~al.}(2014)\citenamefont
  {Benalcazar}, \citenamefont {Teo},\ and\ \citenamefont
  {Hughes}}]{benalcazar2014}%
  \BibitemOpen
  \bibfield  {author} {\bibinfo {author} {\bibfnamefont {Wladimir~A.}\
  \bibnamefont {Benalcazar}}, \bibinfo {author} {\bibfnamefont {Jeffrey C.~Y.}\
  \bibnamefont {Teo}}, \ and\ \bibinfo {author} {\bibfnamefont {Taylor~L.}\
  \bibnamefont {Hughes}},\ }\bibfield  {title} {\enquote {\bibinfo {title}
  {{Classification of two-dimensional topological crystalline superconductors
  and Majorana bound states at disclinations}},}\ }\href {\doibase
  10.1103/PhysRevB.89.224503} {\bibfield  {journal} {\bibinfo  {journal} {Phys.
  Rev. B}\ }\textbf {\bibinfo {volume} {89}},\ \bibinfo {pages} {224503}
  (\bibinfo {year} {2014})}\BibitemShut {NoStop}%
\bibitem [{\citenamefont {Benalcazar}\ \emph
  {et~al.}(2017{\natexlab{a}})\citenamefont {Benalcazar}, \citenamefont
  {Bernevig},\ and\ \citenamefont {Hughes}}]{benalcazar2017quad}%
  \BibitemOpen
  \bibfield  {author} {\bibinfo {author} {\bibfnamefont {Wladimir~A.}\
  \bibnamefont {Benalcazar}}, \bibinfo {author} {\bibfnamefont {B.~Andrei}\
  \bibnamefont {Bernevig}}, \ and\ \bibinfo {author} {\bibfnamefont
  {Taylor~L.}\ \bibnamefont {Hughes}},\ }\bibfield  {title} {\enquote {\bibinfo
  {title} {{Quantized electric multipole insulators}},}\ }\href {\doibase
  10.1126/science.aah6442} {\bibfield  {journal} {\bibinfo  {journal}
  {Science}\ }\textbf {\bibinfo {volume} {357}},\ \bibinfo {pages} {61}
  (\bibinfo {year} {2017}{\natexlab{a}})}\BibitemShut {NoStop}%
\bibitem [{\citenamefont {Benalcazar}\ \emph
  {et~al.}(2017{\natexlab{b}})\citenamefont {Benalcazar}, \citenamefont
  {Bernevig},\ and\ \citenamefont {Hughes}}]{benalcazar2017quadPRB}%
  \BibitemOpen
  \bibfield  {author} {\bibinfo {author} {\bibfnamefont {Wladimir~A.}\
  \bibnamefont {Benalcazar}}, \bibinfo {author} {\bibfnamefont {B.~Andrei}\
  \bibnamefont {Bernevig}}, \ and\ \bibinfo {author} {\bibfnamefont
  {Taylor~L.}\ \bibnamefont {Hughes}},\ }\bibfield  {title} {\enquote {\bibinfo
  {title} {{Electric multipole moments, topological multipole moment pumping,
  and chiral hinge states in crystalline insulators}},}\ }\href {\doibase
  10.1103/PhysRevB.96.245115} {\bibfield  {journal} {\bibinfo  {journal} {Phys.
  Rev. B}\ }\textbf {\bibinfo {volume} {96}},\ \bibinfo {pages} {245115}
  (\bibinfo {year} {2017}{\natexlab{b}})}\BibitemShut {NoStop}%
\bibitem [{\citenamefont {Song}\ \emph {et~al.}(2017)\citenamefont {Song},
  \citenamefont {Fang},\ and\ \citenamefont {Fang}}]{song2017}%
  \BibitemOpen
  \bibfield  {author} {\bibinfo {author} {\bibfnamefont {Zhida}\ \bibnamefont
  {Song}}, \bibinfo {author} {\bibfnamefont {Zhong}\ \bibnamefont {Fang}}, \
  and\ \bibinfo {author} {\bibfnamefont {Chen}\ \bibnamefont {Fang}},\
  }\bibfield  {title} {\enquote {\bibinfo {title}
  {{$(d\ensuremath{-}2)$-Dimensional Edge States of Rotation Symmetry Protected
  Topological States}},}\ }\href {\doibase 10.1103/PhysRevLett.119.246402}
  {\bibfield  {journal} {\bibinfo  {journal} {Phys. Rev. Lett.}\ }\textbf
  {\bibinfo {volume} {119}},\ \bibinfo {pages} {246402} (\bibinfo {year}
  {2017})}\BibitemShut {NoStop}%
\bibitem [{\citenamefont {Langbehn}\ \emph {et~al.}(2017)\citenamefont
  {Langbehn}, \citenamefont {Peng}, \citenamefont {Trifunovic}, \citenamefont
  {von Oppen},\ and\ \citenamefont {Brouwer}}]{langbehn2017}%
  \BibitemOpen
  \bibfield  {author} {\bibinfo {author} {\bibfnamefont {Josias}\ \bibnamefont
  {Langbehn}}, \bibinfo {author} {\bibfnamefont {Yang}\ \bibnamefont {Peng}},
  \bibinfo {author} {\bibfnamefont {Luka}\ \bibnamefont {Trifunovic}}, \bibinfo
  {author} {\bibfnamefont {Felix}\ \bibnamefont {von Oppen}}, \ and\ \bibinfo
  {author} {\bibfnamefont {Piet~W.}\ \bibnamefont {Brouwer}},\ }\bibfield
  {title} {\enquote {\bibinfo {title} {{Reflection-Symmetric Second-Order
  Topological Insulators and Superconductors}},}\ }\href {\doibase
  10.1103/PhysRevLett.119.246401} {\bibfield  {journal} {\bibinfo  {journal}
  {Phys. Rev. Lett.}\ }\textbf {\bibinfo {volume} {119}},\ \bibinfo {pages}
  {246401} (\bibinfo {year} {2017})}\BibitemShut {NoStop}%
\bibitem [{\citenamefont {Yan}\ \emph {et~al.}(2018)\citenamefont {Yan},
  \citenamefont {Song},\ and\ \citenamefont {Wang}}]{YanZB2018}%
  \BibitemOpen
  \bibfield  {author} {\bibinfo {author} {\bibfnamefont {Zhongbo}\ \bibnamefont
  {Yan}}, \bibinfo {author} {\bibfnamefont {Fei}\ \bibnamefont {Song}}, \ and\
  \bibinfo {author} {\bibfnamefont {Zhong}\ \bibnamefont {Wang}},\ }\bibfield
  {title} {\enquote {\bibinfo {title} {{Majorana Corner Modes in a
  High-Temperature Platform}},}\ }\href {\doibase
  10.1103/PhysRevLett.121.096803} {\bibfield  {journal} {\bibinfo  {journal}
  {Phys. Rev. Lett.}\ }\textbf {\bibinfo {volume} {121}},\ \bibinfo {pages}
  {096803} (\bibinfo {year} {2018})}\BibitemShut {NoStop}%
\bibitem [{\citenamefont {Wang}\ \emph
  {et~al.}(2018{\natexlab{b}})\citenamefont {Wang}, \citenamefont {Liu},
  \citenamefont {Lu},\ and\ \citenamefont {Zhang}}]{WangQY2018}%
  \BibitemOpen
  \bibfield  {author} {\bibinfo {author} {\bibfnamefont {Qiyue}\ \bibnamefont
  {Wang}}, \bibinfo {author} {\bibfnamefont {Cheng-Cheng}\ \bibnamefont {Liu}},
  \bibinfo {author} {\bibfnamefont {Yuan-Ming}\ \bibnamefont {Lu}}, \ and\
  \bibinfo {author} {\bibfnamefont {Fan}\ \bibnamefont {Zhang}},\ }\bibfield
  {title} {\enquote {\bibinfo {title} {{High-Temperature Majorana Corner
  States}},}\ }\href {\doibase 10.1103/PhysRevLett.121.186801} {\bibfield
  {journal} {\bibinfo  {journal} {Phys. Rev. Lett.}\ }\textbf {\bibinfo
  {volume} {121}},\ \bibinfo {pages} {186801} (\bibinfo {year}
  {2018}{\natexlab{b}})}\BibitemShut {NoStop}%
\bibitem [{\citenamefont {Hsu}\ \emph {et~al.}(2018)\citenamefont {Hsu},
  \citenamefont {Stano}, \citenamefont {Klinovaja},\ and\ \citenamefont
  {Loss}}]{chen2018}%
  \BibitemOpen
  \bibfield  {author} {\bibinfo {author} {\bibfnamefont {Chen-Hsuan}\
  \bibnamefont {Hsu}}, \bibinfo {author} {\bibfnamefont {Peter}\ \bibnamefont
  {Stano}}, \bibinfo {author} {\bibfnamefont {Jelena}\ \bibnamefont
  {Klinovaja}}, \ and\ \bibinfo {author} {\bibfnamefont {Daniel}\ \bibnamefont
  {Loss}},\ }\bibfield  {title} {\enquote {\bibinfo {title} {Majorana kramers
  pairs in higher-order topological insulators},}\ }\href {\doibase
  10.1103/PhysRevLett.121.196801} {\bibfield  {journal} {\bibinfo  {journal}
  {Phys. Rev. Lett.}\ }\textbf {\bibinfo {volume} {121}},\ \bibinfo {pages}
  {196801} (\bibinfo {year} {2018})}\BibitemShut {NoStop}%
\bibitem [{\citenamefont {Wang}\ \emph
  {et~al.}(2018{\natexlab{c}})\citenamefont {Wang}, \citenamefont {Lin},\ and\
  \citenamefont {Hughes}}]{WangYX2018}%
  \BibitemOpen
  \bibfield  {author} {\bibinfo {author} {\bibfnamefont {Yuxuan}\ \bibnamefont
  {Wang}}, \bibinfo {author} {\bibfnamefont {Mao}\ \bibnamefont {Lin}}, \ and\
  \bibinfo {author} {\bibfnamefont {Taylor~L.}\ \bibnamefont {Hughes}},\
  }\bibfield  {title} {\enquote {\bibinfo {title} {{Weak-pairing higher order
  topological superconductors}},}\ }\href {\doibase 10.1103/PhysRevB.98.165144}
  {\bibfield  {journal} {\bibinfo  {journal} {Phys. Rev. B}\ }\textbf {\bibinfo
  {volume} {98}},\ \bibinfo {pages} {165144} (\bibinfo {year}
  {2018}{\natexlab{c}})}\BibitemShut {NoStop}%
\bibitem [{\citenamefont {Zhu}(2018)}]{zhu2018tunable}%
  \BibitemOpen
  \bibfield  {author} {\bibinfo {author} {\bibfnamefont {Xiaoyu}\ \bibnamefont
  {Zhu}},\ }\bibfield  {title} {\enquote {\bibinfo {title} {{Tunable Majorana
  corner states in a two-dimensional second-order topological superconductor
  induced by magnetic fields}},}\ }\href {\doibase 10.1103/PhysRevB.97.205134}
  {\bibfield  {journal} {\bibinfo  {journal} {Phys. Rev. B}\ }\textbf {\bibinfo
  {volume} {97}},\ \bibinfo {pages} {205134} (\bibinfo {year}
  {2018})}\BibitemShut {NoStop}%
\bibitem [{\citenamefont {Geier}\ \emph {et~al.}(2018)\citenamefont {Geier},
  \citenamefont {Trifunovic}, \citenamefont {Hoskam},\ and\ \citenamefont
  {Brouwer}}]{geier2018second}%
  \BibitemOpen
  \bibfield  {author} {\bibinfo {author} {\bibfnamefont {Max}\ \bibnamefont
  {Geier}}, \bibinfo {author} {\bibfnamefont {Luka}\ \bibnamefont
  {Trifunovic}}, \bibinfo {author} {\bibfnamefont {Max}\ \bibnamefont
  {Hoskam}}, \ and\ \bibinfo {author} {\bibfnamefont {Piet~W.}\ \bibnamefont
  {Brouwer}},\ }\bibfield  {title} {\enquote {\bibinfo {title} {{Second-order
  topological insulators and superconductors with an order-two crystalline
  symmetry}},}\ }\href {\doibase 10.1103/PhysRevB.97.205135} {\bibfield
  {journal} {\bibinfo  {journal} {Phys. Rev. B}\ }\textbf {\bibinfo {volume}
  {97}},\ \bibinfo {pages} {205135} (\bibinfo {year} {2018})}\BibitemShut
  {NoStop}%
\bibitem [{\citenamefont {Khalaf}(2018)}]{khalaf2018higher}%
  \BibitemOpen
  \bibfield  {author} {\bibinfo {author} {\bibfnamefont {Eslam}\ \bibnamefont
  {Khalaf}},\ }\bibfield  {title} {\enquote {\bibinfo {title} {{Higher-order
  topological insulators and superconductors protected by inversion
  symmetry}},}\ }\href {\doibase 10.1103/PhysRevB.97.205136} {\bibfield
  {journal} {\bibinfo  {journal} {Phys. Rev. B}\ }\textbf {\bibinfo {volume}
  {97}},\ \bibinfo {pages} {205136} (\bibinfo {year} {2018})}\BibitemShut
  {NoStop}%
\bibitem [{\citenamefont {Pan}\ \emph {et~al.}(2019)\citenamefont {Pan},
  \citenamefont {Yang}, \citenamefont {Chen}, \citenamefont {Xu}, \citenamefont
  {Liu},\ and\ \citenamefont {Liu}}]{pan2018lattice}%
  \BibitemOpen
  \bibfield  {author} {\bibinfo {author} {\bibfnamefont {Xiao-Hong}\
  \bibnamefont {Pan}}, \bibinfo {author} {\bibfnamefont {Kai-Jie}\ \bibnamefont
  {Yang}}, \bibinfo {author} {\bibfnamefont {Li}~\bibnamefont {Chen}}, \bibinfo
  {author} {\bibfnamefont {Gang}\ \bibnamefont {Xu}}, \bibinfo {author}
  {\bibfnamefont {Chao-Xing}\ \bibnamefont {Liu}}, \ and\ \bibinfo {author}
  {\bibfnamefont {Xin}\ \bibnamefont {Liu}},\ }\bibfield  {title} {\enquote
  {\bibinfo {title} {{Lattice-Symmetry-Assisted Second-Order Topological
  Superconductors and Majorana Patterns}},}\ }\href@noop {} {\bibfield
  {journal} {\bibinfo  {journal} {Phys. Rev. Lett.}\ }\textbf {\bibinfo
  {volume} {123}},\ \bibinfo {pages} {156801} (\bibinfo {year}
  {2019})}\BibitemShut {NoStop}%
\bibitem [{\citenamefont {Wu}\ \emph {et~al.}(2019{\natexlab{a}})\citenamefont
  {Wu}, \citenamefont {Yan},\ and\ \citenamefont {Huang}}]{wu2019higher}%
  \BibitemOpen
  \bibfield  {author} {\bibinfo {author} {\bibfnamefont {Zhigang}\ \bibnamefont
  {Wu}}, \bibinfo {author} {\bibfnamefont {Zhongbo}\ \bibnamefont {Yan}}, \
  and\ \bibinfo {author} {\bibfnamefont {Wen}\ \bibnamefont {Huang}},\
  }\bibfield  {title} {\enquote {\bibinfo {title} {{Higher-order topological
  superconductivity: Possible realization in Fermi gases and
  ${\mathrm{Sr}}_{2}{\mathrm{RuO}}_{4}$}},}\ }\href {\doibase
  10.1103/PhysRevB.99.020508} {\bibfield  {journal} {\bibinfo  {journal} {Phys.
  Rev. B}\ }\textbf {\bibinfo {volume} {99}},\ \bibinfo {pages} {020508}
  (\bibinfo {year} {2019}{\natexlab{a}})}\BibitemShut {NoStop}%
\bibitem [{\citenamefont {Zhang}\ \emph
  {et~al.}(2019{\natexlab{b}})\citenamefont {Zhang}, \citenamefont {Cole},\
  and\ \citenamefont {Das~Sarma}}]{ZhangRX2019PRL}%
  \BibitemOpen
  \bibfield  {author} {\bibinfo {author} {\bibfnamefont {Rui-Xing}\
  \bibnamefont {Zhang}}, \bibinfo {author} {\bibfnamefont {William~S.}\
  \bibnamefont {Cole}}, \ and\ \bibinfo {author} {\bibfnamefont
  {S.}~\bibnamefont {Das~Sarma}},\ }\bibfield  {title} {\enquote {\bibinfo
  {title} {{Helical Hinge Majorana Modes in Iron-Based Superconductors}},}\
  }\href {\doibase 10.1103/PhysRevLett.122.187001} {\bibfield  {journal}
  {\bibinfo  {journal} {Phys. Rev. Lett.}\ }\textbf {\bibinfo {volume} {122}},\
  \bibinfo {pages} {187001} (\bibinfo {year} {2019}{\natexlab{b}})}\BibitemShut
  {NoStop}%
\bibitem [{\citenamefont {Volpez}\ \emph {et~al.}(2019)\citenamefont {Volpez},
  \citenamefont {Loss},\ and\ \citenamefont {Klinovaja}}]{Volpez2019}%
  \BibitemOpen
  \bibfield  {author} {\bibinfo {author} {\bibfnamefont {Yanick}\ \bibnamefont
  {Volpez}}, \bibinfo {author} {\bibfnamefont {Daniel}\ \bibnamefont {Loss}}, \
  and\ \bibinfo {author} {\bibfnamefont {Jelena}\ \bibnamefont {Klinovaja}},\
  }\bibfield  {title} {\enquote {\bibinfo {title} {{Second-Order Topological
  Superconductivity in $\ensuremath{\pi}$-Junction Rashba Layers}},}\ }\href
  {\doibase 10.1103/PhysRevLett.122.126402} {\bibfield  {journal} {\bibinfo
  {journal} {Phys. Rev. Lett.}\ }\textbf {\bibinfo {volume} {122}},\ \bibinfo
  {pages} {126402} (\bibinfo {year} {2019})}\BibitemShut {NoStop}%
\bibitem [{\citenamefont {Ghorashi}\ \emph {et~al.}(2019)\citenamefont
  {Ghorashi}, \citenamefont {Hu}, \citenamefont {Hughes},\ and\ \citenamefont
  {Rossi}}]{Ghorashi2019}%
  \BibitemOpen
  \bibfield  {author} {\bibinfo {author} {\bibfnamefont {Sayed Ali~Akbar}\
  \bibnamefont {Ghorashi}}, \bibinfo {author} {\bibfnamefont {Xiang}\
  \bibnamefont {Hu}}, \bibinfo {author} {\bibfnamefont {Taylor~L.}\
  \bibnamefont {Hughes}}, \ and\ \bibinfo {author} {\bibfnamefont {Enrico}\
  \bibnamefont {Rossi}},\ }\bibfield  {title} {\enquote {\bibinfo {title}
  {Second-order dirac superconductors and magnetic field induced majorana hinge
  modes},}\ }\href@noop {} {\bibfield  {journal} {\bibinfo  {journal} {Phys.
  Rev. B}\ }\textbf {\bibinfo {volume} {100}},\ \bibinfo {pages} {020509}
  (\bibinfo {year} {2019})}\BibitemShut {NoStop}%
\bibitem [{\citenamefont {Wu}\ \emph {et~al.}(2020)\citenamefont {Wu},
  \citenamefont {Hou}, \citenamefont {Li}, \citenamefont {Luo}, \citenamefont
  {Shi},\ and\ \citenamefont {Zhang}}]{wu2019inplane}%
  \BibitemOpen
  \bibfield  {author} {\bibinfo {author} {\bibfnamefont {Ya-Jie}\ \bibnamefont
  {Wu}}, \bibinfo {author} {\bibfnamefont {Junpeng}\ \bibnamefont {Hou}},
  \bibinfo {author} {\bibfnamefont {Yun-Mei}\ \bibnamefont {Li}}, \bibinfo
  {author} {\bibfnamefont {Xi-Wang}\ \bibnamefont {Luo}}, \bibinfo {author}
  {\bibfnamefont {Xiaoyan}\ \bibnamefont {Shi}}, \ and\ \bibinfo {author}
  {\bibfnamefont {Chuanwei}\ \bibnamefont {Zhang}},\ }\bibfield  {title}
  {\enquote {\bibinfo {title} {{In-Plane Zeeman-Field-Induced Majorana Corner
  and Hinge Modes in an $s$-Wave Superconductor Heterostructure}},}\ }\href
  {\doibase 10.1103/PhysRevLett.124.227001} {\bibfield  {journal} {\bibinfo
  {journal} {Phys. Rev. Lett.}\ }\textbf {\bibinfo {volume} {124}},\ \bibinfo
  {pages} {227001} (\bibinfo {year} {2020})}\BibitemShut {NoStop}%
\bibitem [{\citenamefont {Zhang}\ \emph
  {et~al.}(2019{\natexlab{c}})\citenamefont {Zhang}, \citenamefont {Cole},
  \citenamefont {Wu},\ and\ \citenamefont {Das~Sarma}}]{ZhangRX-PRL20192}%
  \BibitemOpen
  \bibfield  {author} {\bibinfo {author} {\bibfnamefont {Rui-Xing}\
  \bibnamefont {Zhang}}, \bibinfo {author} {\bibfnamefont {William~S.}\
  \bibnamefont {Cole}}, \bibinfo {author} {\bibfnamefont {Xianxin}\
  \bibnamefont {Wu}}, \ and\ \bibinfo {author} {\bibfnamefont {S.}~\bibnamefont
  {Das~Sarma}},\ }\bibfield  {title} {\enquote {\bibinfo {title} {{Higher-Order
  Topology and Nodal Topological Superconductivity in Fe(Se,Te)
  Heterostructures}},}\ }\href {\doibase 10.1103/PhysRevLett.123.167001}
  {\bibfield  {journal} {\bibinfo  {journal} {Phys. Rev. Lett.}\ }\textbf
  {\bibinfo {volume} {123}},\ \bibinfo {pages} {167001} (\bibinfo {year}
  {2019}{\natexlab{c}})}\BibitemShut {NoStop}%
\bibitem [{\citenamefont {Wu}\ \emph {et~al.}(2019{\natexlab{b}})\citenamefont
  {Wu}, \citenamefont {Liu}, \citenamefont {Thomale},\ and\ \citenamefont
  {Liu}}]{Wufetese2019}%
  \BibitemOpen
  \bibfield  {author} {\bibinfo {author} {\bibfnamefont {Xianxin}\ \bibnamefont
  {Wu}}, \bibinfo {author} {\bibfnamefont {Xin}\ \bibnamefont {Liu}}, \bibinfo
  {author} {\bibfnamefont {Ronny}\ \bibnamefont {Thomale}}, \ and\ \bibinfo
  {author} {\bibfnamefont {Chao-Xing}\ \bibnamefont {Liu}},\ }\bibfield
  {title} {\enquote {\bibinfo {title} {{High-$T_c$ Superconductor Fe(Se,Te)
  Monolayer: an Intrinsic, Scalable and Electrically-tunable Majorana
  Platform}},}\ }\href@noop {} {\ ,\ \bibinfo {pages} {arXiv:1905.10648}
  (\bibinfo {year} {2019}{\natexlab{b}})}\BibitemShut {NoStop}%
\bibitem [{\citenamefont {Khalaf}\ \emph {et~al.}(2019)\citenamefont {Khalaf},
  \citenamefont {Benalcazar}, \citenamefont {Hughes},\ and\ \citenamefont
  {Queiroz}}]{khalaf2019boundary}%
  \BibitemOpen
  \bibfield  {author} {\bibinfo {author} {\bibfnamefont {Eslam}\ \bibnamefont
  {Khalaf}}, \bibinfo {author} {\bibfnamefont {Wladimir~A}\ \bibnamefont
  {Benalcazar}}, \bibinfo {author} {\bibfnamefont {Taylor~L}\ \bibnamefont
  {Hughes}}, \ and\ \bibinfo {author} {\bibfnamefont {Raquel}\ \bibnamefont
  {Queiroz}},\ }\bibfield  {title} {\enquote {\bibinfo {title}
  {Boundary-obstructed topological phases},}\ }\href@noop {} {\ ,\ \bibinfo
  {pages} {arXiv:1908.00011} (\bibinfo {year} {2019})}\BibitemShut {NoStop}%
\bibitem [{\citenamefont {Katayama}\ \emph {et~al.}(2013)\citenamefont
  {Katayama}, \citenamefont {Kudo}, \citenamefont {Onari}, \citenamefont
  {Mizukami}, \citenamefont {Sugawara}, \citenamefont {Sugiyama}, \citenamefont
  {Kitahama}, \citenamefont {Iba}, \citenamefont {Fujimura}, \citenamefont
  {Nishimoto}, \citenamefont {Nohara},\ and\ \citenamefont
  {Sawa}}]{Katayama2013}%
  \BibitemOpen
  \bibfield  {author} {\bibinfo {author} {\bibfnamefont {Naoyuki}\ \bibnamefont
  {Katayama}}, \bibinfo {author} {\bibfnamefont {Kazutaka}\ \bibnamefont
  {Kudo}}, \bibinfo {author} {\bibfnamefont {Seiichiro}\ \bibnamefont {Onari}},
  \bibinfo {author} {\bibfnamefont {Tasuku}\ \bibnamefont {Mizukami}}, \bibinfo
  {author} {\bibfnamefont {Kento}\ \bibnamefont {Sugawara}}, \bibinfo {author}
  {\bibfnamefont {Yuki}\ \bibnamefont {Sugiyama}}, \bibinfo {author}
  {\bibfnamefont {Yutaka}\ \bibnamefont {Kitahama}}, \bibinfo {author}
  {\bibfnamefont {Keita}\ \bibnamefont {Iba}}, \bibinfo {author} {\bibfnamefont
  {Kazunori}\ \bibnamefont {Fujimura}}, \bibinfo {author} {\bibfnamefont
  {Naoki}\ \bibnamefont {Nishimoto}}, \bibinfo {author} {\bibfnamefont
  {Minoru}\ \bibnamefont {Nohara}}, \ and\ \bibinfo {author} {\bibfnamefont
  {Hiroshi}\ \bibnamefont {Sawa}},\ }\bibfield  {title} {\enquote {\bibinfo
  {title} {{Superconductivity in Ca$_{1-x}$La$_x$FeAs$_2$: A Novel 112-Type
  Iron Pnictide with Arsenic Zigzag Bonds}},}\ }\href {\doibase
  10.7566/JPSJ.82.123702} {\bibfield  {journal} {\bibinfo  {journal} {J. Phys.
  Soc. Jpn.}\ }\textbf {\bibinfo {volume} {82}},\ \bibinfo {pages} {123702}
  (\bibinfo {year} {2013})}\BibitemShut {NoStop}%
\bibitem [{\citenamefont {Yakita}\ \emph {et~al.}(2014)\citenamefont {Yakita},
  \citenamefont {Ogino}, \citenamefont {Okada}, \citenamefont {Yamamoto},
  \citenamefont {Kishio}, \citenamefont {Tohei}, \citenamefont {Ikuhara},
  \citenamefont {Gotoh}, \citenamefont {Fujihisa}, \citenamefont {Kataoka},
  \citenamefont {Eisaki},\ and\ \citenamefont {Shimoyama}}]{Yakita2013}%
  \BibitemOpen
  \bibfield  {author} {\bibinfo {author} {\bibfnamefont {Hiroyuki}\
  \bibnamefont {Yakita}}, \bibinfo {author} {\bibfnamefont {Hiraku}\
  \bibnamefont {Ogino}}, \bibinfo {author} {\bibfnamefont {Tomoyuki}\
  \bibnamefont {Okada}}, \bibinfo {author} {\bibfnamefont {Akiyasu}\
  \bibnamefont {Yamamoto}}, \bibinfo {author} {\bibfnamefont {Kohji}\
  \bibnamefont {Kishio}}, \bibinfo {author} {\bibfnamefont {Tetsuya}\
  \bibnamefont {Tohei}}, \bibinfo {author} {\bibfnamefont {Yuichi}\
  \bibnamefont {Ikuhara}}, \bibinfo {author} {\bibfnamefont {Yoshito}\
  \bibnamefont {Gotoh}}, \bibinfo {author} {\bibfnamefont {Hiroshi}\
  \bibnamefont {Fujihisa}}, \bibinfo {author} {\bibfnamefont {Kunimitsu}\
  \bibnamefont {Kataoka}}, \bibinfo {author} {\bibfnamefont {Hiroshi}\
  \bibnamefont {Eisaki}}, \ and\ \bibinfo {author} {\bibfnamefont {Jun-ichi}\
  \bibnamefont {Shimoyama}},\ }\bibfield  {title} {\enquote {\bibinfo {title}
  {{A New Layered Iron Arsenide Superconductor: (Ca,Pr)FeAs$_2$}},}\ }\href
  {\doibase 10.1021/ja410845b} {\bibfield  {journal} {\bibinfo  {journal} {J.
  Am. Chem. Soc.}\ }\textbf {\bibinfo {volume} {136}},\ \bibinfo {pages} {846}
  (\bibinfo {year} {2014})}\BibitemShut {NoStop}%
\bibitem [{\citenamefont {Kudo}\ \emph {et~al.}(2014)\citenamefont {Kudo},
  \citenamefont {Kitahama}, \citenamefont {Fujimura}, \citenamefont {Mizukami},
  \citenamefont {Ota},\ and\ \citenamefont {Nohara}}]{Kudo2014}%
  \BibitemOpen
  \bibfield  {author} {\bibinfo {author} {\bibfnamefont {Kazutaka}\
  \bibnamefont {Kudo}}, \bibinfo {author} {\bibfnamefont {Yutaka}\ \bibnamefont
  {Kitahama}}, \bibinfo {author} {\bibfnamefont {Kazunori}\ \bibnamefont
  {Fujimura}}, \bibinfo {author} {\bibfnamefont {Tasuku}\ \bibnamefont
  {Mizukami}}, \bibinfo {author} {\bibfnamefont {Hiromi}\ \bibnamefont {Ota}},
  \ and\ \bibinfo {author} {\bibfnamefont {Minoru}\ \bibnamefont {Nohara}},\
  }\bibfield  {title} {\enquote {\bibinfo {title} {{Superconducting Transition
  Temperatures of up to 47 K from Simultaneous Rare-Earth Element and Antimony
  Doping of 112-Type CaFeAs$_2$}},}\ }\href {\doibase 10.7566/JPSJ.83.093705}
  {\bibfield  {journal} {\bibinfo  {journal} {J. Phys. Soc. Jpn.}\ }\textbf
  {\bibinfo {volume} {83}},\ \bibinfo {pages} {093705} (\bibinfo {year}
  {2014})}\BibitemShut {NoStop}%
\bibitem [{\citenamefont {Wu}\ \emph {et~al.}(2014)\citenamefont {Wu},
  \citenamefont {Le}, \citenamefont {Liang}, \citenamefont {Qin}, \citenamefont
  {Fan},\ and\ \citenamefont {Hu}}]{WuXX2014}%
  \BibitemOpen
  \bibfield  {author} {\bibinfo {author} {\bibfnamefont {Xianxin}\ \bibnamefont
  {Wu}}, \bibinfo {author} {\bibfnamefont {Congcong}\ \bibnamefont {Le}},
  \bibinfo {author} {\bibfnamefont {Yi}~\bibnamefont {Liang}}, \bibinfo
  {author} {\bibfnamefont {Shengshan}\ \bibnamefont {Qin}}, \bibinfo {author}
  {\bibfnamefont {Heng}\ \bibnamefont {Fan}}, \ and\ \bibinfo {author}
  {\bibfnamefont {Jiangping}\ \bibnamefont {Hu}},\ }\bibfield  {title}
  {\enquote {\bibinfo {title} {{Effect of As-chain layers in
  ${\text{CaFeAs}}_{2}$}},}\ }\href {\doibase 10.1103/PhysRevB.89.205102}
  {\bibfield  {journal} {\bibinfo  {journal} {Phys. Rev. B}\ }\textbf {\bibinfo
  {volume} {89}},\ \bibinfo {pages} {205102} (\bibinfo {year}
  {2014})}\BibitemShut {NoStop}%
\bibitem [{\citenamefont {Wu}\ \emph {et~al.}(2015{\natexlab{a}})\citenamefont
  {Wu}, \citenamefont {Qin}, \citenamefont {Liang}, \citenamefont {Le},
  \citenamefont {Fan},\ and\ \citenamefont {Hu}}]{WuXX2015-PRB}%
  \BibitemOpen
  \bibfield  {author} {\bibinfo {author} {\bibfnamefont {Xianxin}\ \bibnamefont
  {Wu}}, \bibinfo {author} {\bibfnamefont {Shengshan}\ \bibnamefont {Qin}},
  \bibinfo {author} {\bibfnamefont {Yi}~\bibnamefont {Liang}}, \bibinfo
  {author} {\bibfnamefont {Congcong}\ \bibnamefont {Le}}, \bibinfo {author}
  {\bibfnamefont {Heng}\ \bibnamefont {Fan}}, \ and\ \bibinfo {author}
  {\bibfnamefont {Jiangping}\ \bibnamefont {Hu}},\ }\bibfield  {title}
  {\enquote {\bibinfo {title} {{${\mathrm{CaFeAs}}_{2}$: A staggered
  intercalation of quantum spin Hall and high-temperature
  superconductivity}},}\ }\href {\doibase 10.1103/PhysRevB.91.081111}
  {\bibfield  {journal} {\bibinfo  {journal} {Phys. Rev. B}\ }\textbf {\bibinfo
  {volume} {91}},\ \bibinfo {pages} {081111} (\bibinfo {year}
  {2015}{\natexlab{a}})}\BibitemShut {NoStop}%
\bibitem [{\citenamefont {Thomale}\ \emph {et~al.}(2011)\citenamefont
  {Thomale}, \citenamefont {Platt}, \citenamefont {Hanke},\ and\ \citenamefont
  {Bernevig}}]{Thomale2011}%
  \BibitemOpen
  \bibfield  {author} {\bibinfo {author} {\bibfnamefont {Ronny}\ \bibnamefont
  {Thomale}}, \bibinfo {author} {\bibfnamefont {Christian}\ \bibnamefont
  {Platt}}, \bibinfo {author} {\bibfnamefont {Werner}\ \bibnamefont {Hanke}}, \
  and\ \bibinfo {author} {\bibfnamefont {B.~Andrei}\ \bibnamefont {Bernevig}},\
  }\bibfield  {title} {\enquote {\bibinfo {title} {Mechanism for explaining
  differences in the order parameters of feas-based and fep-based pnictide
  superconductors},}\ }\href {\doibase 10.1103/PhysRevLett.106.187003}
  {\bibfield  {journal} {\bibinfo  {journal} {Phys. Rev. Lett.}\ }\textbf
  {\bibinfo {volume} {106}},\ \bibinfo {pages} {187003} (\bibinfo {year}
  {2011})}\BibitemShut {NoStop}%
\bibitem [{\citenamefont {Seo}\ \emph {et~al.}(2008)\citenamefont {Seo},
  \citenamefont {Bernevig},\ and\ \citenamefont {Hu}}]{Seo2008}%
  \BibitemOpen
  \bibfield  {author} {\bibinfo {author} {\bibfnamefont {Kangjun}\ \bibnamefont
  {Seo}}, \bibinfo {author} {\bibfnamefont {B.~Andrei}\ \bibnamefont
  {Bernevig}}, \ and\ \bibinfo {author} {\bibfnamefont {Jiangping}\
  \bibnamefont {Hu}},\ }\bibfield  {title} {\enquote {\bibinfo {title}
  {{Pairing Symmetry in a Two-Orbital Exchange Coupling Model of
  Oxypnictides}},}\ }\href@noop {} {\bibfield  {journal} {\bibinfo  {journal}
  {Phys. Rev. Lett.}\ }\textbf {\bibinfo {volume} {101}},\ \bibinfo {pages}
  {206404} (\bibinfo {year} {2008})}\BibitemShut {NoStop}%
\bibitem [{\citenamefont {Li}\ \emph {et~al.}(2015)\citenamefont {Li},
  \citenamefont {Liu}, \citenamefont {Zhou}, \citenamefont {Yang},
  \citenamefont {Shen}, \citenamefont {Li}, \citenamefont {Jiang},
  \citenamefont {Niu}, \citenamefont {Xie}, \citenamefont {Sun}, \citenamefont
  {Fan}, \citenamefont {Yao}, \citenamefont {Liu}, \citenamefont {Shi},\ and\
  \citenamefont {Xie}}]{LiMY2015}%
  \BibitemOpen
  \bibfield  {author} {\bibinfo {author} {\bibfnamefont {M.~Y.}\ \bibnamefont
  {Li}}, \bibinfo {author} {\bibfnamefont {Z.~T.}\ \bibnamefont {Liu}},
  \bibinfo {author} {\bibfnamefont {W.}~\bibnamefont {Zhou}}, \bibinfo {author}
  {\bibfnamefont {H.~F.}\ \bibnamefont {Yang}}, \bibinfo {author}
  {\bibfnamefont {D.~W.}\ \bibnamefont {Shen}}, \bibinfo {author}
  {\bibfnamefont {W.}~\bibnamefont {Li}}, \bibinfo {author} {\bibfnamefont
  {J.}~\bibnamefont {Jiang}}, \bibinfo {author} {\bibfnamefont {X.~H.}\
  \bibnamefont {Niu}}, \bibinfo {author} {\bibfnamefont {B.~P.}\ \bibnamefont
  {Xie}}, \bibinfo {author} {\bibfnamefont {Y.}~\bibnamefont {Sun}}, \bibinfo
  {author} {\bibfnamefont {C.~C.}\ \bibnamefont {Fan}}, \bibinfo {author}
  {\bibfnamefont {Q.}~\bibnamefont {Yao}}, \bibinfo {author} {\bibfnamefont
  {J.~S.}\ \bibnamefont {Liu}}, \bibinfo {author} {\bibfnamefont {Z.~X.}\
  \bibnamefont {Shi}}, \ and\ \bibinfo {author} {\bibfnamefont {X.~M.}\
  \bibnamefont {Xie}},\ }\bibfield  {title} {\enquote {\bibinfo {title}
  {Significant contribution of as $4p$ orbitals to the low-lying electronic
  structure of the 112-type iron-based superconductor
  $\mathrm{Ca}{}_{0.9}\mathrm{La}{}_{0.1}\mathrm{FeAs}{}_{2}$},}\ }\href
  {\doibase 10.1103/PhysRevB.91.045112} {\bibfield  {journal} {\bibinfo
  {journal} {Phys. Rev. B}\ }\textbf {\bibinfo {volume} {91}},\ \bibinfo
  {pages} {045112} (\bibinfo {year} {2015})}\BibitemShut {NoStop}%
\bibitem [{\citenamefont {Jiang}\ \emph {et~al.}(2016)\citenamefont {Jiang},
  \citenamefont {Liu}, \citenamefont {Cao}, \citenamefont {Birol},
  \citenamefont {Allred}, \citenamefont {Tian}, \citenamefont {Liu},
  \citenamefont {Cho}, \citenamefont {Krogstad}, \citenamefont {Ma},
  \citenamefont {Taddei}, \citenamefont {Tanatar}, \citenamefont {Hoesch},
  \citenamefont {Prozorov}, \citenamefont {Rosenkranz}, \citenamefont {Uemura},
  \citenamefont {Kotliar},\ and\ \citenamefont {Ni}}]{JiangS2016}%
  \BibitemOpen
  \bibfield  {author} {\bibinfo {author} {\bibfnamefont {Shan}\ \bibnamefont
  {Jiang}}, \bibinfo {author} {\bibfnamefont {Chang}\ \bibnamefont {Liu}},
  \bibinfo {author} {\bibfnamefont {Huibo}\ \bibnamefont {Cao}}, \bibinfo
  {author} {\bibfnamefont {Turan}\ \bibnamefont {Birol}}, \bibinfo {author}
  {\bibfnamefont {Jared~M.}\ \bibnamefont {Allred}}, \bibinfo {author}
  {\bibfnamefont {Wei}\ \bibnamefont {Tian}}, \bibinfo {author} {\bibfnamefont
  {Lian}\ \bibnamefont {Liu}}, \bibinfo {author} {\bibfnamefont {Kyuil}\
  \bibnamefont {Cho}}, \bibinfo {author} {\bibfnamefont {Matthew~J.}\
  \bibnamefont {Krogstad}}, \bibinfo {author} {\bibfnamefont {Jie}\
  \bibnamefont {Ma}}, \bibinfo {author} {\bibfnamefont {Keith~M.}\ \bibnamefont
  {Taddei}}, \bibinfo {author} {\bibfnamefont {Makariy~A.}\ \bibnamefont
  {Tanatar}}, \bibinfo {author} {\bibfnamefont {Moritz}\ \bibnamefont
  {Hoesch}}, \bibinfo {author} {\bibfnamefont {Ruslan}\ \bibnamefont
  {Prozorov}}, \bibinfo {author} {\bibfnamefont {Stephan}\ \bibnamefont
  {Rosenkranz}}, \bibinfo {author} {\bibfnamefont {Yasutomo~J.}\ \bibnamefont
  {Uemura}}, \bibinfo {author} {\bibfnamefont {Gabriel}\ \bibnamefont
  {Kotliar}}, \ and\ \bibinfo {author} {\bibfnamefont {Ni}~\bibnamefont {Ni}},\
  }\bibfield  {title} {\enquote {\bibinfo {title} {{Structural and magnetic
  phase transitions in
  ${\mathrm{Ca}}_{0.73}{\mathrm{La}}_{0.27}{\mathrm{FeAs}}_{2}$ with
  electron-overdoped FeAs layers}},}\ }\href {\doibase
  10.1103/PhysRevB.93.054522} {\bibfield  {journal} {\bibinfo  {journal} {Phys.
  Rev. B}\ }\textbf {\bibinfo {volume} {93}},\ \bibinfo {pages} {054522}
  (\bibinfo {year} {2016})}\BibitemShut {NoStop}%
\bibitem [{\citenamefont {Liu}\ \emph {et~al.}(2016)\citenamefont {Liu},
  \citenamefont {Xing}, \citenamefont {Li}, \citenamefont {Zhou}, \citenamefont
  {Sun}, \citenamefont {Fan}, \citenamefont {Yang}, \citenamefont {Liu},
  \citenamefont {Yao}, \citenamefont {Li}, \citenamefont {Shi}, \citenamefont
  {Shen},\ and\ \citenamefont {Wang}}]{LiuZT2016}%
  \BibitemOpen
  \bibfield  {author} {\bibinfo {author} {\bibfnamefont {Z.~T.}\ \bibnamefont
  {Liu}}, \bibinfo {author} {\bibfnamefont {X.~Z.}\ \bibnamefont {Xing}},
  \bibinfo {author} {\bibfnamefont {M.~Y.}\ \bibnamefont {Li}}, \bibinfo
  {author} {\bibfnamefont {W.}~\bibnamefont {Zhou}}, \bibinfo {author}
  {\bibfnamefont {Y.}~\bibnamefont {Sun}}, \bibinfo {author} {\bibfnamefont
  {C.~C.}\ \bibnamefont {Fan}}, \bibinfo {author} {\bibfnamefont {H.~F.}\
  \bibnamefont {Yang}}, \bibinfo {author} {\bibfnamefont {J.~S.}\ \bibnamefont
  {Liu}}, \bibinfo {author} {\bibfnamefont {Q.}~\bibnamefont {Yao}}, \bibinfo
  {author} {\bibfnamefont {W.}~\bibnamefont {Li}}, \bibinfo {author}
  {\bibfnamefont {Z.~X.}\ \bibnamefont {Shi}}, \bibinfo {author} {\bibfnamefont
  {D.~W.}\ \bibnamefont {Shen}}, \ and\ \bibinfo {author} {\bibfnamefont
  {Z.}~\bibnamefont {Wang}},\ }\bibfield  {title} {\enquote {\bibinfo {title}
  {{Observation of the anisotropic Dirac cone in the band dispersion of
  112-structured iron-based superconductor Ca$_{0.9}$La$_{0.1}$FeAs$_2$}},}\
  }\href {\doibase 10.1063/1.4960164} {\bibfield  {journal} {\bibinfo
  {journal} {Appl. Phys. Lett.}\ }\textbf {\bibinfo {volume} {109}},\ \bibinfo
  {pages} {042602} (\bibinfo {year} {2016})}\BibitemShut {NoStop}%
\bibitem [{\citenamefont {Zhang}\ and\ \citenamefont
  {Trauzettel}(2020)}]{ZhangSB2019}%
  \BibitemOpen
  \bibfield  {author} {\bibinfo {author} {\bibfnamefont {Song-Bo}\ \bibnamefont
  {Zhang}}\ and\ \bibinfo {author} {\bibfnamefont {Bj\"orn}\ \bibnamefont
  {Trauzettel}},\ }\bibfield  {title} {\enquote {\bibinfo {title} {Detection of
  second-order topological superconductors by josephson junctions},}\
  }\href@noop {} {\bibfield  {journal} {\bibinfo  {journal} {Phys. Rev.
  Research}\ }\textbf {\bibinfo {volume} {2}},\ \bibinfo {pages} {012018}
  (\bibinfo {year} {2020})}\BibitemShut {NoStop}%
\bibitem [{\citenamefont {Chiu}\ \emph {et~al.}(2016)\citenamefont {Chiu},
  \citenamefont {Teo}, \citenamefont {Schnyder},\ and\ \citenamefont
  {Ryu}}]{Chiu2016}%
  \BibitemOpen
  \bibfield  {author} {\bibinfo {author} {\bibfnamefont {Ching-Kai}\
  \bibnamefont {Chiu}}, \bibinfo {author} {\bibfnamefont {Jeffrey C.~Y.}\
  \bibnamefont {Teo}}, \bibinfo {author} {\bibfnamefont {Andreas~P.}\
  \bibnamefont {Schnyder}}, \ and\ \bibinfo {author} {\bibfnamefont {Shinsei}\
  \bibnamefont {Ryu}},\ }\bibfield  {title} {\enquote {\bibinfo {title}
  {{Classification of topological quantum matter with symmetries}},}\
  }\href@noop {} {\bibfield  {journal} {\bibinfo  {journal} {Rev. Mod. Phys.}\
  }\textbf {\bibinfo {volume} {88}},\ \bibinfo {pages} {035005} (\bibinfo
  {year} {2016})}\BibitemShut {NoStop}%
\bibitem [{\citenamefont {Gray}\ \emph {et~al.}(2019)\citenamefont {Gray},
  \citenamefont {Freudenstein}, \citenamefont {Zhao}, \citenamefont
  {O’Connor}, \citenamefont {Jenkins}, \citenamefont {Kumar}, \citenamefont
  {Hoek}, \citenamefont {Kopec}, \citenamefont {Huh}, \citenamefont
  {Taniguchi}, \citenamefont {Watanabe}, \citenamefont {Zhong}, \citenamefont
  {Kim}, \citenamefont {Gu},\ and\ \citenamefont {Burch}}]{Gray2019}%
  \BibitemOpen
  \bibfield  {author} {\bibinfo {author} {\bibfnamefont {Mason~J.}\
  \bibnamefont {Gray}}, \bibinfo {author} {\bibfnamefont {Josef}\ \bibnamefont
  {Freudenstein}}, \bibinfo {author} {\bibfnamefont {Shu Yang~F.}\ \bibnamefont
  {Zhao}}, \bibinfo {author} {\bibfnamefont {Ryan}\ \bibnamefont {O’Connor}},
  \bibinfo {author} {\bibfnamefont {Samuel}\ \bibnamefont {Jenkins}}, \bibinfo
  {author} {\bibfnamefont {Narendra}\ \bibnamefont {Kumar}}, \bibinfo {author}
  {\bibfnamefont {Marcel}\ \bibnamefont {Hoek}}, \bibinfo {author}
  {\bibfnamefont {Abigail}\ \bibnamefont {Kopec}}, \bibinfo {author}
  {\bibfnamefont {Soonsang}\ \bibnamefont {Huh}}, \bibinfo {author}
  {\bibfnamefont {Takashi}\ \bibnamefont {Taniguchi}}, \bibinfo {author}
  {\bibfnamefont {Kenji}\ \bibnamefont {Watanabe}}, \bibinfo {author}
  {\bibfnamefont {Ruidan}\ \bibnamefont {Zhong}}, \bibinfo {author}
  {\bibfnamefont {Changyoung}\ \bibnamefont {Kim}}, \bibinfo {author}
  {\bibfnamefont {G.~D.}\ \bibnamefont {Gu}}, \ and\ \bibinfo {author}
  {\bibfnamefont {K.~S.}\ \bibnamefont {Burch}},\ }\bibfield  {title} {\enquote
  {\bibinfo {title} {Evidence for helical hinge zero modes in an fe-based
  superconductor},}\ }\href@noop {} {\bibfield  {journal} {\bibinfo  {journal}
  {Nano Letters}\ }\textbf {\bibinfo {volume} {19}},\ \bibinfo {pages} {4890}
  (\bibinfo {year} {2019})}\BibitemShut {NoStop}%
\bibitem [{\citenamefont {Graser}\ \emph {et~al.}(2009)\citenamefont {Graser},
  \citenamefont {Maier}, \citenamefont {Hirschfeld},\ and\ \citenamefont
  {Scalapino}}]{Graser2009}%
  \BibitemOpen
  \bibfield  {author} {\bibinfo {author} {\bibfnamefont {S}~\bibnamefont
  {Graser}}, \bibinfo {author} {\bibfnamefont {T~A}\ \bibnamefont {Maier}},
  \bibinfo {author} {\bibfnamefont {P~J}\ \bibnamefont {Hirschfeld}}, \ and\
  \bibinfo {author} {\bibfnamefont {D~J}\ \bibnamefont {Scalapino}},\
  }\bibfield  {title} {\enquote {\bibinfo {title} {Near-degeneracy of several
  pairing channels in multiorbital models for the fe pnictides},}\ }\href@noop
  {} {\bibfield  {journal} {\bibinfo  {journal} {New Journal of Physics}\
  }\textbf {\bibinfo {volume} {11}},\ \bibinfo {pages} {025016} (\bibinfo
  {year} {2009})}\BibitemShut {NoStop}%
\bibitem [{\citenamefont {Berk}\ and\ \citenamefont
  {Schrieffer}(1966)}]{Berk1966}%
  \BibitemOpen
  \bibfield  {author} {\bibinfo {author} {\bibfnamefont {N.~F.}\ \bibnamefont
  {Berk}}\ and\ \bibinfo {author} {\bibfnamefont {J.~R.}\ \bibnamefont
  {Schrieffer}},\ }\bibfield  {title} {\enquote {\bibinfo {title} {Effect of
  ferromagnetic spin correlations on superconductivity},}\ }\href@noop {}
  {\bibfield  {journal} {\bibinfo  {journal} {Phys. Rev. Lett.}\ }\textbf
  {\bibinfo {volume} {17}},\ \bibinfo {pages} {433--435} (\bibinfo {year}
  {1966})}\BibitemShut {NoStop}%
\bibitem [{\citenamefont {Scalapino}\ \emph {et~al.}(1986)\citenamefont
  {Scalapino}, \citenamefont {Loh},\ and\ \citenamefont
  {Hirsch}}]{Scalapino1986}%
  \BibitemOpen
  \bibfield  {author} {\bibinfo {author} {\bibfnamefont {D.~J.}\ \bibnamefont
  {Scalapino}}, \bibinfo {author} {\bibfnamefont {E.}~\bibnamefont {Loh}}, \
  and\ \bibinfo {author} {\bibfnamefont {J.~E.}\ \bibnamefont {Hirsch}},\
  }\bibfield  {title} {\enquote {\bibinfo {title} {$d$-wave pairing near a
  spin-density-wave instability},}\ }\href@noop {} {\bibfield  {journal}
  {\bibinfo  {journal} {Phys. Rev. B}\ }\textbf {\bibinfo {volume} {34}},\
  \bibinfo {pages} {8190--8192} (\bibinfo {year} {1986})}\BibitemShut {NoStop}%
\bibitem [{\citenamefont {Kemper}\ \emph {et~al.}(2010)\citenamefont {Kemper},
  \citenamefont {Maier}, \citenamefont {Graser}, \citenamefont {Cheng},
  \citenamefont {Hirschfeld},\ and\ \citenamefont {Scalapino}}]{Kemper2010}%
  \BibitemOpen
  \bibfield  {author} {\bibinfo {author} {\bibfnamefont {A~F}\ \bibnamefont
  {Kemper}}, \bibinfo {author} {\bibfnamefont {T~A}\ \bibnamefont {Maier}},
  \bibinfo {author} {\bibfnamefont {S}~\bibnamefont {Graser}}, \bibinfo
  {author} {\bibfnamefont {H-P}\ \bibnamefont {Cheng}}, \bibinfo {author}
  {\bibfnamefont {P~J}\ \bibnamefont {Hirschfeld}}, \ and\ \bibinfo {author}
  {\bibfnamefont {D~J}\ \bibnamefont {Scalapino}},\ }\bibfield  {title}
  {\enquote {\bibinfo {title} {Sensitivity of the superconducting state and
  magnetic susceptibility to key aspects of electronic structure in
  ferropnictides},}\ }\href@noop {} {\bibfield  {journal} {\bibinfo  {journal}
  {New Journal of Physics}\ }\textbf {\bibinfo {volume} {12}},\ \bibinfo
  {pages} {073030} (\bibinfo {year} {2010})}\BibitemShut {NoStop}%
\bibitem [{\citenamefont {Wu}\ \emph {et~al.}(2015{\natexlab{b}})\citenamefont
  {Wu}, \citenamefont {Yang}, \citenamefont {Le}, \citenamefont {Fan},\ and\
  \citenamefont {Hu}}]{WuXX2015}%
  \BibitemOpen
  \bibfield  {author} {\bibinfo {author} {\bibfnamefont {Xianxin}\ \bibnamefont
  {Wu}}, \bibinfo {author} {\bibfnamefont {Fan}\ \bibnamefont {Yang}}, \bibinfo
  {author} {\bibfnamefont {Congcong}\ \bibnamefont {Le}}, \bibinfo {author}
  {\bibfnamefont {Heng}\ \bibnamefont {Fan}}, \ and\ \bibinfo {author}
  {\bibfnamefont {Jiangping}\ \bibnamefont {Hu}},\ }\bibfield  {title}
  {\enquote {\bibinfo {title} {Triplet ${p}_{z}$-wave pairing in
  quasi-one-dimensional ${A}_{2}{\mathrm{cr}}_{3}{\mathrm{as}}_{3}$
  superconductors $(a=\mathrm{K},\mathrm{Rb},\mathrm{Cs})$},}\ }\href@noop {}
  {\bibfield  {journal} {\bibinfo  {journal} {Phys. Rev. B}\ }\textbf {\bibinfo
  {volume} {92}},\ \bibinfo {pages} {104511} (\bibinfo {year}
  {2015}{\natexlab{b}})}\BibitemShut {NoStop}%
\bibitem [{\citenamefont {Takimoto}\ \emph {et~al.}(2004)\citenamefont
  {Takimoto}, \citenamefont {Hotta},\ and\ \citenamefont
  {Ueda}}]{Takimoto2004}%
  \BibitemOpen
  \bibfield  {author} {\bibinfo {author} {\bibfnamefont {Tetsuya}\ \bibnamefont
  {Takimoto}}, \bibinfo {author} {\bibfnamefont {Takashi}\ \bibnamefont
  {Hotta}}, \ and\ \bibinfo {author} {\bibfnamefont {Kazuo}\ \bibnamefont
  {Ueda}},\ }\bibfield  {title} {\enquote {\bibinfo {title} {Strong-coupling
  theory of superconductivity in a degenerate hubbard model},}\ }\href@noop {}
  {\bibfield  {journal} {\bibinfo  {journal} {Phys. Rev. B}\ }\textbf {\bibinfo
  {volume} {69}},\ \bibinfo {pages} {104504} (\bibinfo {year}
  {2004})}\BibitemShut {NoStop}%
\bibitem [{\citenamefont {Kubo}(2007)}]{Kubo2007}%
  \BibitemOpen
  \bibfield  {author} {\bibinfo {author} {\bibfnamefont {Katsunori}\
  \bibnamefont {Kubo}},\ }\bibfield  {title} {\enquote {\bibinfo {title}
  {Pairing symmetry in a two-orbital hubbard model on a square lattice},}\
  }\href@noop {} {\bibfield  {journal} {\bibinfo  {journal} {Phys. Rev. B}\
  }\textbf {\bibinfo {volume} {75}},\ \bibinfo {pages} {224509} (\bibinfo
  {year} {2007})}\BibitemShut {NoStop}%
\bibitem [{\citenamefont {Qi}\ \emph {et~al.}(2010)\citenamefont {Qi},
  \citenamefont {Hughes},\ and\ \citenamefont {Zhang}}]{QiXL2010}%
  \BibitemOpen
  \bibfield  {author} {\bibinfo {author} {\bibfnamefont {Xiao-Liang}\
  \bibnamefont {Qi}}, \bibinfo {author} {\bibfnamefont {Taylor~L.}\
  \bibnamefont {Hughes}}, \ and\ \bibinfo {author} {\bibfnamefont {Shou-Cheng}\
  \bibnamefont {Zhang}},\ }\bibfield  {title} {\enquote {\bibinfo {title}
  {Topological invariants for the fermi surface of a time-reversal-invariant
  superconductor},}\ }\href@noop {} {\bibfield  {journal} {\bibinfo  {journal}
  {Phys. Rev. B}\ }\textbf {\bibinfo {volume} {81}},\ \bibinfo {pages} {134508}
  (\bibinfo {year} {2010})}\BibitemShut {NoStop}%
\bibitem [{\citenamefont {Sato}\ \emph {et~al.}(2011)\citenamefont {Sato},
  \citenamefont {Tanaka}, \citenamefont {Yada},\ and\ \citenamefont
  {Yokoyama}}]{Sato2011}%
  \BibitemOpen
  \bibfield  {author} {\bibinfo {author} {\bibfnamefont {Masatoshi}\
  \bibnamefont {Sato}}, \bibinfo {author} {\bibfnamefont {Yukio}\ \bibnamefont
  {Tanaka}}, \bibinfo {author} {\bibfnamefont {Keiji}\ \bibnamefont {Yada}}, \
  and\ \bibinfo {author} {\bibfnamefont {Takehito}\ \bibnamefont {Yokoyama}},\
  }\bibfield  {title} {\enquote {\bibinfo {title} {Topology of andreev bound
  states with flat dispersion},}\ }\href@noop {} {\bibfield  {journal}
  {\bibinfo  {journal} {Phys. Rev. B}\ }\textbf {\bibinfo {volume} {83}},\
  \bibinfo {pages} {224511} (\bibinfo {year} {2011})}\BibitemShut {NoStop}%
\end{thebibliography}%

\clearpage

\renewcommand{\theequation}{S\arabic{equation}}
\renewcommand{\thefigure}{S\arabic{figure}}
\renewcommand{\bibnumfmt}[1]{[S#1]}
\renewcommand{\citenumfont}[1]{S#1}

  \begin{widetext}

\section{band structure and tight-binding model for FeAs layers in CaFeAs$_2$}
We adopted to a 2D five-band model capture the low-energy electronic structure of CaFeAs$_2$. The hopping parameters are given in Table \ref{hopping_parameters} and the obtained band structure is displayed in Fig.\ref{DFTFeAs}(a), which fit well with DFT. In
general, a ten-band model should be used to describe the low-energy
physics of CaFeAs$_2$ due to the absence of the glide mirror symmetry and four-fold rotational symmetry. But this five-band model can already provide a good description for the Fermi surfaces and therefore we adopt it in the following RPA calculations. It is noted that the largest hole pockets around $\Gamma$ point are attributed to Ca $d$ and As1 $p_z$ orbital and the small pockets around X point are attributed to As1 $p_{x,y}$ orbitals\cite{WuXX2014,WuXX2015-PRB}. As the correlation effect in As and Ca atoms is relatively weak, we neglect the pockets from them in our model and focus on the Fermi surfaces derived from the FeAs layers.

\begin{table}[bt]
\caption{\label{hopping_parameters} The hopping parameters  to fit the DFT band structures with a five-orbital model. The definition of the kinetic energy terms $\xi_{mn}(k)$ follow those in Ref.\onlinecite{Graser2009}. The $x$ direction is along the Fe-Fe bond. The onsite energies $\epsilon_i$ (in eV) $\epsilon_1=-0.0411$, $\epsilon_3=-0.329$, $\epsilon_4=0.122$ and $\epsilon_5=-0.182$, where $i$=1 corresponds to the $d_{xz}$, $i$=2 to the $d_{yz}$, $i$=3 to the $d_{x^2-y^2}$, $i$=4 to the $d_{xy}$ and $i$=5 to the $d_{3z^2-r^2}$.}
\begin{ruledtabular}
\begin{tabular}{ccccccccc}
$t^{mn}_i$ & $i$=x & $i$=y & $i$=xy & $i$=xx & $i$=yy & $i$=xxy & $i$=xyy & $i$=xxyy \\
 \colrule
$mn$=11 & -0.08  & -0.40 & 0.28 & 0.02 & -0.01 & -0.040 &  & 0.035  \\
$mn$=33 & 0.375  &   +$t_x$   & -0.075 & -0.022 &  +$t_{xx}$ &   &    & 0.013  \\
$mn$=44 & 0.172  &   +$t_x$    & 0.125 & -0.025 &  +$t_{xx}$  & -0.032  & +$t_{xyy}$  & -0.02  \\
$mn$=55 & -0.061  &   +$t_x$    & -0.095 & -0.042 & +$t_{xx}$  & 0.01  & +$t_{xxy}$  & -0.006  \\
$mn$=12 &     &       & 0.121 &   &    & -0.022  &  +$t_{xxy}$ & 0.043  \\
$mn$=13 &    & -0.414  & 0.112 &   &    & 0.022  & -$t_{xxy}$  &   \\
$mn$=14 & -0.32  &       & -0.015 & -0.007 &   & -0.018  &   &   \\
$mn$=15 & -0.093  &       & -0.103 &   &    &    &    & -0.021  \\
$mn$=34 &    &       &   &   &    & 0.01  & -$t_{xxy}$  &    \\
$mn$=35 & -0.334  &  -$t_x$    &    &    &     & -0.018  &  -$t_{xxy}$ &   \\
$mn$=45 &    &       & 0.121 &    &     &     &    & -0.012  \\
\end{tabular}
\end{ruledtabular}

\end{table}

\begin{figure}[tb]
\centerline{\includegraphics[width=1.0\columnwidth]{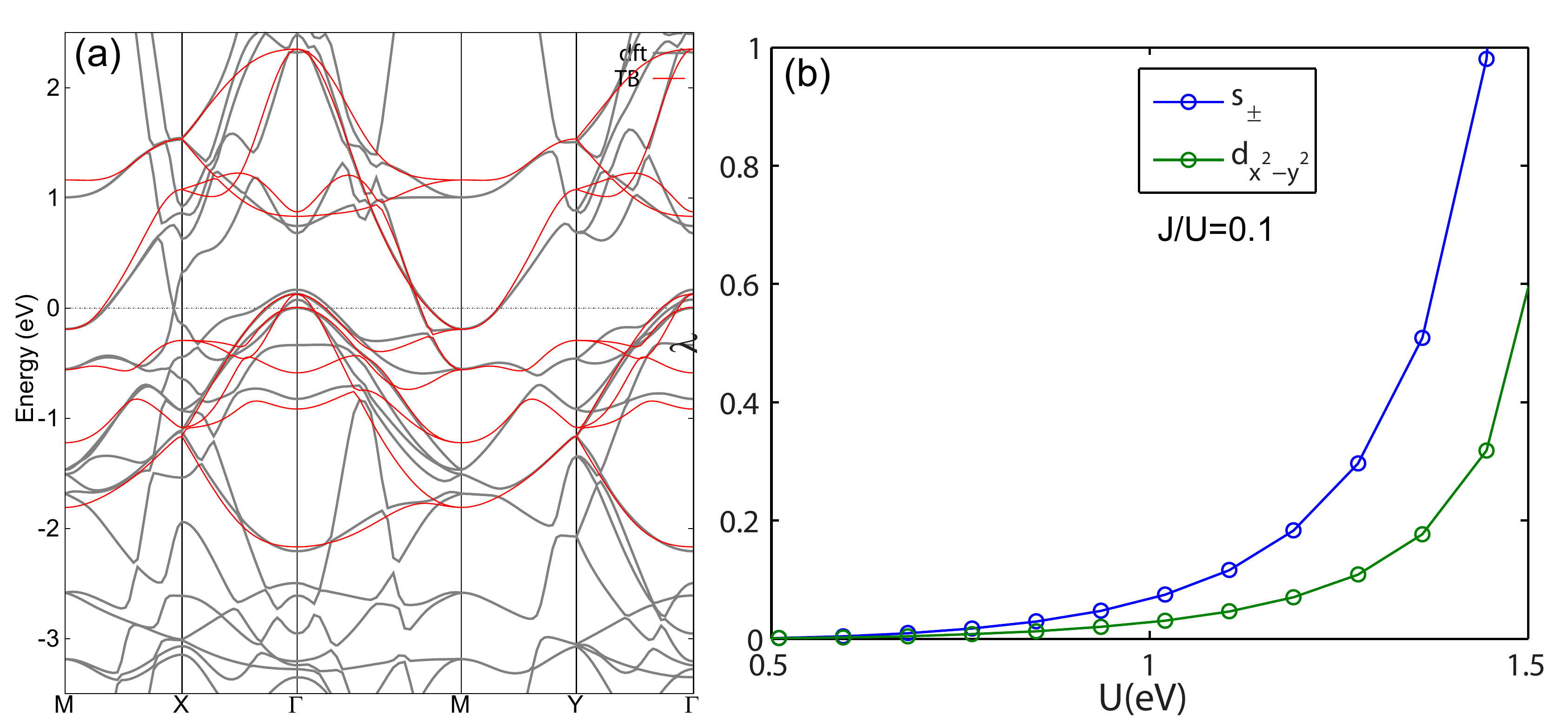}}
\caption{(color online) (a) Band structures for CaFeAs$_2$ without SOC from  DFT (gray lines, $k_z=0$ plane) and five-band tight-binding model (red lines). (b) Pairing eigen values $\lambda$ as a function of interaction $U$ in RPA calculations and we adopted $J/U=0.1$ with Kanamori relations $U=U'+2J$ and $J=J'$.
 \label{DFTFeAs} }
\end{figure}

\section{RPA calculations}
In this section, we explain the formalism of the multiorbital RPA approach\cite{Berk1966,Scalapino1986,Graser2009,Kemper2010,WuXX2015}, adopted in the main text. The adopted onsite Coulomb interaction terms in CaFeAs$_2$ are,
\begin{equation}
\begin{aligned}
H_{int}=U\sum_{i,\alpha}n_{i\alpha\uparrow}n_{i\alpha\downarrow}+U'\sum_{i,\alpha<\beta}n_{i\alpha}n_{i\beta}    +J_{H}\sum_{i,\alpha<\beta,\sigma\sigma^{'}}c^{\dagger}_{i\alpha\sigma}c^{\dagger}_{i\beta\sigma^{'}}c_{i\alpha\sigma^{'}}c_{i\beta\sigma}
         +J'\sum_{i,\alpha\neq\beta}c^{\dagger}_{i\alpha\uparrow}c^{\dagger}_{i\alpha\downarrow}c_{i\beta\downarrow}c_{i\beta\uparrow},
\end{aligned}
\label{hh}
\end{equation}
where $n_{i,\alpha}=n_{i,\alpha,\uparrow}+n_{i,\alpha,\downarrow}$. $U$, $U'$, $J$ and $J'$ represent the intra- and inter-orbital repulsion, the Hund's rule and pair-hopping terms. In the following calculations, we use Kanamori relations $U=U'+2J$ and $J=J'$ as requried by the lattice symmetry.

The multi-orbital susceptibility is defined as,
\begin{eqnarray}
\chi_{l_1l_2l_3l_4}(\bm{q},\tau)&=&\frac{1}{N}\sum_{\bm{k}\bm{k}'}\langle T_{\tau} c^{\dag}_{l_3\sigma}(\bm{k}+\bm{q},\tau)c_{l_4\sigma}(\bm{k},\tau)c^{\dag}_{l_2\sigma'}(\bm{k}'-\bm{q},0)c_{l_1\sigma'}(\bm{k}',0) \rangle .
\end{eqnarray}
In momentum-frequency space, the multi-orbital bare susceptibility is given by
\begin{equation}
\begin{aligned}
&\chi^{0}_{l_{1}l_{2}l_{3}l_{4}}(\bm{q},i\omega_{n})=-\frac{1}{N}\sum_{\bm{k}\mu\nu}a^{l_{4}}_{\mu}(\bm{k})a^{l_{2}*}_{\mu}(\bm{k})a^{l_{1}}_{\nu}(\bm{k}+\bm{q})a^{l_{3}*}_{\nu}(\bm{k}+\bm{q})\frac{n_{F}(E_{\mu}(\bm{k}))-n_{F}(E_{\nu}(\bm{k}+\bm{q}))}{i\omega_{n}+E_{\mu}(\bm{k})-E_{\nu}(\bm{k}+\bm{q})},
\end{aligned}
\end{equation}
where $\mu$ and $\nu$ are the band indices, $n_{F}$ is the usual Fermi distribution, $l_{i}$ $(i=1,2,3,4)$ are the orbital indices, $a^{l_{i}}_{\mu}(k)$ is the $l_{i}$ orbital component of the eigenvector for band $\mu$ resulting from the diagonalization of the tight-binding Hamiltonian $H_{0}$ and $E_{\mu}(\bm{k})$ is the corresponding eigenvalue. With interactions, the RPA spin and charge susceptibilities are given by
\begin{equation}
\begin{aligned}
\chi^{RPA}_{s}(\bm{q})=\chi^{0}(\bm{q})[1-\bar{U}^{s}\chi^{0}(\bm{q})]^{-1},\\
\chi^{RPA}_{c}(\bm{q})=\chi^{0}(\bm{q})[1+\bar{U}^{c}\chi^{0}(\bm{q})]^{-1},
\end{aligned}
\end{equation}
where $\bar{U}^{s}$ ($\bar{U}^{c}$) is the spin (charge) interaction matrix,
\begin{eqnarray}
\bar{U}^s_{l_1l_2l_3l_4}(\bm{q})&=&
\begin{cases}
U    & l_1=l_2=l_3=l_4,\\
U'     & l_1=l_3\neq l_2=l_4,\\
J   & l_1=l_2\neq l_3=l_4,\\
J'   & l_1=l_4\neq l_2=l_3,\\
\end{cases}\\
\label{EQ:Us}
\bar{U}^c_{l_1l_2l_3l_4}(\bm{q})&=&
\begin{cases}
U   & l_1=l_2=l_3=l_4,\\
-U'+2J    & l_1=l_3\neq l_2=l_4,\\
2U'-J    & l_1=l_2\neq l_3=l_4,\\
J'   & l_1=l_4\neq l_2=l_3,\\
\end{cases}\\
\label{EQ:Uc1}
\end{eqnarray}
In the main text, we plot the largest eigenvalues of the susceptibility matrix $\chi^{RPA}_{s,l_{1}l_{1}l_{2}l_{2}}(\bm{q},0)$. Within RPA approximation, the effective Cooper scattering interaction on Fermi surfaces is,
\begin{equation}
\begin{aligned}
\Gamma_{ij}(\bm{k},\bm{k}')=&\sum_{l_{1}l_{2}l_{3}l_{4}}a^{l_{2},\ast}_{\emph{v}_{i}}(\bm{k})a^{l_{3},\ast}_{\emph{v}_{i}}(-\bm{k})\emph{Re}\bigg[\Gamma_{l_{1}l_{2}l_{3}l_{4}}(\bm{k},\bm{k}',\omega=0)\bigg]a^{l_{1}}_{\emph{v}_{j}}(\bm{k}')a^{l_{4}}_{\emph{v}_{j}}(-\bm{k}'),
\end{aligned}
\end{equation}
where the momenta $\bm{k}$ and $\bm{k}'$ is restricted to different FSs with $\bm{k}\in C_{i}$ and $\bm{k}'\in C_{j}$. The  orbital vertex function $\Gamma_{l_{1}l_{2}l_{3}l_{4}}$ in spin singlet and triplet channels \cite{Takimoto2004,Kubo2007} are
\begin{equation}
\begin{aligned}
\Gamma^{S}_{l_{1}l_{2}l_{3}l_{4}}(\bm{k},\bm{k}',\omega)=&\bigg[\frac{3}{2}\bar{U}^{s}\chi^{RPA}_{s}(\bm{k}-\bm{k}',\omega)\bar{U}^{s}+\frac{1}{2}\bar{U}^{s}-\frac{1}{2}\bar{U}^{c}\chi^{RPA}_{c}(\bm{k}-\bm{k}',\omega)\bar{U}^{c}+\frac{1}{2}\bar{U}^{c}\bigg]_{l_{1}l_{2}l_{3}l_{4}},\\
\Gamma^{T}_{l_{1}l_{2}l_{3}l_{4}}(\bm{k},\bm{k}',\omega)=&\bigg[-\frac{1}{2}\bar{U}^{s}\chi^{RPA}_{s}(\bm{k}-\bm{k}',\omega)\bar{U}^{s}+\frac{1}{2}\bar{U}^{s}-\frac{1}{2}\bar{U}^{c}\chi^{RPA}_{c}(\bm{k}-\bm{k}',\omega)\bar{U}^{c}+\frac{1}{2}\bar{U}^{c}\bigg]_{l_{1}l_{2}l_{3}l_{4}},
\end{aligned}
\end{equation}
where $\chi^{RPA}_{s}$ and $\chi^{RPA}_{c}$ are the RPA spin and charge susceptibility, respectively. The pairing strength functional for a specific pairing state is given by,
\begin{equation}
\begin{aligned}
\lambda\big[\emph{g}(\bm{k})\big]=-\frac{\sum_{ij}\oint_{C_{i}}\frac{d\bm{k}_{\|}}{\emph{v}_{\emph{F}}(\bm{k})}\oint_{C_{j}}\frac{d\bm{k}'_{\|}}{\emph{v}_{\emph{F}}(\bm{k}')}\emph{g}(\bm{k})\Gamma_{ij}(\bm{k},\bm{k}')\emph{g}(\bm{k}')}{(2\pi)^{2}\sum_{i}\oint_{C_{i}}\frac{d\bm{k}_{\|}}{\emph{v}_{\emph{F}}(\bm{k})}\big[\emph{g}(\bm{k})\big]^{2}},
\end{aligned}
\end{equation}
where $v_{F}(\bm{k})=|\nabla_{k}E_{i}(\bm{k})|$ is the Fermi velocity on a given Fermi surface sheet $C_{i}$. The pairing vertex function in spin singlet and triplet channels are symmetric and antisymmetric parts of the interaction, that is, $\Gamma^{S/T}_{ij}(\bm{k},\bm{k}')=\frac{1}{2}[\Gamma_{ij}(\bm{k},\bm{k}')\pm \Gamma_{ij}(\bm{k},-\bm{k}')]$. Fig.\ref{DFTFeAs}(b) displays the pairing eigen value as a function of interaction parameters for CaFeAs$_2$ in RPA calculations with $J/U=0.1$. We find that the $s_{\pm}$-wave pairing is always dominant, which is attributed to the Fermi surfaces nesting between electron and hole pockets.

\section{tight-binding model for As1 layers}
 The crystal structure of CaFeAs$_2$ is shown Fig.1(a) in the main text and the As layers are shown in Fig.1(b).  A two-dimensional four-band model has been derived  to capture  the band structure attributed to $p_x$ and $p_y$ orbitals of two As-1 in a unit cell. As a unit cell includes  two As1 atoms, we divide the As1 lattices  into two sublattices.  We introduce the operator $\psi^\dag_{\textbf{k}\sigma}=(c^\dag_{Ax\sigma}(\textbf{k}),c^\dag_{Ay\sigma}(\textbf{k}),c^\dag_{Bx\sigma}(\textbf{k}),c^\dag_{By\sigma}(\textbf{k}))$, where $c^\dag_{\alpha \eta\sigma}(\textbf{k})$ is a Fermionic creation operator with  $\sigma$,  $\eta$ and $\alpha$ being spin, orbital and sublattice indices respectively. The tight-binding Hamiltonian can be written as:
 \begin{eqnarray}
 \mathcal{H}_{TB}=\sum_{\textbf{k}\sigma}\psi^\dag_{\textbf{k}\sigma}h(\textbf{k})\psi_{\textbf{k}\sigma}.
 \end{eqnarray}
  The matrix elements in the Hermitian $h(\textbf{k})$ matrix are given by
\begin{eqnarray}
\label{caas_tb}
&&h_{11}=h_{33}=\epsilon_X+2t^{11}_{1}cosk_x +2t^{11}_{2}cosk_y, \nonumber\\
&&h_{13}=(2t^{13}_{1}e^{i(x_0-1)k_x}+2t^{13}_{2}e^{ix_0k_x})cos(k_y/2),\nonumber\\
&&h_{14}=h_{23}=-(2it^{14}_1e^{i(x_0-1)k_x}+2it^{14}_2e^{ix_0k_x})sin(k_y/2),\nonumber\\
&&h_{22}=h_{44}=\epsilon_Y+2t^{11}_{2}cosk_x +2t^{11}_{1}cosk_y,\nonumber\\
&&h_{24}=(2t^{24}_1e^{i(x_0-1)k_x}+2t^{24}_2e^{ix_0k_x})cos(k_y/2),
\end{eqnarray}
with $x_0=0.418$ being the difference between the $x$ components of $B$ and $A$ sublattices.  The corresponding tight binding parameters are specified in unit of $eV$ as,
\begin{eqnarray}
&&\epsilon_X=-0.30, \quad \epsilon_Y=-0.109, \quad t^{11}_1=-0.149,  \nonumber \\
&&t^{11}_2=0.128, \quad t^{13}_1=0.89, \quad t^{13}_2=0.649, \nonumber \\
&&t^{14}_1=1.169,\quad t^{14}_2=-1.740, \quad t^{24}_1=0.567  \nonumber \\
&& t^{24}_2=1.213.
\label{hopping}
\end{eqnarray}
Now we consider the atomic SOC term in the As1 atoms and the SOC Hamiltonian can be written as
\begin{eqnarray}
H_{so}&=&i\lambda/2\sum_{\alpha \sigma\textbf{k}}\sigma c^{\dag}_{\alpha x\sigma}(\textbf{k}) c_{\alpha y\sigma}(\textbf{k})+h.c.\nonumber\\
&=&i\lambda/2\sum_{\alpha\textbf{k}}[ c^{\dag}_{\alpha x\uparrow}(\textbf{k}) c_{\alpha y\uparrow}(\textbf{k})-c^{\dag}_{\alpha x\downarrow}(\textbf{k}) c_{\alpha y\downarrow}(\textbf{k})]+h.c.\\
&=&i\lambda/2\sum_{\bm{k}}(\psi^\dag_{\bm{k}\uparrow}h_{so,\uparrow}\psi_{\bm{k}\uparrow}+\psi^\dag_{\bm{k}\downarrow}h_{so,\downarrow}\psi_{\bm{k}\downarrow})
\end{eqnarray}
where the spin-orbit coupling strength for As atoms is $\lambda=0.19$ eV and $h_{so,\uparrow}=-h_{so,\downarrow}=\tau_0\otimes-i\sigma_2$. $\bm{\tau}$, $\bm{\sigma}$, $\bm{s}$ are the Pauli matrices in sublattice, orbital and spin space. The total Hamiltonian is $H_0=H_{TB}+H_{so}$. Due to the conservation of $s_z$, this Hamiltonian can be block diagonal $H_0=H_{0\uparrow}+H_{0\downarrow}$, with $h_{0\uparrow/\downarrow}(\bm{k})=h(\bm{k})+i\frac{\lambda}{2} h_{so,\uparrow/\downarrow}$. The obtained band structure with SOC is displayed in Fig.\ref{DFTBand}(a), in agreement with DFT calculations in Fig.\ref{DFTBand}(b)\cite{WuXX2014,WuXX2015-PRB}.

\begin{figure}[tb]
\centerline{\includegraphics[width=1.0\columnwidth]{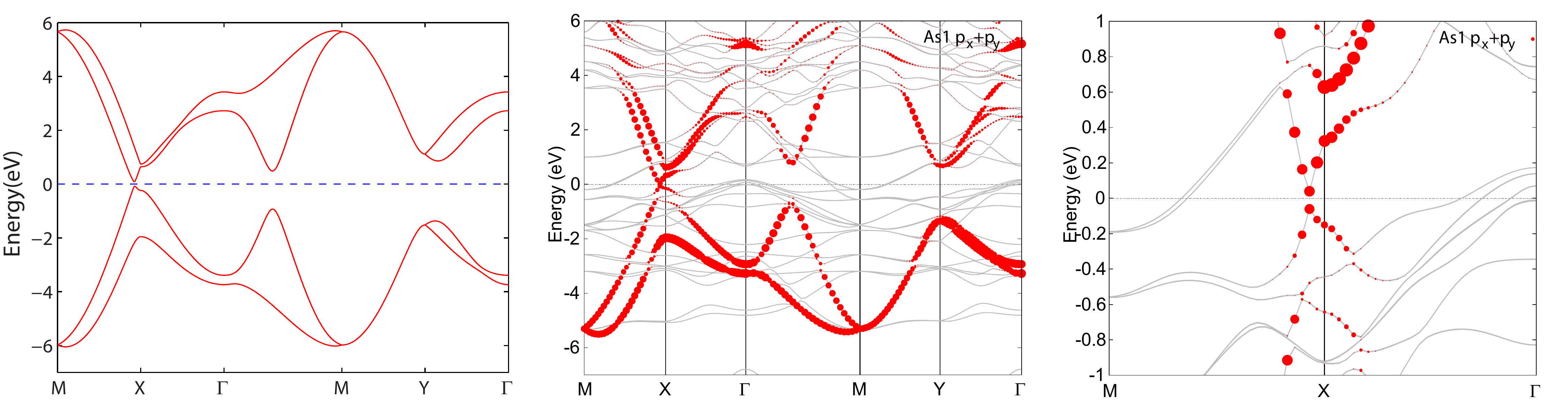}}
\caption{(color online) Band structures for As1 layers in CaFeAs$_2$ with SOC from tight-binding model (a) and DFT (b). (c) the zoom-in bands around X point in DFT calculations. The Dirac cone from As1 layers is gapped by SOC.
 \label{DFTBand} }
\end{figure}

\section{effective model around X point for As1 layers in CaFeAs$_2$}
For a point group operation, the wavefunction at $\bm{k}$ transforms as,
\begin{eqnarray}
\hat{P}_g\psi_{\bm{k},\alpha}(\bm{r})&=&\sum_{\beta}\psi_{g\bm{k},\beta}(\bm{r})D_{\beta\alpha}[g],
\end{eqnarray}
where $D[g]$ is the representation matrix for the point group operation $g$.
If we consider $g\bm{k}=\bm{G}+\bm{k}$, then $\psi_{g\bm{k},\beta}(\bm{r})=V^\dag_{
\beta\alpha}(\bm{G})\psi_{\bm{k},\alpha}(\bm{r})=e^{-i\bm{G}\cdot\tau_\beta}\psi_{\bm{k},\beta}(\bm{r})$. Therefore, we have,
\begin{eqnarray}
\hat{P}_g\psi_{\bm{k},\alpha}(\bm{r})&=&\sum_{\beta}\psi_{\bm{k},\beta}(\bm{r})e^{-i\bm{G}\cdot\tau_\beta}D_{\beta\alpha}[g].
\end{eqnarray}
For As1 lattice, the point group at $X$ is $C_{2v}$. Around $X$ point, the effective Hamiltonian can be written as $\mathcal{H}^{\bm{X}}_{eff}=\sum_{\bm{k}}\tilde{\psi}^\dag_{\bm{k}}h_{eff}(\mathbf{k})\tilde{\psi}_{\bm{k}} $ with $\tilde{\psi}^\dag_{\bm{k}}=(c^\dag_{X\bm{k},-\uparrow},c^\dag_{X\bm{k},+\uparrow},c^\dag_{X\bm{k},-\downarrow},c^\dag_{X\bm{k},+\downarrow})$, where "+/-" denotes the eigenvalue of $C_{2z}$ for eigenstates at the ${\bf X}$ point and $\bm{k}$ is relative to $X$ $(\pi,0)$. The representation matrices for the two generators in $C_{2v}$ are: $D(C_{2z})=is_3\sigma_3$ and $D(M_{xz})=is_2$. Here $\sigma$ and $s$ are Pauli matrices in orbital and spin spaces. The time reversal symmetry operator is $\hat{T}=-is_2\mathcal{K}$. The corresponding Hamiltonian matrix reads,
\begin{eqnarray}
h_{eff}(\mathbf{k})  &=& \epsilon_0(\mathbf{k})+ \left(\begin{array}{cccc}
M(\mathbf{k}) & iA_1k_x+A_2k_y & 0 & 0 \\
-iA_1k_x+A_2k_y  & -M(\mathbf{k}) & 0  & 0 \\
0 & 0 & M(\mathbf{k}) & iA_1k_x-A_2k_y  \\
0 & 0 & -iA_1k_x-A_2k_y & -M(\mathbf{k})  \\
\end{array}\right),\\
&=&\epsilon_0(\mathbf{k}) + M(\mathbf{k})\sigma_z-A_1k_xs_0\sigma_2 + A_2k_ys_3\sigma_1
\label{TI2d}
\end{eqnarray}
where $\epsilon_k=C_0+C_1k^2_x+C_2k^2_y$ and $M(k)=M_0-B_1k^2_x-B_2k^2_y$. This Hamiltonian satisfies all the symmetry operations: $\hat{g}H(\bm{k})\hat{g}^{-1}=H(g\bm{k})$ with $\hat{g}=C_{2z},M_{xz},\hat{T}$. Fitting the above effective model to the TB band structure around X point, the resulting band structure is displayed in Fig.\ref{KPband} and the corresponding parameters are given in Table \ref{parameters}. These bands show good agreement around X point.

\begin{figure}[tb]
\centerline{\includegraphics[width=0.5\columnwidth]{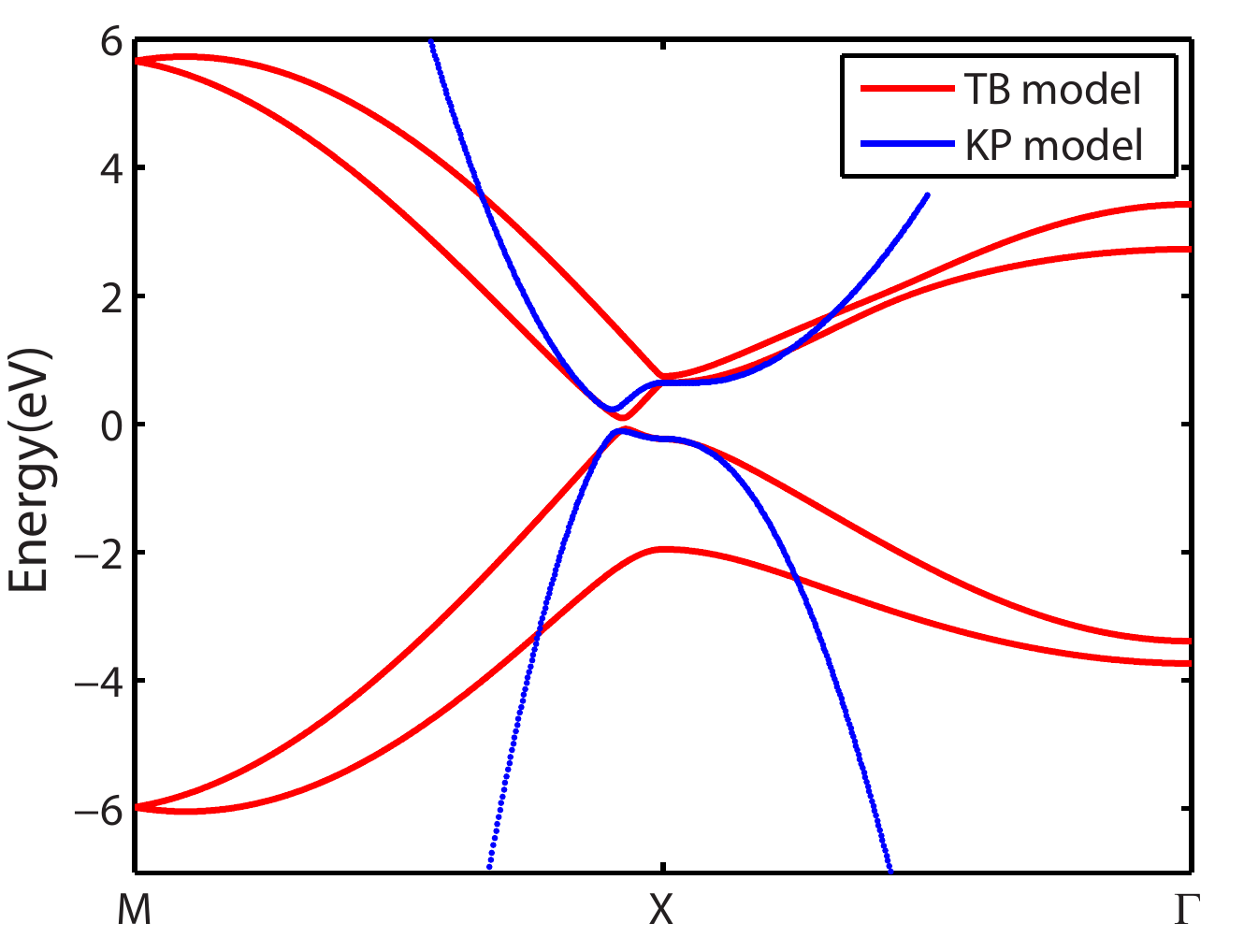}}
\caption{(color online) Band structure around X point from the TB model (red) and the effective model (blue).
 \label{KPband} }
\end{figure}

\begin{table}[bt]
\caption{\label{parameters}%
Obtained parameters from fitting the effective model to the TB band structures around X point. }
\begin{ruledtabular}
\begin{tabular}{cccccccc}
 $C$(eV)  & $M_0$(eV)  & $C_1$(eV \AA$^2$)  & $C_2$(eV \AA$^2$)  & $B_1$(eV \AA$^2$)  & $B_2$(eV \AA$^2$)  &$A_1$(eV \AA) & $A_2$(eV \AA) \\
 \colrule
  0.20 &-0.44 & -19.9& -29.7 & -40.2 & -80.0 & 7.0 & 2.5 \\
\end{tabular}
\end{ruledtabular}
\end{table}

\section{BdG Hamiltonian for bulk states}
As the As1 atoms are located just above the Fe site, we consider the $s_{\pm}$ pairing for As1 layer (intrasublattice pairing) from the proximity effect from the adjacent FeAs layers. In the superconducting phase, the BdG Hamiltonian, for the basis $\Psi^\dag_{\bm{k}}=(\psi^\dag_{\bm{k}\uparrow},\psi^T_{-\bm{k}\downarrow}, \psi^\dag_{\bm{k}\downarrow},\psi^T_{-\bm{k}\uparrow})$, is given by
\begin{eqnarray}
&&\mathcal{H}_{BdG}=\frac{1}{2}\sum_{\bm{k}}\Psi^\dag_{\bm{k}}h_{BdG}(\mathbf{k})\Psi_{\bm{k}}, \nonumber\\
&&h_{BdG}=\left(\begin{array}{cc}
h_{B,\uparrow\downarrow}(\bm{k}) & 0 \\
0 & h_{B,\downarrow\uparrow}(\bm{k}) \\
\end{array}\right)\nonumber\\
\label{BdGH}
&&\quad \quad\quad=\left(\begin{array}{cccc}
h_{0\uparrow}(\bm{k})-\mu & \Delta_{\uparrow\downarrow}(\bm{k}) & 0&0 \\
\Delta_{\uparrow\downarrow}^\dag(\bm{k}) & -h^*_{0\downarrow}(-\bm{k})+\mu & 0 &0\\
0 & 0 & h_{0\downarrow}(\bm{k})-\mu & \Delta_{\downarrow\uparrow}(\bm{k}) \\
0 & 0 & \Delta_{\downarrow\uparrow}^\dag(\bm{k}) & -h^*_{0\uparrow}(-\bm{k})+\mu\\
\end{array}\right).
\label{BdG-block}
\end{eqnarray}
Here we consider the intrasublattice and intraorbital singlet pairing and $\Delta_{\uparrow\downarrow}(\bm{k})=-\Delta_{\downarrow\uparrow}(\bm{k})=[\Delta_0+2\Delta_1(cosk_x+cosk_y)]\tau_0\sigma_0$. $h_{B,\uparrow\downarrow}$ and $h_{B,\downarrow\uparrow}$ are related by time reversal symmetry or particle hole symmetry. However, for $h_{B,\uparrow\downarrow}$, time reversal symmetry and particle hole symmetry are broken but chiral symmetry is preserved. Therefore, it belongs to AIII class and the topological invariant is $\mathbb{Z}$ in 1D.

Around X point, with basis $\tilde{\Psi}^\dag_{\bm{k}}=(\tilde{\psi}^\dag_{\bm{k}},\tilde{\psi}_{-\bm{k}})$ the BdG Hamiltonian can written as $\mathcal{H}_{BdG}=\frac{1}{2}\sum_{\bm{k}}\tilde{\Psi}^\dag_{\bm{k}}\tilde{h}_{BdG}(\mathbf{k})\tilde{\Psi}_{\bm{k}}$. $\tilde{h}_{B,\uparrow\downarrow}(\bm{k})$ can be explicitly written as,
\begin{eqnarray}
\tilde{h}_{B,\uparrow\downarrow}(\bm{k})=\left(\begin{array}{cccc}
\epsilon_0(\bm{k})+M(\mathbf{k})-\mu & iA_1k_x+A_2k_y & \Delta_{X}(\bm{k}) & 0 \\
-iA_1k_x+A_2k_y  & \epsilon_0(\bm{k})-M(\mathbf{k})-\mu & 0  & \Delta_{X}(\bm{k}) \\
\Delta_{X}(\bm{k}) & 0 & -\epsilon_0(\bm{k})-M(\mathbf{k})+\mu & -iA_1k_x+A_2k_y  \\
0 & \Delta_{X}(\bm{k}) & iA_1k_x+A_2k_y & -\epsilon_0(\bm{k})+M(\mathbf{k})+\mu  \\
\end{array}\right),
\end{eqnarray}
 with $\Delta_{X}(\bm{k})=\Delta_0+\Delta_1(k^2_x-k^2_y)$.

 \section{effective pairing for the edge states}
 To calculate the effective pairing on edges of As1 layers, we first
analytically obtain the wavefunctions of edge states by solving $H_{eff}$ with open boundary conditions and then project the
bulk pairing on edge states (omit the $\epsilon_0(\mathbf{k})$ term in the effective model)\cite{YanZB2018,ZhangSB2019}. The obtained pairings on (100) and (010) edge with finite chemical potential are,
\begin{eqnarray}
\Delta^{(100)}_{\text{eff}}&=&\Delta_0+2\Delta_1\frac{M_0}{B_1}-2\Delta_1(1+\frac{B_2}{B_1})\frac{\mu_1^2}{A^2_2},\\
\Delta^{(010)}_{\text{eff}}&=&\Delta_0-2\Delta_1\frac{M_0}{B_2}+2\Delta_1(1+\frac{B_1}{B_2})\frac{\mu_2^2}{A^2_1},
\end{eqnarray}
where $\mu_{1/2}$ is the chemical potential on edge states and is defined relative to the corresponding Dirac points. According to the Fig.2(c) and (d) in the main text and Table II, the chemical potential is in the range from 0.1 to 0.15 eV and thus $|\mu_{1/2}/A_{2/1}|^2\ll 1$. Therefore the finite chemical potential has a negligible effect on the effective pairing on edge states. This will not has a qualitative change on the analysis of pairing on edge states in the main text.

\section{Symmetries of the Hamiltonian and absence of bulk obstructions}

Bulk-obstructed topological phases have gapped Hamiltonians adiabatically disconnected from each other. The adiabatic disconnection manifest in the spectrum by the fact that, when transitioning from one phase to another, the Hamiltonian goes through a gapless point to achieve a band inversion. In the presence of crystalline symmetries, such band inversions typically occur at high-symmetry points of the Brillouin zone and can be tracked by calculating symmetry indicators topological invariants. In fact, whenever a topological phase has nontrivial symmetry indicators, it will necessarily have to close its bulk bands when deformed into the trivial phase. Therefore, a boundary-obstructed phase necessarily has trivial symmetry indicators.

In what follows, we show that the Hamiltonian for the As1 layers in Eq.\ref{BdGH}
has vanishing symmetry indicator invariants and thus it is not constrained to be bulk-obstructed.

For completeness, let us us begin by pointing out the local symmetries. This Hamiltonian obeys particle-hole, time-reversal, and chiral symmetries, which are, respectively,
\begin{align}
\hat{\Xi} h_{BdG}(\kk) \hat{\Xi}^{-1} &= - h_{BdG}(-\kk), \quad \hat{\Xi} = (s_x \otimes \rho_x \otimes \tau_0 \otimes \sigma_0) \mathcal{K}\\
\hat{\Theta} h_{BdG}(\kk) \hat{\Theta}^{-1} &= h_{BdG}(-\kk), \quad \hat{\Theta}=(is_y \otimes \rho_z \otimes \tau_0 \otimes \sigma_0)\mathcal{K}\\
\hat{\Pi} h_{BdG}(\kk) \hat{\Pi}^{-1} &= -h_{BdG}(\kk), \quad \hat{\Pi}=s_z \otimes \rho_y \otimes \tau_0 \otimes \sigma_0,
\end{align}
where $\mathcal{K}$ is complex conjugation. The degrees of freedom in which the Pauli matrices $\bm{s}$, $\bm{\rho}$, $\bm{\tau}$ and $\bm{\sigma}$ operate are: spin, Nambu basis, sublattice, and orbital, respectively.

\subsection{Crystalline symmetries}
In addition to the local symmetries, our material has $C_{2z}$ rotation and refrlection symmetries, as shown in Fig.~\ref{fig:symmetries}. We now show that for both $C_{2z}$ and reflection symmetries, the symmetry indicator invariants vanish due to time reversal symmetry for all phases.

\begin{figure}[h]%
\centering
\includegraphics[width=.35\columnwidth]{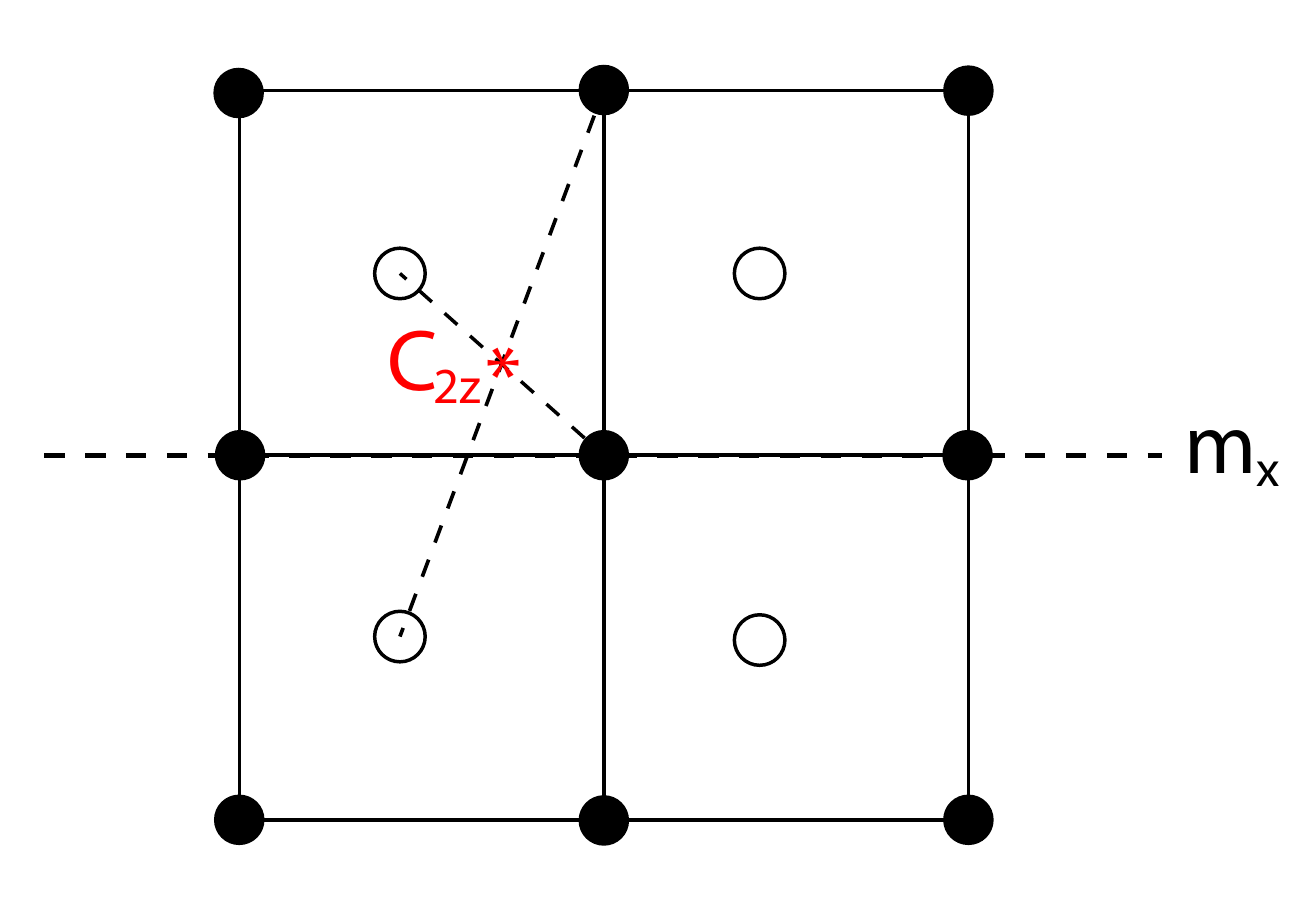}
\caption{Crystalline symmetries of the layers of As1 in CaFeAs2. $C_{2z}$ rotation center is marked by a red asterix, $m_x$ mirror plane is show by the horizontal dashed line. Black and white circles represent As1 atoms of different sublattice. Notice that $C_{2z}$ rotation exchanges the two sublattices, while $m_x$ preserves them.}
\label{fig:symmetries}
\end{figure}

\subsubsection{$C_2$ symmetry}
This Hamiltonian has additional $C_{2z}$ rotational symmetry,
\begin{align}
\hat{r}_2 h_{BdG}(\kk) \hat{r}_2^{-1} = h_{BdG}(-\kk), \quad \hat{r}_2=-is_z \otimes \rho_0 \otimes \tau_x \otimes \sigma_0.
\end{align}
The rotation operator obeys $\hat{r}_2^2=-1$ and $[\hat{r}_2,\hat{\Theta}]=0$.
At the high symmetry points of the Brillouin zone (HSP), ${\bf \Pi}={\bf \Gamma}$, {\bf X}, ${\bf Y}$, ${\bf M}$, we have ${\bf \Pi}=-{\bf \Pi}$, and $h_{BdG}(\kk)$ commutes with both $\hat{\Theta}$ and $\hat{r}_2$. In particular, let us calculate the $C_{2z}$ eigenvalues of the energy eigenstates $\ket{u_n({\bf \Pi})}$,
\begin{align}
\hat{r}_2 \ket{u_n({\bf \Pi})} = r_p({\bf \Pi}) \ket{u_n({\bf \Pi})}.
\end{align}
Here $r_p({\bf \Pi})$, for $p=1$ or $2$ is the rotation eigenvalue at ${\bf \Pi}$, and can be any of the two values  $r_1({\bf \Pi})=+i$ or $r_2({\bf \Pi})=-i$. We then define the rotation symmetry indicator invariants
\begin{align}
[\Pi_p]=\#r_p({\bf \Pi})-\#r_p({\bf \Gamma})\quad \in \mathbb{Z},
\label{eq:RotationInvariants}
\end{align}
where $\#r_p({\bf \Pi})$ is the number of eigenstates in the \emph{negative} energy bands with eigenvalue $r_p({\bf \Pi})$, for ${\bf \Pi}={\bf X}$, ${\bf Y}$, and ${\bf M}$. Due to the fact that the number of negative energy bands is constant across the BZ, we have
\begin{align}
\sum_{p=1,2} \#r_p({\bf \Pi}) = \sum_{p=1,2} \#r_p({\bf \Gamma}) \rightarrow [\Pi_1] + [\Pi_2]=0.
\label{eq:RotInvConstraint1}
\end{align}
Now, due to TRS, we have
\begin{align}
\hat{r}_2 \hat{\Theta} \ket{u_n(\kk)} = \hat{\Theta} \hat{r}_2 \ket{u_n(\kk)} = r^*_p \hat{\Theta} \ket{u_n(\kk)},
\end{align}
which means that the rotation eigenvalues have to come in complex conjugate pairs. For the rotation invariants, Eq. \ref{eq:RotationInvariants}, this implies that
\begin{align}
\#r_1({\bf \Pi})=\#r_2({\bf \Pi})\quad \mbox{and} \quad \#r_1({\bf \Gamma})=\#r_2({\bf \Gamma})
\end{align}
or, in terms of the invariants,
\begin{align}
[\Pi_1] \stackrel{\text{TRS}}{=} [\Pi_2].
\label{eq:RotInvConstraint2}
\end{align}
This, along with the restriction in Eq. \ref{eq:RotInvConstraint1}, results in vanishing rotation invariants at all HSPs,
\begin{align}
[\Pi_p] \stackrel{\text{TRS}}{=}0.
\end{align}
Thus, we see that $C_2$ symmetry does not force the phases of $h_{BdG}({\bf k})$ to be bulk-obstructed.

\subsubsection{Reflection symmetry}
The Hamiltonian also has reflection symmetries,
\begin{align}
\hat{m}_2 h_{BdG}(k_x,k_y) \hat{m}_2^{-1} = h_{BdG}(k_x,-k_y), \quad \hat{m}_2= i s_x \otimes \rho_z \otimes \tau_0 \otimes \sigma_z,
\end{align}
The reflection operator obeys $\hat{m}_x^2=-1$ and $[\hat{m}_x, \hat{\Theta}]=0$. We can proceed as for $C_{2z}$ symmetry to define the symmetry indicator invariants. However, the same constraints as in Eq.~\ref{eq:RotInvConstraint1} and Eq.~\ref{eq:RotInvConstraint2} apply, leading to trivial invariants at all HSPs.

\section{phase transition on the edge and winding number}
For a fixed $\Delta_0$, with increasing $\Delta_1$, there is no gap closing for the bulk state and (100) edge states but a phase transition for the (010) edge states, as displayed in Fig.\ref{edge}. It indicates the resulting high-order topology is derived from the edge state not from bulk states, indicating that the system has boundary topological obstruction. In insulators, boundary topological obstructions have recently been explained in terms of the Wannier centers of the occupied bands. Following this, we calculate the corresponding Wannier centers and Fig.\ref{wannier100} and \ref{wannier010} show the Wannier center spectra as a function of $k_y/k_x$ with a increasing $\Delta_1$ and a fixed $\Delta_0=0.005$ eV (the phase transition point is at $\Delta_1=0.051$ eV). A small nonzero $\Delta_1$ will immediately open a gap around $k_x/k_y=0$ and the Wannier center spectra donot exhibit significant changes across the phase transition on (010) edge states, suggesting that they cannot be utilized  to characterize the edge topological phase transition. Therefore, we need to introduce other topological invariants to identify the observed topological phase transition.

\subsection{winding number}

For a gapped system, we can define the Projection operator at $\bm{k}$ point as,
\begin{eqnarray}
P_{\bm{k}}=\sum^{N_{occ}}_{n=1}|u_{n}(\bm{k})\rangle\langle u_{n}(\bm{k})|,
\end{eqnarray}
where $|u_{n}(\bm{k})\rangle$ is the eigen states for BdG Hamiltonian and $N_{occ}$ is the number of the occupied bands. It is convenient to introduce the $Q$ matrix by,
\begin{eqnarray}
Q_{\bm{k}}=1-2P_{\bm{k}},
\end{eqnarray}
where $Q$ matrix satisfies $Q^\dag_{\bm{k}}=Q_{\bm{k}}$ and $Q^2_{\bm{k}}=\bm{1}$. The eigenvalues of $Q_{\bm{k}}$ are $\pm1$. For systems with chiral symmetry, the $Q$ matrix can be written as a block off-diagonal form in the basis of chiral operator,
\begin{eqnarray}
Q_{\bm{k}}=\left(
\begin{array}{cc}
0 & q_{\bm{k}}  \\
q^\dag_{\bm{k}}  & 0
\end{array}
\right),
\end{eqnarray}
where $q_{\bm{k}}$ satisfies $q_{\bm{k}}q^\dag_{\bm{k}}=q^\dag_{\bm{k}}q_{\bm{k}}=\bm{1}$. Hence, the $q$ matrix defines a map from the base space (Brillouin zone) to the space of unitary matrix $U(N)$. The relevant homotopy group for projectors is $\pi_d[U(N)]=\mathcal{Z}$ for odd $d$ and $N\geq(d+1)/2$\cite{Chiu2016}. The winding number in 1D and 3D are defined as,
\begin{eqnarray}
\nu_1&=&\frac{i}{2\pi}\int_{BZ}dk\text{Tr}[q^{-1}\partial_kq],\\
\nu_3&=&\frac{1}{24\pi^2}\int_{BZ}d^3\bm{k}\epsilon^{ijl}\text{Tr}[(q^{-1}\partial_{k_i}q)(q^{-1}\partial_{k_j}q)(q^{-1}\partial_{k_l}q)].
\end{eqnarray}
In the weak pairing limit, $q$ matrix can be written as,
\begin{eqnarray}
q_{\bm{k}}=\sum_ne^{i\theta_{n\bm{k}}}|n,\bm{k}\rangle\langle n,\bm{k}|,
\end{eqnarray}
with $e^{i\theta_{n\bm{k}}}=(\epsilon_{n\bm{k}}+i\delta_{n\bm{k}})/|\epsilon_{n\bm{k}}+i\delta_{n\bm{k}}|$ and $\delta_{n\bm{k}}=\langle n,\bm{k}|U_T\Delta^\dag_{\bm{k}}|n,\bm{k}\rangle$ is the gap function\cite{QiXL2010} (the time-reversal symmetry $\hat{T}=U_T K$). Here $|n,\bm{k}\rangle$ is the eigen state for normal Hamilton without superconductivity (not the BdG Hamiltonian). To the leading order, near the Fermi surface we have,
\begin{eqnarray}
e^{i\theta_{n\bm{k}}}=\frac{v_F(k_{\perp}-k_F)+i\delta_{n\bm{k}_F}}{\sqrt{v^2_F(k_{\perp}-k_F)^2+\delta^2_{n\bm{k}_F}}}
\end{eqnarray}
In the limit $\delta_{nk_F}\rightarrow 0$, we have $\theta_{n\bm{k}}\rightarrow -\pi sgn(\delta_{k\bm{k}})\eta(k_F-k_\perp)$ with $\eta(x)$ being the step function. The derivative of $\theta$ is given by,
\begin{eqnarray}
\partial_{k_{\perp}}\theta_{n\bm{k}}=-\pi sgn(\delta_{n\bm{k}})\delta(k_{\perp}-k_F).
\end{eqnarray}
Generally, we have $\nabla\theta_{n\bm{k}}=-\pi \bm{v}_{n\bm{k}}sgn(\delta_{n\bm{k}})\delta(\epsilon_{n\bm{k}})$.
 Omitting the interband pairing, the widing number in 1D can be further written as\cite{Sato2011},
\begin{eqnarray}
\nu_1&=&\frac{i}{2\pi}\int_{BZ}dk\text{Tr}[q^{-1}\partial_kq]=
\frac{i}{2\pi}\sum_n\int_{BZ}dke^{-i\theta_{n\bm{k}}}\partial_ke^{i\theta_{n\bm{k}}}=-\sum_n\frac{1}{2\pi}\int^{2\pi}_{0}dk \partial_k \theta_{n\bm{k}}\nonumber\\
&=&\frac{1}{2}\sum_{\epsilon_{nk}=0}sgn[\partial_{k}\epsilon_{nk}]sgn[\delta_{nk}].
\end{eqnarray}
Here $\nu_1$ is a real number. From the definition, only the bands crossing the Fermi level can contribute the winding number. For gapped bulk system with gapless localized edge states, the dominant contribution in the slab model comes from the edge states. In the following, we define the site-resolved winding number to identify the topological phase transition on the edges.

\subsection{spatial dependent winding number}
For surfaces or edges in the slab model, we introduce the spatial dependent winding number to identify the phase transition on the surfaces or edges. Generally we introduce $\bar{\Psi}^\dag_{\bm{k}}=(\psi^\dag_{\bm{k},11},\psi^\dag_{\bm{k},12},...,\psi^\dag_{\bm{k},1m},...,\psi^\dag_{\bm{k},i\gamma},...,\psi^\dag_{\bm{k},N\gamma},\psi_{-\bm{k},1\bar{\gamma}},...,\psi_{-\bm{k},i\bar{\gamma}},...,\psi_{-\bm{k},N\bar{\gamma}})$, and the BdG Hamiltonian for the slab can be written as,
\begin{eqnarray}
&&\mathcal{H}^{slab}_{BdG}=\sum_{\bm{k}}\bar{\Psi}^\dag_{\bm{k}}\bar{h}_{BdG}(\mathbf{k})\bar{\Psi}_{\bm{k}}, \\
&&\bar{h}_{BdG}(\bm{k})=\left(\begin{array}{cc}
\bar{h}_{\sigma}(\bm{k}) & \bar{\Delta}_{\bm{k}} \\
\bar{\Delta}^\dag_{\bm{k}} & -\bar{h}^*_{\bar{\sigma}}(-\bm{k}) \\
\end{array}\right),
\end{eqnarray}
where $i$ denotes the lattice site (sublattice index), $N$ is the lattice size, $\gamma$ denotes the other indices(orbital or spin) and $\bar{\gamma}$ represent the index with flipped spin. The chiral symmetry operator is $S=U_S=\tau_y \otimes \bm{1}_{Nm} $ and it can be expressed as $S\bar{h}_{BdG}(\bm{k})S^{-1}=-\bar{h}_{BdG}(\bm{k})$, which results $\bar{h}_{\sigma}(\bm{k})=\bar{h}^*_{\bar{\sigma}}(-\bm{k})$ and $\bar{\Delta}^\dag_{\bm{k}}=\bar{\Delta}_{\bm{k}}$. Under a unitary transformation $U$, the chiral symmetry matrix become diagonal and the Hamiltonian matrix and $Q$ matrix become off-diagonal,
\begin{eqnarray}
U^\dag U_S U&=&\left(
\begin{array}{cc}
\bm{1} & 0 \\
0 & -\bm{1}
\end{array}
\right)\\
U&=&\frac{1}{\sqrt{2}}\left(
\begin{array}{cc}
\bm{1} & \bm{1}  \\
i\bm{1} & -i\bm{1}
\end{array}
\right),\\
\tilde{\bar{h}}_{BdG}(\bm{k})=U^\dag\bar{h}_{BdG}(\bm{k})U&=&\left(
\begin{array}{cc}
\bm{0} & \bar{h}_{\sigma}(\bm{k})-i\bar{\Delta}^\dag_{\bm{k}}  \\
\bar{h}_{\sigma}(\bm{k})+i\bar{\Delta}_{\bm{k}} & \bm{0}
\end{array}
\right),\\
Q^{\sigma\bar{\sigma}}_{\bm{k}}&=&\left(
\begin{array}{cc}
0 & q^{\sigma\bar{\sigma}}_{\bm{k}}  \\
q^{\sigma\bar{\sigma},\dag}_{\bm{k}}  & 0
\end{array}
\right).
\end{eqnarray}
Here $q^{\sigma\bar{\sigma}}_{\bm{k}}$ is $Nm\times Nm$ matrix and $N$ is lattice size and $m$ is the dimension of tight-binding Hamiltonian. The new basis is  $\tilde{\bar{\Psi}}^\dag_{\bm{k}}=\bar{\Psi}^\dag_{\bm{k}}U=(\psi^\dag_{\bm{k},11}+i\psi_{-\bm{k},1\bar{1}},...,\psi^\dag_{\bm{k},1m}+i\psi_{-\bm{k},1\bar{m}},...,\psi^\dag_{\bm{k},i\gamma}+i\psi_{-\bm{k},i\bar{\gamma}},...,\psi^\dag_{\bm{k},N\gamma}+i\psi_{-\bm{k},N\bar{\gamma}},\psi_{-\bm{k},1\bar{\gamma}}-i\psi^\dag_{\bm{k},1\gamma},...,\psi_{-\bm{k},i\bar{\gamma}}-i\psi^\dag_{\bm{k},i\gamma},...,\psi_{-\bm{k},N\bar{\gamma}}-i\psi^\dag_{\bm{k},N\gamma})$ and each has a definite lattice site index. The total winding number can be written as,
\begin{eqnarray}
\nu_1&=&\frac{i}{2\pi}\int_{BZ}dk\text{Tr}[q^{-1}_{\bm{k}}\partial_kq_{\bm{k}}]\nonumber\\
&=&\frac{i}{2\pi}\sum_{i\gamma}\int_{BZ}dk\text{Tr}[q^{-1}_{\bm{k}}\partial_kq_{\bm{k}}]_{i\gamma,i\gamma}\nonumber\\
&=&\sum_{i=1}^{N}\nu^i_1,
\end{eqnarray}
with site-resolved winding number $\nu^i_1=\frac{i}{2\pi}\sum_{\gamma}\int_{BZ}dk[q^{-1}_{\bm{k}}\partial_kq_{\bm{k}}]_{i\gamma,i\gamma}$.
The integral can be numerically evaluated on a discretized Brillouin zone.

For (010) slab of As1 layers, the total winding number for is always zero as shown in Fig.\ref{totalwinding}(a), therefore it cannot be used to identify the phase transition. We calculated the corresponding spatial dependent winding numbers and Fig.\ref{totalwinding}(b) shows the typical profiles in two topological phases. The profiles in two phases are significant different with each other. To further quantitatively characterize the edge phase transition, we separate the top and bottom edges and define the winding number for them $\nu^T_1$ and $\nu^B_1$ by summing the winding number of the upper half and lower half edges, respectively.
Although $\nu^{T,B}_1$ are generally not an integer, the its change is an integer during the topological phase transition. For the (010) edge, with increasing $\Delta_1$, the sign change of superconducting pairing on edge states will change winding number of top edge and bottom edge by -1 and 1 in the thermodynamic limit (demonstrated in the main text), respectively. Meantime, the winding number of left and right edge for (100) edge donot change. Therefore, the left/right and top/bottom edges are in topologically distinct phases and when they meet at corners, Majorana modes will appear. The representative local density of states at corners and edges are displayed in Fig.\ref{ldos}(b), where Majorana modes will induce a zero-bias peak at corners.

\begin{figure}[tb]
\centerline{\includegraphics[width=1.0\columnwidth]{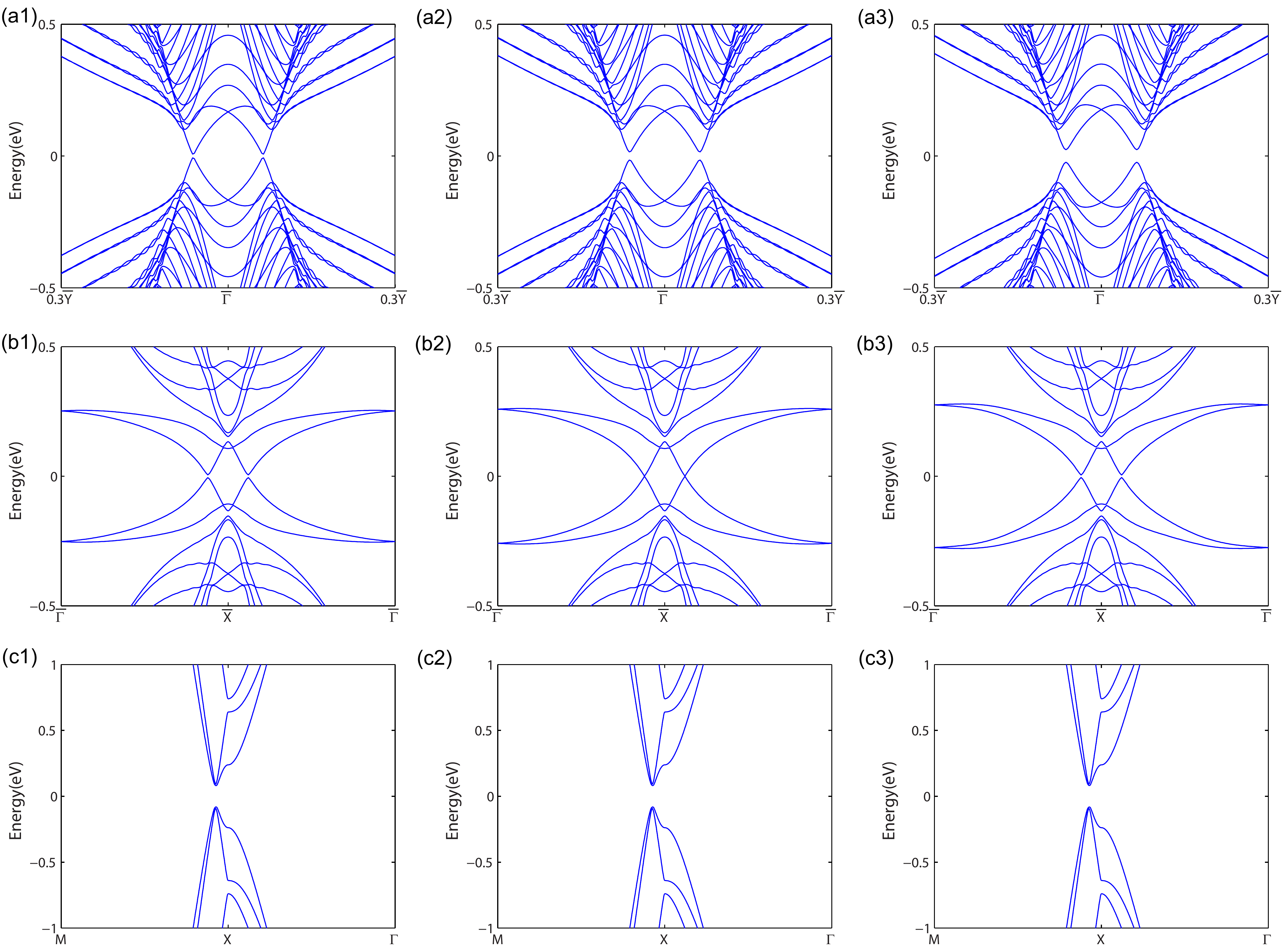}}
\caption{(color online) Evolution of band structures for (100) edge (a), (010) edge (b) and bulk (c). he plots in each column are generated with the same parameters and the adopted parameters in three columns (from left to right) are: $\Delta_1= 0$, $\Delta_1= 51$ meV and $\Delta_1=100$ meV with a fixed $\Delta_0=5$ meV.
 \label{edge} }
\end{figure}

\begin{figure}[tb]
\centerline{\includegraphics[width=0.8\columnwidth]{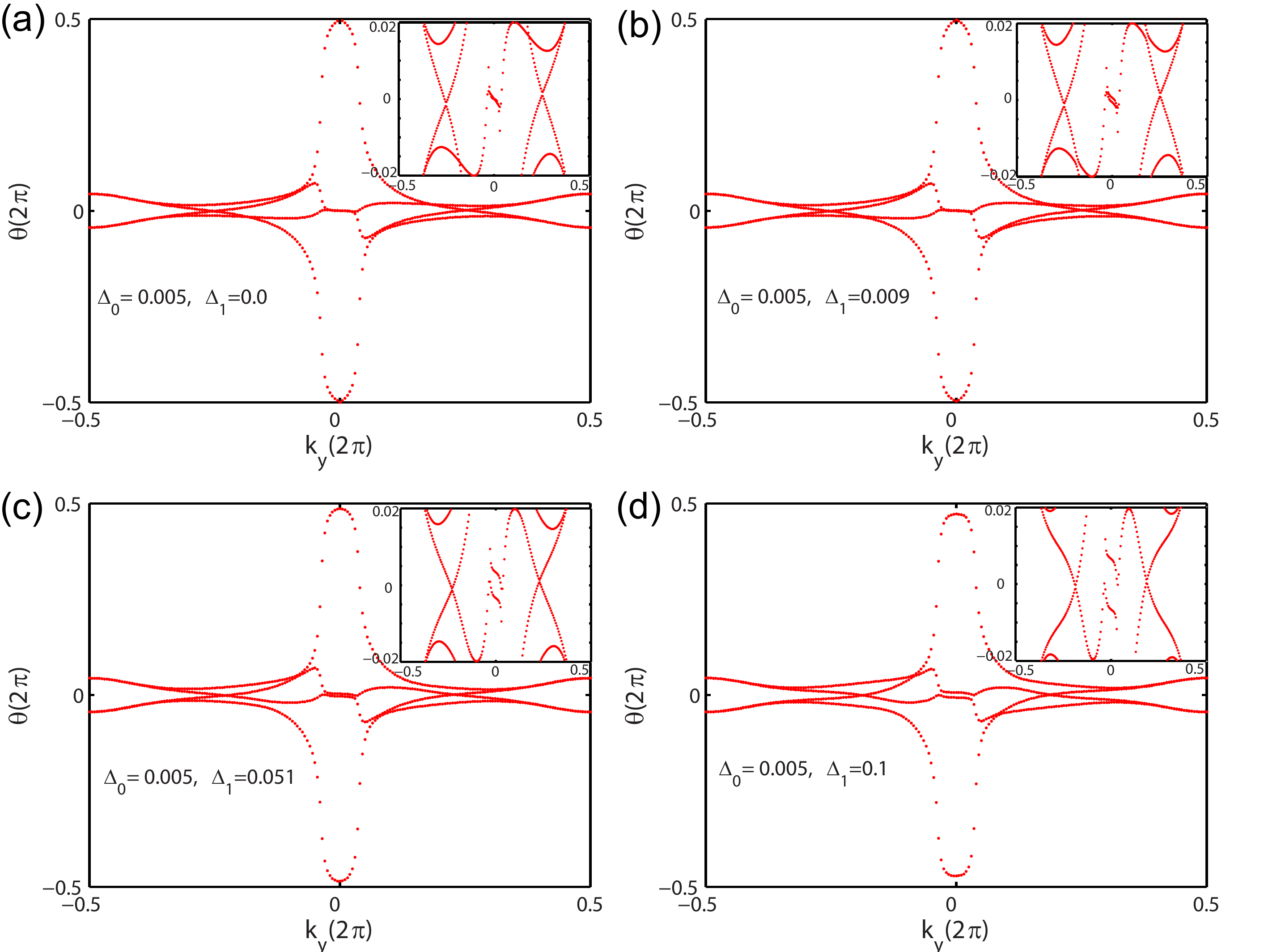}}
\caption{(color online) Wannier center spectra as a function of $k_y$ with $\Delta_0=0.005$ eV and varying $\Delta_1$: (a) $\Delta_1=0$, (b) $\Delta_1=0.009$ eV, (c) $\Delta_1=0.051$ eV, (d) $\Delta_1=0.1$ eV. Insets show zoom-in images of the Wannier center close to zero.
 \label{wannier100} }
\end{figure}

\begin{figure}[tb]
\centerline{\includegraphics[width=0.8\columnwidth]{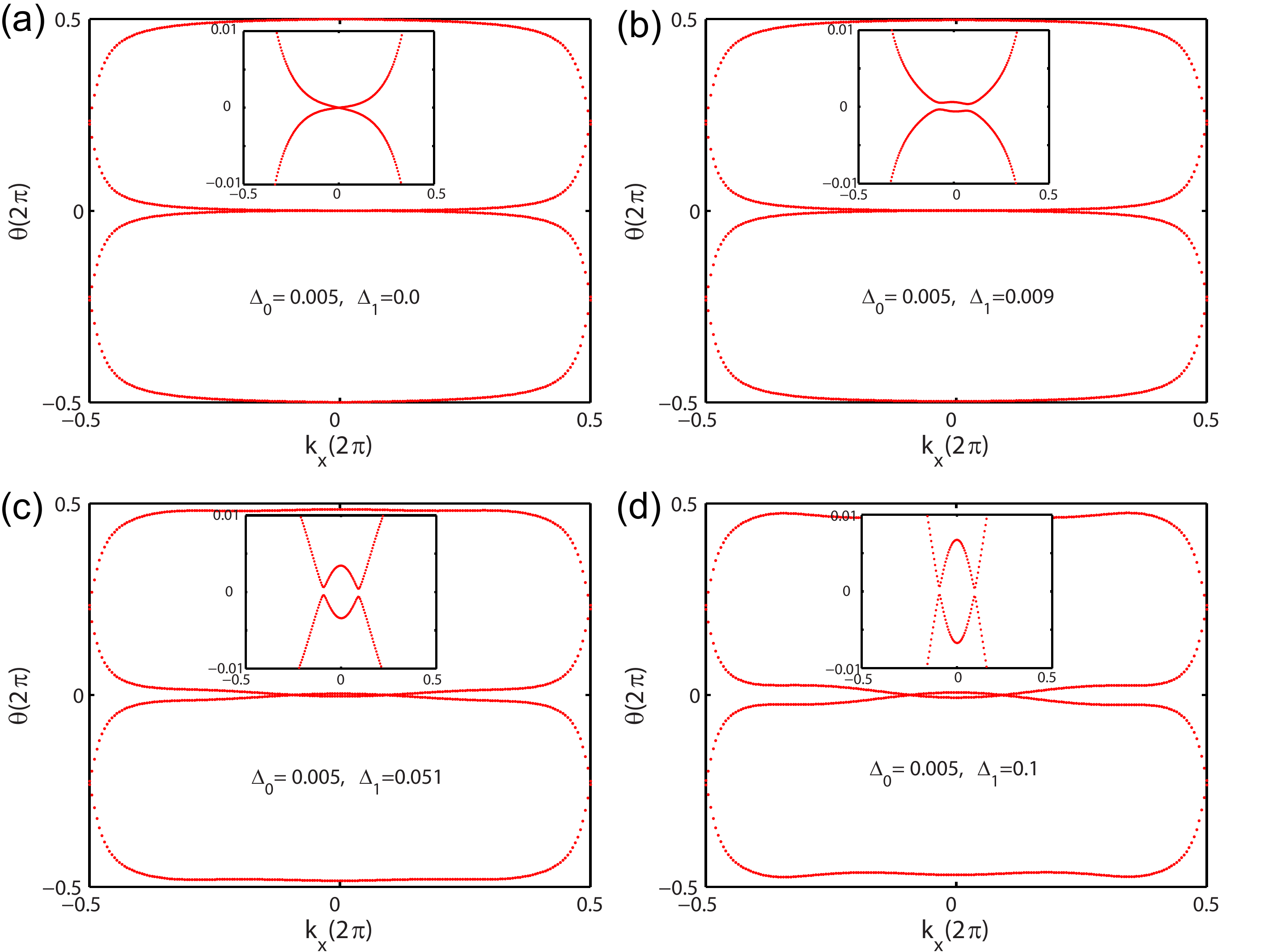}}
\caption{(color online) Wannier center spectra as a function of $k_x$ with $\Delta_0=0.005$ eV and varying $\Delta_1$: (a) $\Delta_1=0$, (b) $\Delta_1=0.009$ eV, (c) $\Delta_1=0.051$ eV, (d) $\Delta_1=0.1$ eV. Insets show zoom-in images of the Wannier center close to zero.
 \label{wannier010} }
\end{figure}

\begin{figure}[tb]
\centerline{\includegraphics[width=1.0\columnwidth]{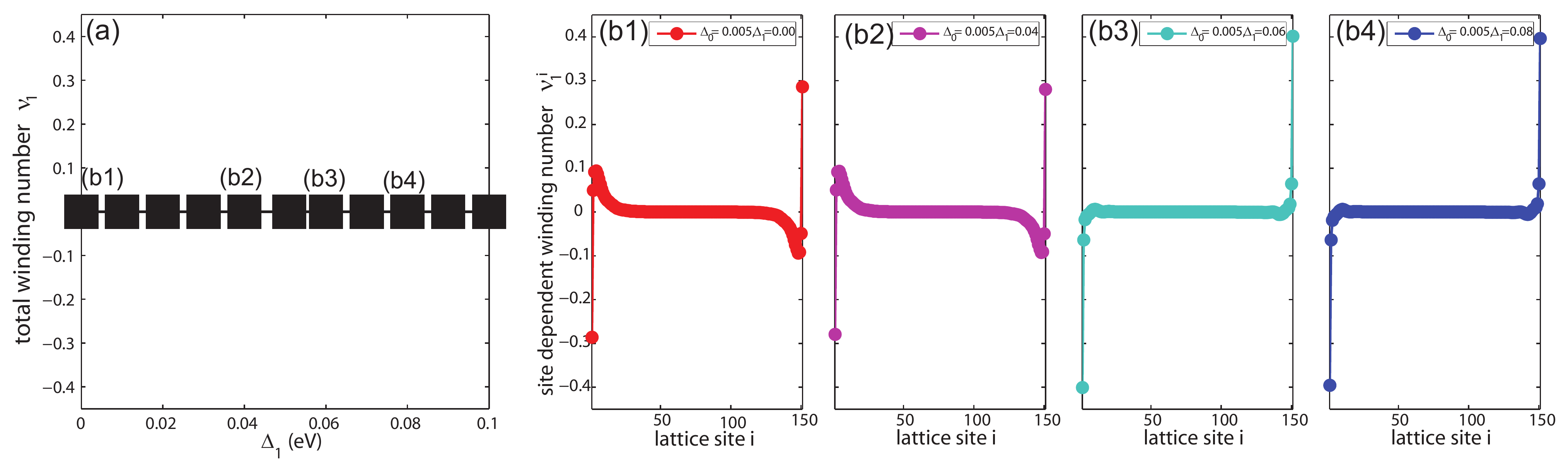}}
\caption{(color online) (a) Total winding number as function of $\Delta_1$ for (010) slab with $\Delta_0=0.005$ eV. Spatial dependent winding number with fixed $\Delta_0=0.005$ and different $\Delta_1$: (b1) $\Delta_1=0.00$, (b2) $\Delta_1=0.04$, (b3) $\Delta_1=0.06$ and (b4) $\Delta_1=0.08$.
 \label{totalwinding} }
\end{figure}

\begin{figure}[tb]
\centerline{\includegraphics[width=0.8\columnwidth]{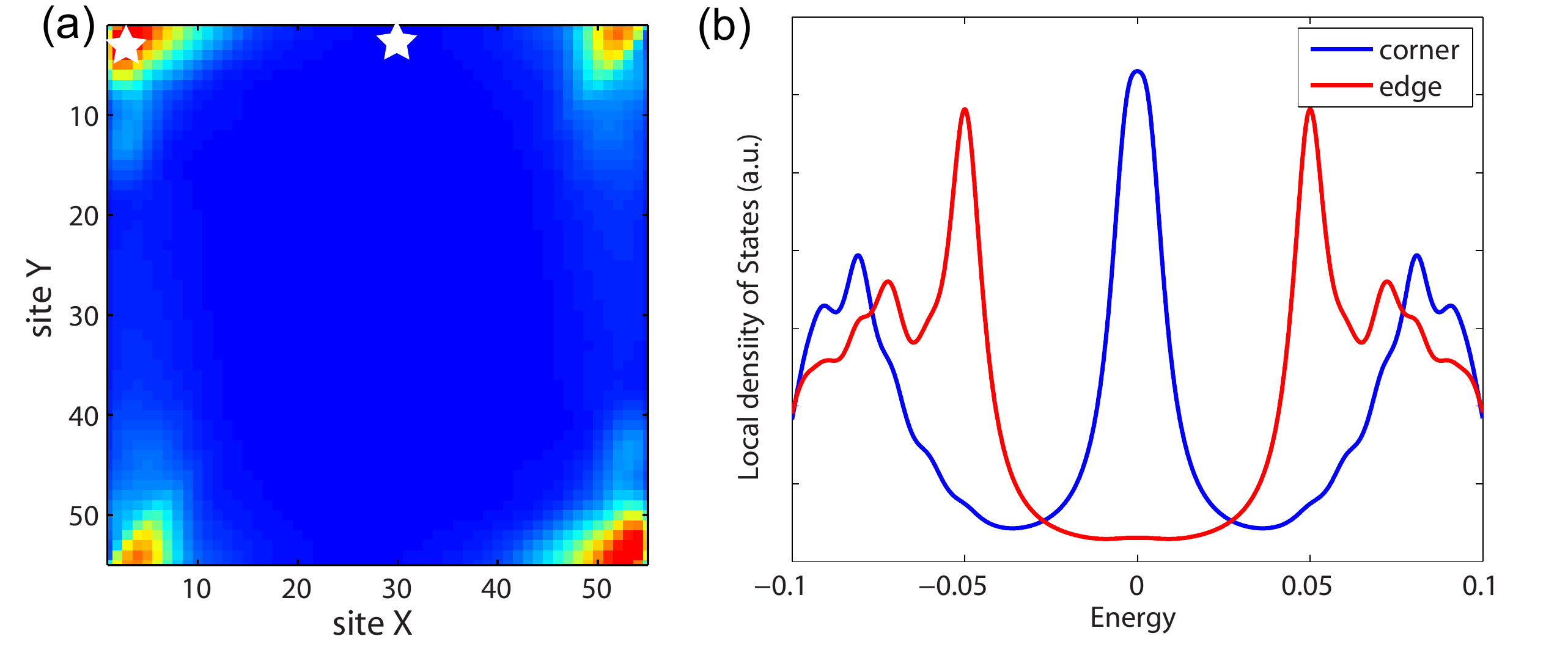}}
\caption{(color online) (a) Spatial profile of Majorana corner modes with $65\times65$ lattice sites. (b) The local density of states at corners and edges (positions labelled by stars in (a)).
 \label{ldos} }
\end{figure}

\end{widetext}

\end{document}